\documentclass{article}

\usepackage{epsf,multicol,ifthen}
\usepackage[dvips]{graphicx}

\title{Bremsstrahlung during $\alpha$-decay: \\
quantum multipolar model}

\author{Sergei~P.~Maydanyuk\thanks{E-mail: maidan@kinr.kiev.ua} \\
\small\emph{Institute for Nuclear Research, National Academy of Science of Ukraine} \\
\small\emph{47, prosp. Nauki, Kiev, 03680, Ukraine}}

\date{\small\today}

\begin{document}

\maketitle
%-----------------------------------------------------------------------------------------------------------------------

%-----------------------------------------------------------------------------------------------------------------------
\begin{abstract}
In this paper the improved multipolar model of bremsstrahlung accompanied the $\alpha$-decay is presented. The angular formalism of calculations of the matrix elements, being enough complicated component of the model, is stated in details. A new definition of the angular (differential) probability of the photon emission in the $\alpha$-decay is proposed where direction of motion of the $\alpha$-particle outside (with its tunneling inside barrier) is defined on the basis of angular distribution of its spacial wave function. In such approach, the model gives values of the angular probability of the photons emission in absolute scale, without its normalization on experimental data. Effectiveness of the proposed definition and accuracy of the spectra calculations of the bremsstrahlung spectra are analyzed in their comparison with experimental data for the $^{210}{\rm Po}$, $^{214}{\rm Po}$, $^{226}{\rm Ra}$ and $^{244}{\rm Cm}$ nuclei, and for some other nuclei predictions are performed (in absolute scale). With a purpose to find characteristics taking influence on the bremsstrahlung probability strongly, a dependence of the bremsstrahlung probability on effective charge of the decaying system is analyzed. As a natural result, as supposed type of decaying system the emission of proton from nucleus is studied, for which the effective charge is essentially larger in a comparison with $\alpha$-decay. For some proton emitters estimations of the bremsstrahlung probability are obtained (at first time, in fully quantum approach). Also the bremsstrahlung in fission of the $^{252}{\rm Cf}$ nucleus is analyzed (at first time, in fully quantum approach).

\end{abstract}
%-----------------------------------------------------------------------------------------------------------------------

%-----------------------------------------------------------------------------------------------------------------------
{\bf PACS numbers:}
23.60.+e, % Alpha decay
41.60.-m, % Radiation by moving charges
23.20.Js, % Multipole matrix elements (in electromagnetic transitions)
03.65.Xp, % Tunneling, traversal time, quantum Zeno dynamics
27.80.+w  % A is greater than or equal to 190 and is less than or equal
%      to 219 (properties of specific nuclei listed by mass ranges)


% \arxiv{nucl-th/0408022}
%-----------------------------------------------------------------------------------------------------------------------

%-----------------------------------------------------------------------------------------------------------------------
% \newpage
% \tableofcontents
% *******************************************************************************************************************

% *******************************************************************************************************************
% \newpage
\section{Introduction
\label{introduction}}

For last two decades many experimental and theoretical efforts have been made to investigate on the nature of the bremsstrahlung emission accompanying $\alpha$-decay of heavy nuclei. A key idea of such researches consists in finding a method of extraction of a new information about dynamics of $\alpha$-decay (and dynamics of tunneling) from measured bremsstrahlung spectra.

The first paper \cite{Batkin.1986.SJNCA} at this topic was published by \emph{I.~S.~Batkin, I.~V.~Kopytin and T.~A.~Churakova} as far back as in 1986 where a general quantum-mechanical formalism of the calculation of the bremsstrahlung spectra in the $\alpha$-decay was constructed and the bremsstrahlung spectrum for $^{210}\mbox{Po}$ was estimated in the relative scale for photons energies up to 200 keV (until the fulfillment of the first experiments).
In 1991 \emph{C.~J.~Luke, C.~A.~Gossett and R.~Vandenbosch} \cite{Luke.1991.PRC} measured $\gamma$-rays for spontaneous fission of the $^{252}{\rm Cm}$ nucleus and established upper limit for photon emission.
In 1994 \emph{A.~D'Arrigo, N.~V.~Eremin, G.~Fazio, G.~Giardina et al.} \cite{D'Arrigo.1994.PHLTA} at first time measured bremsstrahlung emission in $\alpha$-decay obtaining spectra for the $^{214}{\rm Po}$ and $^{226}{\rm Ra}$ nuclei. They analyzed the spectra in frameworks of \emph{instant accelerated models} constructed on the basis of classical electrodynamics with a purpose to study dynamics of $\alpha$-decay.
In 1996 \emph{M.~I.~Dyakonov and I.~V.~Gornyi} in~\cite{Dyakonov.1996.PRLTA} studied bremsstrahlung processes during tunneling of electromagnetic charge inside the potential barrier. They constructed a main basis of semiclassical approach for calculation of the spectra of bremsstrahlung during $\alpha$-decay.
In 1997 \emph{J.~Kasagi, H.~Yamazaki, N. Kasajima, T.~Ohtsuki and H.~Yuki} in~\cite{Kasagi.1997.JPHGB,Kasagi.1997.PRLTA} published the first measurements of the bremsstrahlung accompanying $\alpha$-decay of the $^{210}{\rm Po}$ and $^{244}{\rm Cm}$ nuclei using Ge detectors. They analyzed the spectra on the basis of calculations of the emission probability in the frameworks of model of M.~I.~Dyakonov and I.~V.~Gornyi. They found that the bremsstrahlung probability is mach smaller then estimated by Coulomb accelerated model. A natural and important conclusion from such estimation was the amplitude of emission of photons from the barrier cannot be neglected, that opens importance of direct quantum study of photon emission during tunneling.

In 1998 \emph{T.~Papenbrock and G.~F.~Bertsch} in~\cite{Papenbrock.1998.PRLTA} constructed
directly quantum model of calculation of the bremsstrahlung spectra in the $\alpha$-decay developed on the basis of quantum electrodynamics with use of perturbation theory. Here, wave function (current) of photons is described in the dipole approximation, matrix element is calculated with application of the \emph{Fermi golden rule}. They estimated the bremsstrahlung spectrum for $^{210}{\rm Po}$. This paper essentially improves the previous formalism in \cite{Batkin.1986.SJNCA} and its important achievement is in a possibility to calculate bremsstrahlung spectra with enough good accuracy. And further it was often used as a main basis by other independent research groups.

In 1999 \emph{N.~Takigawa, Y.~Nozawa, K.~Hagino, A.~Ono and D.~M.~Brink} \cite{Takigawa.1999.PHRVA} performed quantum analysis of the bremsstrahlung in the $\alpha$-decay of $^{210}{\rm Po}$ on the basis of semiclassical approach. Introducing idea of mixed region, they studied the contribution from it, from tunneling, outside barrier regions and from the wall of the inner potential well to the total spectrum. They analyzed a possibility to eliminate the ambiguity in the nuclear potential between the $\alpha$-particle and daughter nucleus using the bremsstrahlung spectrum.
Practically, at the same time but independently \emph{E.~V.~Tkalya} \cite{Tkalya.1999.JETP,Tkalya.1999.PHRVA} constructed the multipolar approach in frameworks of the fully quantum model proposing three different ways of calculation of the bremsstrahlung spectra. Such approach opens a possibility to study in this problem the next corrections of bremsstrahlung after dipole one being more accurate in a comparison with the dipole quantum approach of T.~Papenbrock and G.~F.~Bertsch. Calculating spectra for $^{210}{\rm Po}$, $^{214}{\rm Po}$ and $^{226}{\rm Ra}$, he at first time estimated the multipole correction E2.
\emph{C.~A.~Bertulani, D.~T.~de~Paula and V.~G.~Zelevinsky} in \cite{Bertulani.1999.PHRVA} studied tunneling of $\alpha$-particle by directly solving the non-stationary Schr\"{o}dinger equation and on such basis obtained the bremsstrahlung spectrum. They found the large deviation from the classical bremsstrahlung spectrum.
M.~I.~Dyakonov in \cite{Dyakonov.1999.PHRVA} found a classical formula of the bremsstrahlung spectrum in $\alpha$-decay and showed that taking tunneling into account is crucial in accurate description of the bremsstrahlung.
In 2000 \emph{W.~So and Y.~Kim}, calculating the probability of the photons emitted in $\alpha$-decay of the $^{210}{\rm Po}$, $^{214}{\rm Po}$ nuclei by the approach of T.~Papenbrock and G.~F.~Bertsch and at first time predicting the spectra for the $^{212}{\rm Po}$, $^{220}{\rm Rn}$, $^{240}{\rm Cm}$, $^{248}{\rm Cf}$ nuclei (in absolute scale), the first studied a dependence of the bremsstrahlung probability on the $\alpha$-particle energy and the daughter nuclear charge \cite{So_Kim.2000.JKPS}. Such analysis allows to consider deeper the $\alpha$-decay and can be useful in search of nuclei with the maximal bremsstrahlung that can be useful in future experimental study of bremsstrahlung in $\alpha$-decay.

% The emission probabilities for bremsstrahlung at low photon energy agree with those of classical calculations, but differ by an order of magnitude at high photon energy.
% We are able to show that the imaginary part (contribution of the bremsstrahlung from the classical acceleration in the Coulomb field) is dominant to the probability of photon emission.

In~\cite{Eremin.2000.PRLTA,Kasagi.2000.PRLTA} discussions between both experimental groups concerning measurements of the bremsstrahlung spectrum for the $^{210}{\rm Po}$ nucleus were published. Here, a main problem was in the following:
1) spectrum obtained for $^{210}{\rm Po}$ in \cite{Kasagi.1997.JPHGB} is smaller essentially in a comparison with spectra obtained for $^{226}{\rm Ra}$ and $^{214}{\rm Po}$ in \cite{D'Arrigo.1994.PHLTA}; 2) unclear difference between the spectra trends is observed:
while spectra for $^{226}{\rm Ra}$ and $^{214}{\rm Po}$ in \cite{D'Arrigo.1994.PHLTA} are monotonous, spectrum for $^{210}{\rm Po}$ in \cite{Kasagi.1997.JPHGB} has non-monotonous trend with a ``hole'' at 275 keV. The authors found explanation lied in the different values of angle between photon emission and $\alpha$-particle motion used in experiments.
Such discussions had caused an increased interest. If the difference between the bremsstrahlung angular spectra will be confirmed then \emph{the bremsstrahlung intensity should depend on the directions of the photons emission and the $\alpha$-particle motion}. So, a three-dimensional picture of the $\alpha$-decay with the bremsstrahlung in the spatial region of nuclear boundaries has been devised. However, if the bremsstrahlung varies visibly with changing of the angle value, then one can suppose that the photon emission is able to influence essentially on the $\alpha$-decay dynamics and, therefore, to change all its characteristic. From this point of view, the discussions \cite{Eremin.2000.PRLTA,Kasagi.2000.PRLTA} open a way for obtaining a new information about the $\alpha$-decay --- through the \emph{angular analysis of the bremsstrahlung during the $\alpha$-decay}. In such idea we see the first possible indication to study influence of nuclear deformation on the bremsstrahlung spectra and in this direction it can be interesting to study bremsstrahlung in $\alpha$-decay of strongly deformed nuclei. But for such researches a model describing the bremsstrahlung in the $\alpha$-decay, which takes into account the angle between the directions of the $\alpha$-particle propagation (or tunneling) and the photon emission and is based on the realistic $\alpha$-nucleus potential, is needed.

In~\cite{Misicu.2001.JPHGB,Dijk.2003.FBSSE} the next developments in non-stationary description of the $\alpha$-decay with the accompanying bremsstrahlung and calculations were presented.
For this time, one can refer to papers \cite{Serot.1994.NUPHA,Dijk.1999.PRLTA,Dijk.2002.PHRVA,Ivlev.2004.PHRVA} with study of dynamics of tunneling in the $\alpha$-decay; an effect, opened in \cite{Flambaum.1999.PRLTA} and named as \emph{M\"{u}nchhausen effect} which increases the barrier penetrability due to charged-particle emission during its tunneling and which can be interesting for further study of the photon bremsstrahlung during tunneling in the $\alpha$-decay.

% In~\cite{Kurgalin.2004} the first calculations of the bremsstrahlung spectra with realistic $\alpha$-decay barriers were fulfilled.

In~\cite{Maydanyuk.2003.PTP} we developed a multipolar quantum method with taking into account the magnetic component M1. While here there was a difficulty in obtaining convergence in computer calculations, we at first time found a way to study the bremsstrahlung angular distribution in $\alpha$-decay in fully quantum approach.
It turns out that computer calculations of the angular bremsstrahlung spectra are essentially more complicated, if to pass from the $\alpha$-nucleus potential used in~\cite{Papenbrock.1998.PRLTA,Tkalya.1999.PHRVA,Maydanyuk.2003.PTP} to nuclear realistic one.
In \cite{Maydanyuk.2006.EPJA} we constructed a new quantum approach for the calculations of the angular bremsstrahlung spectra where the angle was determined on the basis of a simple idea (started in \cite{Maydanyuk.nucl-th.0404013}) and where at first time the realistic nuclear component of the $\alpha$-nucleus potential was used (in the form~\cite{Denisov.2005.PHRVA}).
With a resolution of the divergence problem (which plays a key role in obtaining of the reliable spectra) by such approach we estimated the spectrum for $^{210}{\rm Po}$ obtaining a little difference with results~\cite{Papenbrock.1998.PRLTA,Tkalya.1999.PHRVA}, and as we find such estimation is in the best agreement (in fully quantum approach) with later obtained experimental data~\cite{Boie.2007.PRL} for this nucleus (see Fig.~1 in~\cite{Maydanyuk.2006.EPJA}, Fig.~5 in~\cite{Boie.2007.PRL}, Fig.~3 in~\cite{Maydanyuk.2008.EPJA}).

Then in~\cite{Amusia.2007.JETP} a new mechanism of formation of electromagnetic radiation that accompanied $\alpha$-decay and was associated with the emission of photons by electrons of atomic shells due to the scattering of $\alpha$-particles by these atoms (polarization bremsstrahlung) was proposed by \emph{M.~Ya.~Amusia, B.~A.~Zon and I.~Yu.~Kretinin}. It has shown that when the photon energy is no higher than the energy of K electrons of an atom, polarized bremsstrahlung makes a significant contribution to the bremsstrahlung in $\alpha$-decay.

With a purpose to study deeper the bremsstrahlung emitted in the $\alpha$-decay of $^{210}{\rm Po}$ discussed in \cite{Eremin.2000.PRLTA,Kasagi.2000.PRLTA}, its new high-statistics measurement \cite{Boie.2007.PRL} was performed for the photon energies up to about 500 keV. Authors found the measured differential probability in good agreement with theoretical results obtained within the semiclassical approximation as well as with the exact quantum mechanical calculation. Here, it was shown that due to a small effective electric dipole charge of radiating system a significant interference between the electric dipole and quadrupole contributions occurs.
With a purpose to find a unified description that incorporates both the radiation during the tunneling through the Coulomb wall and the finite energy $E_{\gamma}$ of the radiated photon up to $E_{\gamma} Q_{\alpha}/\sqrt{\nu}$ (where $Q_{\alpha}$ is the $\alpha$-decay Q-value and $\nu$ is the Sommerfeld parameter), this group proposed in \cite{Jentschura.2008.PRC} a semiclassical theory of $\alpha$-decay accompanied by the bremsstrahlung with a special emphasis on study of $^{210}{\rm Po}$. The corrections with respect to previous semiclassical investigations were found to be substantial, and good agreement with a full quantum mechanical treatment~\cite{Papenbrock.1998.PRLTA} was achieved.
Here, authors found that a dipole-quadrupole interference significantly changes the $\alpha$-$\gamma$ angular correlation, and obtained good agreement between their theoretical predictions and experimental results.

In \cite{Maydanyuk.2008.EPJA} the bremsstrahlung emission has been measured by the $\alpha-\gamma$ coincidence to investigate on the $\alpha$-decay dynamics of the $^{214}{\rm Po}$ nucleus. This experiment was performed using the $^{226}{\rm Ra}$ source and the apparatus with Si-detector for $\alpha$-particles and NaI(Tl)-detector able to collect photons with energies up to about 1 MeV. The experimental data with quantum mechanical calculations are found in the good agreement between theory and experiment for the photon energies up to 765 keV. At present day, this result is in the best agreement between theory and experiment in such problem. Here, at the first time a presence of slight oscillations in the experimental bremsstrahlung spectrum is established.
In \cite{Maydanyuk.2008.MPLA} the spectrum of probability of the bremsstrahlung emission accompanying the $\alpha$-decay of $^{226}{\rm Ra}$ by measuring the $\alpha$-$\gamma$ coincidences  and using the model presented in previous study on the $\alpha-$decay of $^{214}{\rm Po}$ was published. These experimental data are found to be in a good agreement with the quantum mechanical calculations of this group. The differences between the photon spectra connected with the $\alpha$-decay of the $^{226}{\rm Ra}$ and $^{214}{\rm Po}$ nuclei was explained. For two mentioned nuclei the bremsstrahlung emission contributions from the tunneling and external regions into the total spectrum are estimated, and we established the destructive interference between these contributions. A phenomenon of the emission of the bremsstrahlung photons during tunneling of the $\alpha$-particle has been established (such a phenomenon has been confirmed at first time experimentally for the $\alpha$-decay and theoretically with taking into account the realistic $\alpha$-nucleus potential).

%-----------------------------------------------------------------------------------------------------------------------

%-----------------------------------------------------------------------------------------------------------------------
It needs to note that arsenal of experimental data is not rich. A serious difficulty in obtaining of desirable accuracy in measurements lies in small values of the photon emission probability. It is not clear which nucleus (from available ones) should be used in order to obtain the maximal probability and performance of experiment will be easier. One can think that theoretical estimations of the bremsstrahlung spectra should help with finding of nuclei and types of decays where the emission of photons will be maximal (see~\cite{So_Kim.2000.JKPS}).
In aspect of development of the models, a main difficulty in obtaining of reliable values of the probability of the photon emission is concerned with slow convergence in calculations of matrix elements, caused by slowly damping behavior of their integrant functions. A desire to take correctly into account nuclear component of the $\alpha$-nucleus potential, which parameters are determined on the basis of the given nucleus, intensifies this problem. Overcoming of such difficulty gives enough large time of calculations, that often makes further analysis of results hard. This have caused a necessity to revise algorithms of calculations of wave functions and angular formalism of the matrix elements in the multipolar approach~\cite{Maydanyuk.2003.PTP}.

In this paper the improved multipolar model of bremsstrahlung accompanied the $\alpha$-decay is presented. The angular formalism of calculations of the matrix elements, being enough complicated component of the model, is stated in details. A new definition of the angular (differential) probability of the photon emission in the $\alpha$-decay is proposed where direction of motion of the $\alpha$-particle outside (with its tunneling inside barrier) is defined on the basis of angular distribution of its spacial wave function. In such approach, the model gives values of the angular probability of the photons emission in absolute scale, without its normalization on experimental data. Effectiveness of the proposed definition and accuracy of the spectra calculations of the bremsstrahlung spectra are analyzed in their comparison with experimental data for the $^{210}{\rm Po}$, $^{214}{\rm Po}$, $^{226}{\rm Ra}$ and $^{244}{\rm Cm}$ nuclei, and for some other nuclei predictions are performed (in absolute scale). With a purpose to find characteristics taking influence on the bremsstrahlung probability strongly, a dependence of the bremsstrahlung probability on effective charge of the decaying system is analyzed. As a natural result, as supposed type of decaying system the emission of proton from nucleus is studied, for which the effective charge is larger in a comparison with $\alpha$-decay. For some proton emitters estimations of the bremsstrahlung probability are obtained (at first time, in fully quantum approach). Also the bremsstrahlung in fission of the $^{252}{\rm Cf}$ nucleus is analyzed (at first time, in fully quantum approach).

% *******************************************************************************************************************

% *******************************************************************************************************************
\section{Motion of electromagnetic charge inside field of nucleus
\label{sec.2}}

Let's consider a particle with mass $m$ moving (with possible tunneling) inside a field of nucleus with potential $U(\mathbf{r})$. Hamiltonian of such a system is:
\begin{equation}
  \hat{H}_{0} =
    \displaystyle\frac{\mathbf{\hat{p}}^{2}}{2m} + U(\mathbf{r}) =
    - \displaystyle\frac{\hbar^{2}}{2m}\: \triangle + U(\mathbf{r}).
\label{eq.2.1.1}
\end{equation}
If the particle is electrically charged then it is under action of the electromagnetic field with vector potential $\mathbf{A}(\mathbf{r}, t)$ of such nucleus. In this case, we have (see~\cite{Benedetti.1968}, p.~187):
\begin{equation}
  \hat{H} =
    \displaystyle\frac{1}{2m}\:
    \Bigl(\mathbf{\hat{p}} - Z_{\rm eff} \displaystyle\frac{e}{c}\, \mathbf{A}(\mathbf{r},t) \Bigr)^{2} -
      Z_{\rm eff} \displaystyle\frac{e\hbar}{2mc} \;
        {\mathbf\mu} \times {\bf rot}\,\mathbf{A}(\mathbf{r},t) +
      U(\mathbf{r}) =
    \hat{H}_{0} + \hat{W}
\label{eq.2.1.2}
\end{equation}
where
\begin{equation}
  \hat{W} =
    - Z_{\rm eff}\: \displaystyle\frac{e}{2mc}\; \Bigl(\mathbf{\hat{p} A} + \mathbf{A \hat{p}}\Bigr) -
      Z_{\rm eff}\: \displaystyle\frac{e\hbar}{2mc} \; \mathbf{\bf \mu} \times {\bf rot} \,\mathbf{A} +
      Z_{\rm eff}^{2}\, \displaystyle\frac{e^{2}}{2mc^{2}}\; \mathbf{A}^{2}.
\label{eq.2.1.3}
\end{equation}
Here, $Z_{\rm eff}$ is effective charge of the composite system (\emph{particle--nucleus}), $\mu$ is magnetic moment (we assume that it is very small, $\mu \to 0$). Taking into account \emph{Coulomb calibration} (${\rm div}\,\mathbf{A} = 0$), and neglecting item at $\mathbf{A}^{2}/c^{2}$, we obtain form of the operator $\hat{W}$:
\begin{equation}
  \hat{W} = - Z_{\rm eff}\, \displaystyle\frac{e}{mc}\; \mathbf{A \hat{p}}.
\label{eq.2.1.4}
\end{equation}
% *******************************************************************************************************************

% *******************************************************************************************************************
\subsection{Theory of perturbations of quasistationary states
\label{sec.2.2}}

We shall study $\alpha$-decay of nucleus. The decaying nucleus we shall consider as a \emph{composite system: $\alpha$-particle and daughter nucleus}. Let $\Psi_{k}^{0}(t)$ be wave function (WF) of the quasistationary state of this system with decay at level $E_{k}$, $\hat{H}_{0}(t)$ be hamiltonian of the system. Write:
\begin{equation}
  i \hbar\: \displaystyle\frac{\partial \Psi_{k}^{(0)}(t)}{\partial t} =
  \hat{H}_{0}(t)\: \Psi_{k}^{(0)}(t).
\label{eq.2.2.1}
\end{equation}
A solution of the non-stationary wave function (satisfying Schr\"{o}diger equation at $E_{k}$) can be written so:
\begin{equation}
  \Psi^{(0)}(t) = \int\limits_{k_{min}} a_{k} \Psi_{k}^{(0)}(t)\; dk
\label{eq.2.2.2}
\end{equation}
where $\Psi_{k}^{(0)}(t)$ and $a_{k}$ are non-stationary wave function and weight amplitude for the decay at level $E_{k}$.

$\alpha$-particle during its motion inside electro-magnetic field of the daughter nucleus emits photons. We assume that this process is possible also during tunneling of the $\alpha$-particle. Process of the photon emission changes total energy of the studied system and is considered as perturbation of such system. We shall study a spontaneous emission of photons, i.~e. when the field acts on the $\alpha$-particle till the emission of the first photon. \emph{Therefore, we deals with the perturbation acting on the system during \underline{finite period of time}. Here, time moments of beginning and finishing of the action of perturbation on the system are defined by such beginning and finishing of time interval, when we assume that the photon emission by $\alpha$-particle is possible}. Write the total hamiltonian with perturbation so:

\begin{equation}
  \hat{H}(t) = \hat{H}_{0}(t) + \hat{W}(t)
\label{eq.2.2.3}
\end{equation}
where $\hat{W}(t)$ is operator of perturbation dependent on time.

The problem lays in approximated determination of new wave functions $\Psi(t)$ on the basis of wave functions $\Psi^{(0)}_{k}(t)$ of quasistationary states of the unperturbed system. We shall be looking for the unknown wave functions solving the non-stationary Schr\"{o}dinger equation with the perturbed hamiltonian
\begin{equation}
  i \hbar \:\displaystyle\frac{\partial \Psi(t)}{\partial t} = \bigl(\hat{H}_{0}(t) + \hat{W}(t)\bigr)\: \Psi(t)
\label{eq.2.2.4}
\end{equation}
in the following form:
\begin{equation}
  \Psi(t) = \int\limits_{k_{min}} a_{k}(t)\, \Psi_{k}^{(0)}(t)\; dk
\label{eq.2.2.5}
\end{equation}
where coefficients $a_{k}(t)$ are already functions of time (like sec.~40 in~\cite{Landau.v3.1989}, p.~177). Substituting (\ref{eq.2.2.5}) into (\ref{eq.2.2.4}), we have:
\begin{equation}
  i \hbar \int\limits_{k_{min}} \displaystyle\frac{\partial a_{k}(t)}{\partial t} \Psi_{k}^{(0)}\; dk +
  i \hbar \int\limits_{k_{min}} a_{k}(t) \displaystyle\frac{\partial \Psi_{k}^{(0)}}{\partial t}\; dk =
  \hat{H}_{0} \int\limits_{k_{min}} a_{k}(t) \Psi_{k}^{(0)}\; dk +
  \hat{W}(t) \int\limits_{k_{min}} a_{k}(t) \Psi_{k}^{(0)}\; dk.
\label{eq.2.2.6}
\end{equation}
Taking into account (\ref{eq.2.2.1}), we obtain:
\begin{equation}
  i \hbar \int\limits_{k_{min}}  \Psi_{k}^{(0)}\, \displaystyle\frac{\partial a_{k}(t)}{\partial t}\; dk =
  \int\limits_{k_{min}} \hat{W}(t)\; a_{k}(t)\, \Psi_{k}^{(0)}\; dk.
\label{eq.2.2.7}
\end{equation}
Here, we move operator $\hat{W}(t)$ under the integral, using \emph{property of linearity (superposition)} of its action on the set of functions $\Psi_{k}^{(0)}$ with different $k$. Assuming the weight amplitudes $a_{k}(t)$ to be dependent on time, we come to:
\begin{equation}
  i \hbar \int\limits_{k_{min}}  \Psi_{k}^{(0)}\, \displaystyle\frac{\partial a_{k}(t)}{\partial t}\; dk =
  \int\limits_{k_{min}} a_{k}(t)\; \hat{W}(t)\, \Psi_{k}^{(0)}\; dk.
\label{eq.2.2.8}
\end{equation}
Multiplying both parts of this equality by $\Psi^{(0),*}_{m}$ on the left and integrating over whole volume $\mathbf r$, we obtain:
\begin{equation}
  i \hbar \int\limits_{k_{min}}
    \biggl(\int \Psi_{m}^{(0),*} \Psi_{k}^{(0)} \mathbf{dr} \biggr) \cdot
    \displaystyle\frac{\partial a_{k}(t)}{\partial t}\; dk =
  \int\limits_{k_{min}} a_{k}(t) \cdot
    \biggl(\int \Psi_{m}^{(0),*} \hat{W}(t) \Psi_{k}^{(0)} \mathbf{dr} \biggr)\; dk.
\label{eq.2.2.9}
\end{equation}
Let's assume that the decaying system is described by such quasistationary wave functions $\Psi_{k}^{(0)}$ which are satisfied to the following condition of normalization (in continuous energy spectrum; like sec.~5 in~\cite{Landau.v3.1989}, p.~30--35):
\begin{equation}
  \int \Psi_{m}^{(0),*}(t)\, \Psi_{k}^{(0)}(t) \; \mathbf{dr} = \delta (k-m).
\label{eq.2.2.10}
\end{equation}
We rewrite (\ref{eq.2.2.9}) so:
\begin{equation}
  i \hbar\, \displaystyle\frac{\partial a_{m}(t)}{\partial t} =
  \int\limits_{k_{min}} a_{k}(t)\, W_{mk}(t)\; dk
\label{eq.2.2.11}
\end{equation}
where
\begin{equation}
  W_{mk}(t) = \int \Psi_{m}^{(0),*}\, \hat{W}(t)\, \Psi_{k}^{(0)}\; \mathbf{dr}
\label{eq.2.2.12}
\end{equation}
are matrix elements of the non-stationary perturbation.

\vspace{3mm}
As the unperturbed wave function we shall use wave function of the quasistationary $i$-state before the photon emission, which coefficients in (\ref{eq.2.2.5}) correspond to: $a_{i}^{(0)} = \delta(k-i)$ (i.~e. $a_{k}^{(0)} = 0$ at $k \ne i$). To obtain the first correction, we shall be looking for $a_{k}$ in the form $a_{k} = a_{k}^{(0)} + a_{k}^{(1)}$ where we substitute $a_{k} = a_{k}^{(0)}$ into the right part of the equation (\ref{eq.2.2.11}) (already having values $W_{mk}$). This gives:
\begin{equation}
  i \hbar\, \displaystyle\frac{\partial a_{k}^{(1)}(t)}{\partial t} = W_{ki}(t).
\label{eq.2.2.13}
\end{equation}
To point out, which of the unperturbed functions the correction is calculated to, we add the second index to the coefficients $a_{k}$:
\begin{equation}
  \Psi_{i}(t) = \int\limits_{k_{min}} a_{ki}(t)\, \Psi_{k}^{(0)}(t)\; dk.
\label{eq.2.2.14}
\end{equation}
So, we write result of integration of (\ref{eq.2.2.13}) in the form:
\begin{equation}
  a_{ki}^{(1)} (t) = - \displaystyle\frac{i}{\hbar} \displaystyle\int W_{ki}(t) \; dt.
\label{eq.2.2.15}
\end{equation}
By such weight amplitude wave functions in the first correction are determined.
% *******************************************************************************************************************

% *******************************************************************************************************************
% \newpage
\subsection{Matrix element of transition
\label{sec.2.3}}

Let's consider wave packets (WPs) of the form:
\begin{equation}
  \Psi_{i, f} (\mathbf{r}, t) = \int\limits_{0}^{+\infty}
    g(k - k_{i,f}) \,\psi_{i,f}(k, \mathbf{r}) \: e^{-iw(k)t} \; dk.
\label{eq.2.3.1}
\end{equation}
We shall use them as definitions for non-stationary wave functions in the initial and final states. For such packets we have the following properties:
\begin{equation}
\begin{array}{l}
  e^{-i\hat{H}_{0}t} \,\psi_{i}(\mathbf{r}) = \psi_{i}(\mathbf{r})\, e^{-iw_{i}t}, \\
  \psi_{f}^{*}(\mathbf{r}) \: e^{i\hat{H}_{0}t} = \psi_{f}^{*}(\mathbf{r})\, e^{iw_{f}t}.
\end{array}
\label{eq.2.3.2}
\end{equation}

Now we shall assume that the system in the initial and final states is characterized by a number of photons emitted. For accurate description, we come from consideration of the system without emission, which is included into the hamiltonian as a perturbation by the external field, to consideration of the system with presence of emission where the initial and final states are defined by wave function already dependent on numbers of photons. As before, we define the matrix element of transition $i \to f$ of this system as the first correction of perturbation (\ref{eq.2.2.15}) (like sec.~41--42 in~\cite{Landau.v3.1989}):
\begin{equation}
  a_{fi}(t) =
    - \displaystyle\frac{i}{\hbar}
    \displaystyle\int\limits_{t_{0}}^{t}
    \bigl< k_{f}, n_{k}+1 \bigl| \hat{W}(\mathbf{r}, t')| k_{i}, n_{k} \bigr>\;  dt'
\label{eq.2.3.3}
\end{equation}
where $\Psi_{i}(\mathbf{r}, t) = |k_{i}\bigr>$ and $\Psi_{f}(\mathbf{r}, t) = |k_{f}\bigr>$ are non-stationary wave functions in the initial $i$-state and final $f$-state which do not contain number of photons emitted, $\hbar w = E$, $n_{k}$ is number of photons of one sort with impulse $\mathbf{k}$ in the initial $i$-state.
Operator of interaction $\hat{W}$ has the form (\ref{eq.2.1.4}), $Z_{\rm eff}$ is effective charge of the system, $m$ is reduced mass of the system and $\mathbf{A}$ is vector potential of the electro-magnetic field of the daughter nucleus.
If $n_{k}=0$ than emission is named as \emph{spontaneous}. If $n_{k}>0$ than emission is named as \emph{induced}. A presence of photons in the initial state stimulates additional emission of other photons of the same sort (this follows from formula of the emission probability). The matrix element (\ref{eq.2.3.3}) is defined in the first approximation of the perturbation theory. One can develop formalism simpler in the system of units where $\hbar = 1$ and $c = 1$, but we shall write constants $\hbar$ and $c$ explicitly. Rewrite the matrix element $a_{fi}$ so:
\[
\begin{array}{lcl}
  & & a_{fi}(t) =
    - \displaystyle\frac{i}{\hbar}
    \displaystyle\int\limits_{t_{0}}^{t}
    \biggl<
      \Psi_{f} (\mathbf{r}, t'),\;  n_{k}+1
    \biggl|\,
      \hat{W}(\mathbf{r}, t')\,
    \biggr|\,
      \Psi_{i} (\mathbf{r}, t'),\; n_{k}
    \biggr> \;  dt' = \\

  & = &
    - \displaystyle\frac{i}{\hbar}
    \displaystyle\int\limits_{t_{0}}^{t}
    \biggl<
      \displaystyle\int\limits_{0}^{+\infty} g(k_{2} - k_{f}) \,\psi_{f}(k_{2}, \mathbf{r})\:
      e^{-iw(k_{2})t}\; dk_{2},\;  n_{k}+1
    \biggl|\,
      \hat{W}(\mathbf{r}, t')
    \biggr|
      \displaystyle\int\limits_{0}^{+\infty} g(k_{1} - k_{i}) \,\psi_{i}(k_{1}, \mathbf{r})\:
      e^{-iw(k_{1})t}\; dk_{1},\; n_{k}
    \biggr> \;  dt' = \\

  & = &
    - \displaystyle\frac{i}{\hbar}
    \displaystyle\int\limits_{0}^{+\infty} dk_{2}
    \displaystyle\int\limits_{0}^{+\infty} dk_{1}\:
      g^{*}(k_{2} - k_{f})\: g(k_{1} - k_{i}) \cdot
    \biggl<
      \psi_{f}(k_{2}, \mathbf{r}),\;
      n_{k}+1
    \biggl|\,
      \displaystyle\int\limits_{t_{0}}^{t}
      e^{iw(k_{2})t'}\:
      \hat{W}(\mathbf{r}, t')\:
      e^{-iw(k_{1})t'}\;
      dt'
    \biggr|\,
      \psi_{i}(k_{1}, \mathbf{r}),\; n_{k}
    \biggr> = \\

  & = &
    \displaystyle\int\limits_{0}^{+\infty} dk_{2}
    \displaystyle\int\limits_{0}^{+\infty} dk_{1}\:
      g^{*}(k_{2} - k_{f})\: g(k_{1} - k_{i}) \cdot
    \biggl<
      f,\: k_{2},\; n_{k}+1
    \biggl|\,
      - \displaystyle\frac{i}{\hbar}
      \displaystyle\int\limits_{t_{0}}^{t}  e^{iw(k_{2})t'}\:  \hat{W}(\mathbf{r}, t')\:  e^{-iw(k_{1})t'}\; dt'
    \biggr|\,
      i,\: k_{1},\; n_{k}
    \biggr>
\end{array}
\]
or
\begin{equation}
\begin{array}{l}
  a_{fi}(t) =
  \displaystyle\int\limits_{0}^{+\infty} dk_{2}
  \displaystyle\int\limits_{0}^{+\infty} dk_{1}\cdot
  g^{*}(k_{2}-k_{f}) \, g(k_{1}-k_{i}) \cdot
  \bigl< k_{2}, n_{k}+1 \bigl| \,\tilde{W}(\mathbf{r}, t) \,\bigr|\, k_{1}, n_{k} \bigr>
\end{array}
\label{eq.2.3.4}
\end{equation}
where
\begin{equation}
  \tilde{W} (\mathbf{r}, t) =
    -\displaystyle\frac{i}{\hbar} \displaystyle\int\limits_{t_{0}}^{t}
    e^{iw_{2}t'} \,\hat{W}(\mathbf{r}, t') \,e^{-iw_{1}t'} \; dt'.
\label{eq.2.3.5}
\end{equation}

Now let's use the following form of the vector potential $\mathbf{A}(\mathbf{r}, t)$ (see p.~22, 28 in~\cite{Berestetsky.1989}):
\begin{equation}
\begin{array}{ll}
  \mathbf{A}(\mathbf{r}, t) =
        \sum\limits_{\mathbf{k}, \alpha}
        \biggl(\hat{c}_{\mathbf{k}, \alpha} \mathbf{A}_{\mathbf{k}, \alpha} +
        \hat{c}^{+}_{\mathbf{k}, \alpha} \mathbf{A}^{*}_{\mathbf{k}, \alpha}
        \biggr), &
  \mathbf{A}_{\mathbf{k}, \alpha} =
        \sqrt{\displaystyle\frac{2\pi\hbar c^{2}}{w}}
        \mathbf{e}^{(\alpha)}e^{i(\mathbf{kr}-wt)}
\end{array}
\label{eq.2.3.6}
\end{equation}
where $\mathbf{e}^{(\alpha)}$ are unit vectors of polarization of the photon emitted, $\mathbf{k}$ is wave vector of the photon and $w = k = \bigl| \mathbf{k}\bigr|$. Vectors $\mathbf{e}^{(\alpha)}$ are perpendicular to $\mathbf{k}$ in Coulomb calibration. We have two independent polarizations $\mathbf{e}^{(1)}$ and $\mathbf{e}^{(2)}$ for the photon with impulse $\mathbf{k}$ ($\alpha=1,2$).
Taking into account (\ref{eq.2.3.6}), rewrite (\ref{eq.2.1.4}) so:
\begin{equation}
\begin{array}{ll}
  \hat{W} (\mathbf{r}, t) =
    - Z_{eff}\, \displaystyle\frac{e}{mc} \,
    \sum\limits_{k, \alpha}
      \sqrt{\displaystyle\frac{2\pi\hbar c^{2}}{w}}
        \biggl(\hat{c}_{\mathbf{k}, \alpha} \mathbf{e}^{(\alpha)} e^{i(\mathbf{kr}-wt)} +
               \hat{c}_{\mathbf{k}, \alpha}^{+} \mathbf{e}^{(\alpha),*} e^{-i(\mathbf{kr}-wt)}
        \biggr) \:
        (-i\hbar)\, \displaystyle\frac{\partial}{\partial \mathbf{r}}.
\end{array}
\label{eq.2.3.7}
\end{equation}
Substituting this expression into (\ref{eq.2.3.5}), we obtain:
\begin{equation}
\begin{array}{ccl}
  \tilde{W} (\mathbf{r}, t) & = &
    Z_{eff}\, \displaystyle\frac{e}{mc} \,
    \sum\limits_{k, \alpha}
    \sqrt{\displaystyle\frac{2\pi\hbar c^{2}}{w}} \cdot
    \displaystyle\int\limits_{t_{0}}^{t}
      e^{iw_{2}t'} \;
        \biggl(\hat{c}_{\mathbf{k}, \alpha} \mathbf{e}^{(\alpha)} e^{i(\mathbf{kr}-wt)} +
               \hat{c}_{\mathbf{k}, \alpha}^{+} \mathbf{e}^{(\alpha),*} e^{-i(\mathbf{kr}-wt)}
        \biggr) \;
        e^{-iw_{1}t'} \; dt' \cdot
        \displaystyle\frac{\partial}{\partial \mathbf{r}}.
\end{array}
\label{eq.2.3.8}
\end{equation}
The transition matrix element transforms to:
\begin{equation}
\begin{array}{l}
  \bigl< k_{2}, n_{k}+1 \bigl| \,\tilde{W}(t) \,\bigr| \, k_{1}, n_{k} \bigr> =

  - \displaystyle\frac{i}{\hbar}\,
  \Biggl< k_{2}, n_{k}+1 \Biggl| \,
    \displaystyle\int\limits_{t_{0}}^{t} e^{iw_{2}t'} \, \hat{V} (\mathbf{r}, t') \,
    e^{-iw_{1}t'} \; dt' \,\Biggr| \,k_{1}, n_{k}
  \Biggr> = \\

  = Z_{eff}\, \displaystyle\frac{e}{mc} \,
  \sum\limits_{k, \alpha}  \sqrt{\displaystyle\frac{2\pi\hbar c^{2}}{w}} \cdot
  \Biggl< k_{2}, n_{k}+1 \Biggl| \,
    \hat{c}_{\mathbf{k}, \alpha} \mathbf{e}^{(\alpha)} e^{i\mathbf{kr}}
    \displaystyle\int\limits_{t_{0}}^{t} e^{i(w_{2}-w_{1}-w)t'} \, \; dt' \;
    \displaystyle\frac{\partial}{\partial \mathbf{r}} \; + \\
    \; + \;
    \hat{c}_{\mathbf{k}, \alpha}^{+} \mathbf{e}^{(\alpha),*} e^{-i\mathbf{kr}}
    \displaystyle\int\limits_{t_{0}}^{t} e^{i(w_{2}-w_{1}+w)t'} \, \; dt' \;
    \displaystyle\frac{\partial}{\partial \mathbf{r}}
  \,\Biggr| \,k_{1}, n_{k} \Biggr>.
\end{array}
\label{eq.2.3.9}
\end{equation}
Photons are under Bose statistics, and we have:
\begin{equation}
\begin{array}{l}
\vspace{2mm}
  \bigl< n_{k}+1 \bigl| \,\hat{c}_{k,\alpha}^{+} \,\bigr| \, n_{k} \bigr> = \sqrt{n_{k}+1}, \\
  \bigl< n_{k}+1 \bigl| \,\hat{c}_{k,\alpha} \,    \bigr| \, n_{k} \bigr> = 0.
\end{array}
\label{eq.2.3.10}
\end{equation}
According to this, we exclude the first item from (\ref{eq.2.3.9}) and obtain:
\begin{equation}
  \bigl< k_{2}, n_{k}+1 \bigl| \,\tilde{W}(t) \,\bigr| \, k_{1}, n_{k} \bigr> =
  Z_{eff}\, \displaystyle\frac{e}{m} \,
  \sum\limits_{k, \alpha}  \sqrt{\displaystyle\frac{2\pi\hbar}{w}} \cdot
  \biggl< k_{2} \biggl| \,
    \mathbf{e}^{(\alpha),*} e^{-i\mathbf{kr}}
    \displaystyle\frac{\partial}{\partial \mathbf{r}}
  \,\biggr| \,k_{1} \biggr> \cdot
  \sqrt{n_{k}+1} \, \displaystyle\int\limits_{t_{0}}^{t} e^{i(w_{2}-w_{1}+w)t'} \, \; dt'.
\label{eq.2.3.11}
\end{equation}
We shall study the \emph{spontaneous emission}, i.e. emission of one photon with impulse $\mathbf{k}$, before emission of which there were no other photons of this sort ($n_{k}=0$). We omit summation by $n_{k}$:
\begin{equation}
  \bigl< k_{2}, 1 \bigl| \,\tilde{W}(t) \,\bigr| \, k_{1}, 0 \bigr> =
  Z_{eff}\, \displaystyle\frac{e}{m} \,
  \sum\limits_{\alpha=1,2}  \sqrt{\displaystyle\frac{2\pi\hbar}{w}} \cdot
  \biggl< k_{2} \biggl| \,
    \mathbf{e}^{(\alpha),*} e^{-i\mathbf{kr}}
    \displaystyle\frac{\partial}{\partial \mathbf{r}}
  \,\biggr| \,k_{1} \biggr> \cdot
  \displaystyle\int\limits_{t_{0}}^{t} e^{i(w_{2}-w_{1}+w)t'} \, \; dt'.
\label{eq.2.3.12}
\end{equation}
% *******************************************************************************************************************

% *******************************************************************************************************************
\subsection{Direct indication on a presence of oscillations in matrix element in non-stationary approach
\label{sec.2.4}}

We note a physically interesting result, following from the formula (\ref{eq.2.3.12}). If we have the perturbation acting on the system ($\alpha$-particle and daughter nucleus) during the finite period of time, then we have the finite upper limit $t$ of time integral and can use its lower limit $t_{0}=0$, and obtain (in case $w_{1} - w_{2} \ne w$):
\[
\begin{array}{lcl}
  & & \bigl< k_{2}, 1 \bigl| \,\tilde{W}(t) \,\bigr| \, k_{1}, 0 \bigr> =
    Z_{eff}\, \displaystyle\frac{e}{m} \,
    \displaystyle\sum\limits_{\alpha=1,2}  \sqrt{\displaystyle\frac{2\pi\hbar}{w}} \cdot
    \biggl< k_{2} \biggl| \,
      \mathbf{e}^{(\alpha),*} e^{-i\mathbf{kr}}
      \displaystyle\frac{\partial}{\partial \mathbf{r}}\,
    \biggr| \,k_{1} \biggr> \cdot
    \displaystyle\frac{e^{i(w_{2}-w_{1}+w)t'}}{i(w_{2}-w_{1}+w)} \bigg|_{t'=0}^{t'=t} = \\

  & = &
    Z_{eff}\, \displaystyle\frac{e}{m} \,
    \displaystyle\sum\limits_{\alpha=1,2}  \sqrt{\displaystyle\frac{2\pi\hbar}{w}} \cdot
    \biggl< k_{2} \biggl| \,
      \mathbf{e}^{(\alpha),*} e^{-i\mathbf{kr}}
      \displaystyle\frac{\partial}{\partial \mathbf{r}}\,
    \biggr| \,k_{1} \biggr> \cdot
    \displaystyle\frac{\sin{(w_{2}-w_{1}+w)\,t} - i\cos{(w_{2}-w_{1}+w)\,t} + i} {w_{2}-w_{1}+w}
\end{array}
\]
or
\begin{equation}
\begin{array}{lcl}
  & & \bigl< k_{2}, 1 \bigl| \,\tilde{W}(t) \,\bigr| \, k_{1}, 0 \bigr> =
    Z_{eff}\, \displaystyle\frac{e}{m} \,
    \displaystyle\sum\limits_{\alpha=1,2}  \sqrt{\displaystyle\frac{2\pi\hbar}{w}} \cdot
    \biggl< k_{2} \biggl| \,
      \mathbf{e}^{(\alpha),*} e^{-i\mathbf{kr}}
      \displaystyle\frac{\partial}{\partial \mathbf{r}}\,
    \biggr| \,k_{1} \biggr> \cdot
    \displaystyle\frac{i} {w_{2}-w_{1}+w}\; + \\
  & + &
    Z_{eff}\, \displaystyle\frac{e}{m} \,
    \displaystyle\sum\limits_{\alpha=1,2}  \sqrt{\displaystyle\frac{2\pi\hbar}{w}} \cdot
    \biggl< k_{2} \biggl| \,
      \mathbf{e}^{(\alpha),*} e^{-i\mathbf{kr}}
      \displaystyle\frac{\partial}{\partial \mathbf{r}}\,
    \biggr| \,k_{1} \biggr> \cdot
    \displaystyle\frac{\sin{(w_{2}-w_{1}+w)\,t} - i\cos{(w_{2}-w_{1}+w)\,t}} {w_{2}-w_{1}+w}.
\end{array}
\label{eq.2.4.1}
\end{equation}
Here, the first item is stationary function (independent of duration $t$ of the perturbation action) which usually determined completely probability of the photon emission in the $\alpha$-decay in stationary approaches (which at such definition has monotonic behavior).
But the second item is oscillating function in dependence on $w$ and it includes oscillations into the total matrix element. Period of such oscillations is connected directly with duration $t$ of the perturbation action. The probability defined on the basis of such total matrix element has both monotonic and oscillating components.
In result, \emph{we theoretically have obtained (at first time) a direct indication on a presence of oscillations in the bremsstrahlung spectra in $\alpha$-decay!} In such context, period of these oscillations has information about duration of location of the $\alpha$-particle inside the region of electromagnetic forces of the daughter nucleus.

% *******************************************************************************************************************

% *******************************************************************************************************************
\subsection{Stationary approximation
\label{sec.2.5}}

Now we use the following limits of time integrals:
\begin{equation}
\begin{array}{cc}
  t_{0} = -\infty, & t_{1} = +\infty.
\end{array}
\label{eq.2.5.1}
\end{equation}
Taking into account the property:
\begin{equation}
  \displaystyle\int\limits_{-\infty}^{+\infty} e^{i\alpha t} \: dt = 2\pi\:\delta(\alpha),
\label{eq.2.5.2}
\end{equation}
from (\ref{eq.2.3.12}) we obtain:
\begin{equation}
  \bigl< k_{2}, 1 \bigl| \,\tilde{W} \,\bigr| \, k_{1}, 0 \bigr> =
  Z_{eff}\, \displaystyle\frac{e}{m} \,
  \sum\limits_{\alpha=1,2}
  \sqrt{\displaystyle\frac{2\pi\hbar}{w}} \cdot
  \biggl< k_{2} \biggl| \,
    \mathbf{e}^{(\alpha),*} e^{-i\mathbf{kr}}
    \displaystyle\frac{\partial}{\partial \mathbf{r}}
  \,\biggr| \,k_{1} \biggr> \cdot
  2\pi\delta(w_{2}-w_{1}+w).
\label{eq.2.5.3}
\end{equation}
Using the following notations:
\begin{equation}
\begin{array}{cclcl}
  F_{fi} =
    Z_{eff}\, \displaystyle\frac{e}{m} \,
    \sqrt{\displaystyle\frac{2\pi\hbar}{w}} \cdot p(k_{i},k_{f}), &

\hspace{3mm}
  p(k_{i}, k_{f}) = \displaystyle\sum\limits_{\alpha=1,2} \mathbf{e}^{(\alpha),*} \mathbf{p}(k_{i}, k_{f}), &

\hspace{1mm}
  \mathbf{p}(k_{i}, k_{f}) =
    \biggl< k_{2} \biggl| \,  e^{-i\mathbf{kr}} \displaystyle\frac{\partial}{\partial \mathbf{r}} \,
    \biggr| \,k_{1} \biggr>,
\end{array}
\label{eq.2.5.4}
\end{equation}
we rewrite expression (\ref{eq.2.5.3}) so:
\begin{equation}
  \bigl< k_{2}, 1 \bigl| \,\tilde{W} \,\bigr| \, k_{1}, 0 \bigr> =
  F_{21} \cdot 2\pi\delta(w_{2}-w_{1}+w)
\label{eq.2.5.5}
\end{equation}
and the total matrix element (\ref{eq.2.3.4}) has obtained the following form:
\begin{equation}
  a_{fi} =
  \displaystyle\int dk_{2} \; \displaystyle\int dk_{1} \;
    g^{*}(k_{2} - k_{f})\,  g(k_{1} - k_{i}) \cdot
    F_{21} \cdot 2\pi\delta(w_{2}-w_{1}+w).
\label{eq.2.5.6}
\end{equation}

For \emph{quasimonochromatic packets} we have:
\begin{equation}
  a_{fi} = (\Delta k)^{2} |C|^{2} \cdot F_{fi} \cdot 2\pi\, \delta(w_{2}-w_{1}+w)
\label{eq.2.5.7}
\end{equation}
where $C$ is constant. We define it, using the following normalization for the quasimonocromatic packet:
\begin{equation}
  (\Delta k)^{2} C^{2} = 1.
\label{eq.2.5.8}
\end{equation}
Then we obtain the matrix element in the following form:
\begin{equation}
  a_{fi} = F_{fi} \cdot 2\pi \,\delta(w_{f}-w_{i}+w).
\label{eq.2.5.9}
\end{equation}
This expression coincides (up to factor $2\pi$) with a general definition of the matrix element in approach of quantum field theory (for example, see \cite{Bogoliubov.1980} (21.2) sec.~21, p.~168--169 where function $F_{fi}$ for \emph{bound} diagrams is smooth and hasn't other $\delta$-functions). Singular factor in (\ref{eq.2.5.9}) corresponds to conservation law of total energy of the system with emission.

Now we shall be interesting in \emph{probability of transition} defined on the basis of square of the matrix element $a_{fi}$. Usually, in quantum mechanics the probabilities of transitions are defined in time unit and in unit of spacial volume. In quantum field theory they are defined if the matrix element has 4-dimensional $\delta$-function. In our case, the matrix element (\ref{eq.2.5.9}) has one-dimensional $\delta$-function and direct calculation of its square does not give the probability in space unit. However, one can resolve this question introducing further absolute probability (see sec.~\ref{sec.6}) like passing from the probability to cross-section in collisions theory.

In calculation of square of the matrix element we deals with product of two delta-functions which are singular functions. Such product we calculate using approach of quantum field theory (we take it from \cite{Bogoliubov.1980}, sec.~21, p.~169) used for determination of the transition probability on the basis of matrix element. Taking into account that one-dimensional $\delta$-function appears in result of integration by whole time interval ($T \to \infty$), we find \emph{formula of power reduction of $\delta$-function}:
\begin{equation}
  [\delta(w)]^{2} = \delta(w)\: \delta(0) = \delta(w) \: (2\pi)^{-1} \int dt =
  \delta(w)\: (2\pi)^{-1}\, T
\label{eq.2.5.10}
\end{equation}
and obtain:
\begin{equation}
  |a_{fi}|^{2} = 2\pi\: T\: |F_{fi}|^{2} \cdot \delta(w_{f}-w_{i}+w)
\label{eq.2.5.11}
\end{equation}
that looks like (4.21) in \cite{Bogoliubov.1980} (with accuracy up to factor $(2\pi)^{2}$, see sec.~21, p.~169). Singular factor $T$ disappears after passing from (\ref{eq.2.5.11}) to the transition probability in time unit.

\vspace{3mm}
We can find square of the matrix element (\ref{eq.2.5.9}) by another way --- in approach of pure quantum mechanics (like sec.~42, p.~188-189 in~\cite{Landau.v3.1989}). Let's write the matrix element before application of the stationary approximation so:
\begin{equation}
  a_{fi} = - \displaystyle\frac{i}{\hbar}\; \int\limits_{0}^{T} W_{fi}(t) \; dt.
\label{eq.2.5.12}
\end{equation}
We shall find $W_{fi}(t)$ for quasimonochromatic packets with the spontaneous emission. From (\ref{eq.2.3.4}), (\ref{eq.2.3.12}), (\ref{eq.2.5.8}) we obtain:
\begin{equation}
\begin{array}{ccl}
\vspace{2mm}
  W_{fi}(t) & = &
    i\,\hbar \cdot F_{fi} \cdot e^{i\,(w_{2}-w_{1}+w)\,t} = \\
  & = &
    i\,\hbar \cdot
    Z_{eff}\, \displaystyle\frac{e}{m} \,
    \sum\limits_{\alpha=1,2}  \sqrt{\displaystyle\frac{2\pi\hbar}{w}} \cdot
    \biggl< k_{2} \biggl| \,
      \mathbf{e}^{(\alpha),*} e^{-i\mathbf{kr}}
      \displaystyle\frac{\partial}{\partial \mathbf{r}}
    \,\biggr| \,k_{1} \biggr> \cdot  e^{i\,(w_{2}-w_{1}+w)\,t}
\end{array}
\label{eq.2.5.13}
\end{equation}
where the function $F_{fi}$ is defined in (\ref{eq.2.5.4}). Substituting $W_{fi}$ into (\ref{eq.2.5.12}), we obtain (at $t_{0}=0$, $t_{1}=T$):
\begin{equation}
  a_{fi} = -i\, F_{fi}\: \displaystyle\frac{e^{i\,(w_{f}-w_{i}+w)\,T} - 1} {w_{f}-w_{i}+w}
\label{eq.2.5.14}
\end{equation}
and for square of the matrix element we find:
\begin{equation}
  |a_{fi}|^{2} =
  |F_{fi}|^{2}\; \displaystyle\frac{4 \sin^{2}{\displaystyle\frac{(w_{f}-w_{i}+w)\,T}{2}}} {(w_{f}-w_{i}+w)^{2}}.
\label{eq.2.5.15}
\end{equation}
We use the following formula (see p.~188 in~\cite{Landau.v3.1989}):
\begin{equation}
  \lim_{T \to \infty} \displaystyle\frac{\sin^{2}{\alpha T}} {\alpha^{2} T} = \pi\, \delta(\alpha)
\label{eq.2.5.16}
\end{equation}
and at large $T \to +\infty$ we obtain:
\begin{equation}
  |a_{fi}|^{2} = \pi\, T \; |F_{fi}|^{2} \cdot\, \delta \biggl( \displaystyle\frac{w_{f}-w_{i}+w}{2} \biggr).
\label{eq.2.5.17}
\end{equation}
Using formula:
\begin{equation}
  \delta(ax) = \displaystyle\frac{\delta(x)}{|a|},
\label{eq.2.5.18}
\end{equation}
we obtain:
\[
  |a_{fi}|^{2} = 2\pi\: T\: |F_{fi}|^{2} \cdot \delta(w_{f}-w_{i}+w).
\]
So, we have obtained formula (\ref{eq.2.5.11}) which coincides with (42,5) in \cite{Landau.v3.1989} (see p.~189) with accuracy up to factor $\hbar$. However, factor $\hbar$ (introduced after use of the emission operator (\ref{eq.2.3.7})) is included explicitly into the matrix element $F_{fi}$ which gives coincidence between (\ref{eq.2.5.11}) and formalism in sec.~42 from \cite{Landau.v3.1989} complete.

% *******************************************************************************************************************

% *******************************************************************************************************************
% \newpage
\section{Multipole method of calculation of matrix element $p\,(k_{i}, k_{f})$
\label{sec.3}}

\subsection{Linear and circular polarizations of the photon emitted
\label{sec.3.1}}

Now let's consider the matrix element $p\,(k_{i}, k_{f})$. We have:
\begin{equation}
  \mathbf{p}\,(k_{i}, k_{f}) =
    \biggl< k_{2} \biggl| \,  e^{-i\mathbf{kr}} \displaystyle\frac{\partial}{\partial \mathbf{r}} \,
    \biggr| \,k_{1} \biggr> =
    \int
      \psi^{*}_{f}(\mathbf{r}) \:
      e^{-i\mathbf{kr}} \displaystyle\frac{\partial}{\partial \mathbf{r}}\:
      \psi_{i}(\mathbf{r}) \;
      \mathbf{dr}
\label{eq.3.1.1}
\end{equation}
and
\begin{equation}
  p\,(k_{i}, k_{f}) =
  \displaystyle\sum\limits_{\alpha=1,2} \mathbf{e}^{(\alpha),*} \mathbf{p}(k_{i}, k_{f}) =
  \sum\limits_{\alpha = 1, 2} \mathbf{e}^{(\alpha),*}
  \int
    \psi^{*}_{f}(\mathbf{r}) \:
    e^{-i\mathbf{kr}} \displaystyle\frac{\partial}{\partial \mathbf{r}}\:
    \psi_{i}(\mathbf{r}) \;
    \mathbf{dr}.
\label{eq.3.1.2}
\end{equation}
Rewrite vectors of \emph{linear} polarization
$\mathbf{e}^{(\alpha)}$ through \emph{vectors of circular
polarization} $\mathbf{\xi}_{\mu}$ with opposite directions of
rotation (see~\cite{Eisenberg.1973}, (2.39), p.~42):
\begin{equation}
\begin{array}{ccc}
  \mathbf{\xi}_{-1} = \displaystyle\frac{1}{\sqrt{2}}
                      (\mathbf{e}^{(1)} - i\mathbf{e}^{(2)}), &
  \mathbf{\xi}_{+1} = -\displaystyle\frac{1}{\sqrt{2}}
                      (\mathbf{e}^{(1)} + i\mathbf{e}^{(2)}), &
  \mathbf{\xi}_{0} = \mathbf{e}^{(3)} = 0.
\end{array}
\label{eq.3.1.3}
\end{equation}
For vectors $\mathbf{\xi}_{\pm 1}$ we obtain:
\begin{equation}
\left\{
\begin{array}{l}
  \mathbf{\xi}_{+1} - \mathbf{\xi}_{-1} = -\sqrt{2}\mathbf{e}^{(1)} \\
  \mathbf{\xi}_{+1} + \mathbf{\xi}_{-1} = -i\sqrt{2}\mathbf{e}^{(2)}
\end{array}
\right)
\to
\left\{
\begin{array}{l}
  \mathbf{e}^{(1)} =
    -\displaystyle\frac{1}{\sqrt{2}}
    \Bigl(\mathbf{\xi}_{+1} - \mathbf{\xi}_{-1}\Bigr) \\
  \mathbf{e}^{(2)} =
    \displaystyle\frac{i}{\sqrt{2}}
    \Bigl(\mathbf{\xi}_{+1} + \mathbf{\xi}_{-1}\Bigr)
\end{array}
\right)
\to
\left\{
\begin{array}{l}
  \mathbf{e}^{(1),*} =
    -\displaystyle\frac{1}{\sqrt{2}}
    \Bigl(\mathbf{\xi}_{+1}^{*} - \mathbf{\xi}_{-1}^{*}\Bigr) \\
  \mathbf{e}^{(2),*} =
    -\displaystyle\frac{i}{\sqrt{2}}
    \Bigl(\mathbf{\xi}_{+1}^{*} + \mathbf{\xi}_{-1}^{*}\Bigr)
\end{array}
\right).
\label{eq.3.1.4}
\end{equation}
We find:
\begin{equation}
  \sum\limits_{\alpha = 1,2} \mathbf{e}^{(\alpha),*} =
  -\displaystyle\frac{1}{\sqrt{2}}
    \Bigl(\mathbf{\xi}_{+1}^{*} - \mathbf{\xi}_{-1}^{*}\Bigr)
    -\displaystyle\frac{i}{\sqrt{2}}
    \Bigl(\mathbf{\xi}_{+1}^{*} + \mathbf{\xi}_{-1}^{*}\Bigr) =
  \displaystyle\frac{1-i}{\sqrt{2}} \mathbf{\xi}_{-1}^{*} -
    \displaystyle\frac{1+i}{\sqrt{2}} \mathbf{\xi}_{+1}^{*} =
  h_{-1} \mathbf{\xi}_{-1}^{*} + h_{+1} \mathbf{\xi}_{+1}^{*}
\label{eq.3.1.5}
\end{equation}
where
\begin{equation}
\begin{array}{ll}
  h_{-1} = \displaystyle\frac{1-i}{\sqrt{2}}, &
  h_{1}  = -\displaystyle\frac{1+i}{\sqrt{2}}.
\end{array}
\label{eq.3.1.6}
\end{equation}
For numbers $h_{\pm 1}$ we have:
\begin{equation}
  h_{-1} + h_{+1} = -i\sqrt{2}.
\label{eq.3.1.7}
\end{equation}
Then one can rewrite $p\,(k_{i},k_{f})$ so:
\begin{equation}
  p\,(k_{i}, k_{f}) =
    \sum\limits_{\mu = -1, 1}  h_{\mu}\mathbf{\xi}^{*}_{\mu}
    \int
      \psi^{*}_{f}(\mathbf{r})\:
      e^{-i\mathbf{kr}} \displaystyle\frac{\partial}{\partial \mathbf{r}} \:
      \psi_{i}(\mathbf{r}) \;
    \mathbf{dr}.
\label{eq.3.1.8}
\end{equation}
In further calculations of the matrix element $p\,(k_{i}, k_{f})$ the different expansions of function $e^{-i\mathbf{kr}}$ connected with the vector potential $\mathbf{A}$ of the electro-magnetic field of the daughter nucleus are used. On such basis one can construct different approaches for calculation of the bremsstrahlung spectra. Important reason of the expansion of $e^{-i\mathbf{kr}}$ lays in transition from the initial integral in (\ref{eq.3.1.8}), convergence of which is less possible, to series of higher convergent integrals. Usually, the first such integrals are used in estimation of the spectra of the photons emitted.

%-----------------------------------------------------------------------------------------------------------------------

%-----------------------------------------------------------------------------------------------------------------------
\subsection{Expansion of the vector potential $\mathbf{A}$ by multipoles
\label{sec.3.2}}

We expand the vector potential $\mathbf{A}$ into multipoles.
According to \cite{Eisenberg.1973} (see~(2.106) in p.~58), we have:
\begin{equation}
  \mathbf{\xi}_{\mu}\, e^{i \mathbf{kr}} =
    \mu\, \sqrt{2\pi}\, \sum_{l, \nu}\,
    (2l+1)^{1/2}\, i^{l}\,  D_{\nu\mu}^{l} (\varphi,\theta,0) \cdot
    \Bigl[ \mathbf{A}_{l\nu} (\mathbf{r}, M) +
    i\mu\, \mathbf{A}_{l\nu} (\mathbf{r}, E) \Bigr]
\label{eq.3.2.1}
\end{equation}
where (see~\cite{Eisenberg.1973}, (2.73) in p.~49, (2.80) in p.~51)
\begin{equation}
\begin{array}{lcl}
  \mathbf{A}_{l\nu}(\mathbf{r}, M) & = &
        j_{l}(kr) \: \mathbf{T}_{ll,\nu} ({\mathbf n}_{ph}), \\
  \mathbf{A}_{l\nu}(\mathbf{r}, E) & = &
        \sqrt{\displaystyle\frac{l+1}{2l+1}}
        j_{l-1}(kr) \: \mathbf{T}_{ll-1,\nu}({\mathbf n}_{ph}) -
        \sqrt{\displaystyle\frac{l}{2l+1}}
        j_{l+1}(kr) \: \mathbf{T}_{ll+1,\nu}({\mathbf n}_{ph}).
\end{array}
\label{eq.3.2.2}
\end{equation}
Here, $\mathbf{A}_{l\nu}(\textbf{r}, M)$ and $\mathbf{A}_{l\nu}(\textbf{r}, E)$ are \emph{magnetic} and \emph{electric multipoles}, $j_{l}(kr)$ are \emph{spherical Bessel functions of order $l$}, $\mathbf{T}_{ll',\nu}(\mathbf{n})$ are \emph{vector spherical harmonics},
% form of matrix $D_{m' m}^{l} (\theta_{1},\theta_{2},\theta_{3})$ is presented in Appendix,
$\theta_{1}$, $\theta_{2}$, $\theta_{3}$ are angles defining direction of vector $\mathbf{k}$ relatively axis $z$ in selected frame system. According to \cite{Eisenberg.1973} (see p.~45), the functions $\mathbf{T}_{ll',\nu}(\mathbf{n})$ have the following form (${\mathbf \xi}_{0} = 0$):
\begin{equation}
  \mathbf{T}_{jl,m} (\mathbf{n}) =
  \sum\limits_{\mu = \pm 1} (l, 1, j \,\big| \,m-\mu, \mu, m) \; Y_{l,m-\mu}(\mathbf{n}) \; \mathbf{\xi}_{\mu}
\label{eq.3.2.3}
\end{equation}
where $(l, 1, j \,\bigl| \, m-\mu, \mu, m)$ are \emph{Clebsh-Gordon coefficients}.

Formula (\ref{eq.3.2.1}) is defined when the vector $\mathbf{k}$ is oriented arbitrary concerning the selected frame system. If to use such frame system where axis $z$ is parallel to the vector $\mathbf{k}$ then from (\ref{eq.3.2.1}) we obtain (see~\cite{Eisenberg.1973}, (2.105) p.~57):
\begin{equation}
  \mathbf{\xi}_{\mu}\, e^{i \mathbf{kr}} =
    \mu\, \sqrt{2\pi}\, \sum_{l}\,
    (2l+1)^{1/2}\, i^{l}\,
    \Bigl[ \mathbf{A}_{l\mu} (\mathbf{r}, M) +
    i\mu\, \mathbf{A}_{l\mu} (\mathbf{r}, E) \Bigr].
\label{eq.3.2.4}
\end{equation}
This formula is useful when it is not important to orient the frame system relatively the decaying nucleus. This is the case when the nucleus and decay are spherically symmetric. If the decay is asymmetric relatively axis $z$ then it needs to orient the frame system relatively the decaying nucleus. Here, the vector $\mathbf{k}$ and axis $z$ are not parallel and it needs to use (\ref{eq.3.2.1}).

%-----------------------------------------------------------------------------------------------------------------------

%-----------------------------------------------------------------------------------------------------------------------
\subsection{$\alpha$-decay in a general case
\label{sec.3.3}}

At first, we shall consider a case when the decaying nucleus has a general form. From (\ref{eq.3.1.8}) and (\ref{eq.3.2.1}) we find:
\[
\begin{array}{lcl}
  p(k_{i}, k_{f}) & = &
    \displaystyle\sum\limits_{\mu = -1, 1} h_{\mu}
    \displaystyle\int
        \psi^{*}_{f}(\mathbf{r}) \cdot
        \mu\, \sqrt{2\pi}\, \sum\limits_{l, \nu}\: (-i)^{l}\, \sqrt{2l+1} \cdot
        D_{\nu\mu}^{l,*}(\varphi,\theta,0) \times \\
  & \times &
        \Bigl[ \mathbf{A}_{l\nu}^{*} (\mathbf{r}, M) - i\mu \mathbf{A}_{l\nu}^{*} (\mathbf{r}, E) \Bigr] \cdot
        \biggl( \displaystyle\frac{\partial}{\partial \mathbf{r}} \,\psi_{i}(\mathbf{r}) \biggr) \cdot \;
        \mathbf{dr}.
\end{array}
\]
Here, matrix-function $D_{\nu\mu}^{l,*}(\varphi,\theta,0)$ defines direction of vector $\mathbf{k}$ relatively axis $z$ in the frame system for $\mathbf{r}$: angles $\varphi$ and $ \theta$ point to direction of vector $\mathbf{k}$, but not the vector $\mathbf{r}$. So, we obtain:
\begin{equation}
  p(k_{i}, k_{f}) =
  \sqrt{2\pi}\;
      \displaystyle\sum\limits_{l,\nu}\: (-i)^{l}\, \sqrt{2l+1}
      \displaystyle\sum\limits_{\mu = -1, 1} h_{\mu} \; \mu \cdot
      D_{\nu\mu}^{l,*}(\varphi,\theta,0) \cdot
      \Bigl[ p_{l\nu}^{M} - i\mu\: p_{l\nu}^{E} \Bigr]
\label{eq.3.3.1}
\end{equation}
where
\begin{equation}
\begin{array}{lcl}
  p_{l\nu}^{M} & = &
    \displaystyle\int
        \psi^{*}_{f}(\mathbf{r}) \,
        \biggl( \displaystyle\frac{\partial}{\partial \mathbf{r}}\, \psi_{i}(\mathbf{r}) \biggr) \,
        \mathbf{A}_{l\nu}^{*} (\mathbf{r}, M) \;
        \mathbf{dr}, \\

  p_{l\nu}^{E} & = &
    \displaystyle\int
        \psi^{*}_{f}(\mathbf{r}) \,
        \biggl( \displaystyle\frac{\partial}{\partial \mathbf{r}}\, \psi_{i}(\mathbf{r}) \biggr)\,
        \mathbf{A}_{l\nu}^{*} (\mathbf{r}, E) \;
        \mathbf{dr}.
\end{array}
\label{eq.3.3.2}
\end{equation}
%-----------------------------------------------------------------------------------------------------------------------

%-----------------------------------------------------------------------------------------------------------------------
\subsection{Approximation of the spherically symmetric $\alpha$-decay
\label{sec.3.4}}

Now we shall study a case when the decaying nucleus can be considered as spherically symmetric.
In such spherically symmetric approximation, wave functions of the decaying system in the initial and final states are separated into radial and angular components:
\begin{equation}
\begin{array}{lcl}
  \psi_{i} (\mathbf{r}) & = & \varphi_{i} (r) \: Y_{00}({\mathbf n}_{r}^{i}), \\
  \psi_{f} (\mathbf{r}) & = & \varphi_{f} (r) \: Y_{l_{f}m}({\mathbf n}_{r}^{f}).
\end{array}
\label{eq.3.4.1}
\end{equation}
We rewrite integrals (\ref{eq.3.3.2}) so:
\begin{equation}
\begin{array}{lcl}
  p_{l\nu}^{M} & = &
        \displaystyle\int\limits^{+\infty}_{0} dr
        \displaystyle\int d\Omega \: r^{2} \,
        \psi^{*}_{f}(\mathbf{r}) \,
        \biggl( \displaystyle\frac{\partial}{\partial \mathbf{r}}\, \psi_{i}(\mathbf{r}) \biggr) \,
        \mathbf{A}_{l\nu}^{*} (\mathbf{r}, M), \\

  p_{l\nu}^{E} & = &
        \displaystyle\int\limits^{+\infty}_{0} dr
        \displaystyle\int d\Omega \: r^{2} \,
        \psi^{*}_{f}(\mathbf{r}) \,
        \biggl( \displaystyle\frac{\partial}{\partial \mathbf{r}}\, \psi_{i}(\mathbf{r}) \biggr)\,
        \mathbf{A}_{l\nu}^{*} (\mathbf{r}, E).
\end{array}
\label{eq.3.4.2}
\end{equation}
In expansion of the potential $\mathbf{A}$ one can use formulas (\ref{eq.3.2.4}) and (\ref{eq.3.2.2}). Taking into account:
\begin{equation}
  D_{\nu\mu}^{l}(\varphi,\theta,0) = \delta_{\mu\nu},
\label{eq.3.4.3}
\end{equation}
from (\ref{eq.3.3.1}) we obtain:
\begin{equation}
  p(k_{i}, k_{f}) = \sqrt{2\pi}\: \sum\limits_{l} \: (-i)^{l}\, \sqrt{2l+1} \: \Bigl[ p_{l}^{M} -ip_{l}^{E} \Bigr]
\label{eq.3.4.4}
\end{equation}
where
\begin{equation}
\begin{array}{ll}
  p_{l}^{M} = \sum\limits_{\mu = -1, 1} \mu\, h_{\mu}\: p_{l\nu}^{M}, &
  p_{l}^{E} = \sum\limits_{\mu = -1, 1} \mu^{2} h_{\mu}\: p_{l\nu}^{E}.
\end{array}
\label{eq.3.4.5}
\end{equation}

Using the following formula (see (2.56), p.~46 in~\cite{Eisenberg.1973}):
\begin{equation}
  \displaystyle\frac{\partial}{\partial \mathbf{r}}\:
    f(r)\, Y_{lm}({\mathbf n}_{r}) =
  \sqrt{\displaystyle\frac{l}{2l+1}}\:
    \biggl( \displaystyle\frac{df}{dr} + \displaystyle\frac{l+1}{r} f \biggr)\,
    \mathbf{T}_{l l-1, m}({\mathbf n}_{r}) -
  \sqrt{\displaystyle\frac{l+1}{2l+1}}\:
    \biggl( \displaystyle\frac{df}{dr} - \displaystyle\frac{l}{r} f \biggr)\,
    \mathbf{T}_{l l+1, m}({\mathbf n}_{r})
\label{eq.3.4.6}
\end{equation}
and taking into account (\ref{eq.3.4.1}), we obtain:
\begin{equation}
  \displaystyle\frac{\partial}{\partial \mathbf{r}}\: \psi_{i}(\mathbf{r}) =
  -\,\displaystyle\frac{d\,\varphi_{i}(r)}{dr}\: \mathbf{T}_{01,0}(\mathbf{n}^{i}_{r}).
\label{eq.3.4.7}
\end{equation}

Now we shall calculate expressions (\ref{eq.3.4.5}). For the magnetic component $p_{l}^{M}$ we obtain:
\[
\begin{array}{lcl}
  p_{l}^{M} & = &
  \displaystyle\sum\limits_{\mu = -1, 1}
    \mu h_{\mu}
    \displaystyle\int\limits^{+\infty}_{0} dr
    \displaystyle\int d\Omega \:
    r^{2} \psi^{*}_{f}(\mathbf{r}) \:
    \biggl( \displaystyle\frac{\partial}{\partial \mathbf{r}}
            \psi_{i}(\mathbf{r}) \biggr) \:
    \mathbf{A}_{l\mu}^{*} (\mathbf{r}, M) = \\

  & = &
  \displaystyle\sum\limits_{\mu = \pm 1}
    \mu h_{\mu}
    \displaystyle\int\limits^{+\infty}_{0} dr
    \displaystyle\int d\Omega \:
    r^{2} \varphi^{*}_{f}(r) \: Y_{l_{f}m}^{*}({\mathbf n}_{r}^{f}) \:
    \biggl(
      -\displaystyle\frac{d\varphi_{i}(r)}{dr}
      \mathbf{T}_{01,0}(\mathbf{n}^{i}_{r})
    \biggr) \:
    j_{l_{ph}}(kr) \: \mathbf{T}_{l_{ph}l_{ph},\mu}^{*} ({\mathbf n}_{ph}) = \\

  & = &
  -\displaystyle\int\limits^{+\infty}_{0}
    \varphi^{*}_{f}(r)
    \displaystyle\frac{d\varphi_{i}(r)}{dr}
    j_{l_{ph}}(kr) \:
    r^{2} dr
  \; \cdot
    \displaystyle\sum\limits_{\mu = \pm 1}
    \mu h_{\mu}
    \displaystyle\int
    Y_{l_{f}m}^{*}({\mathbf n}_{r}^{f}) \:
    \mathbf{T}_{01,0}(\mathbf{n}^{i}_{r}) \:
    \mathbf{T}_{l_{ph}l_{ph},\mu}^{*} ({\mathbf n}_{ph}) \: d\Omega.
\end{array}
\]
For the electrical component $p^{E}_{l_{ph}}$ we obtain:
\[
\begin{array}{lcl}
  p_{l_{ph}}^{E} & = &
  \displaystyle\sum\limits_{\mu = \pm 1}
    \mu^{2} h_{\mu}
    \displaystyle\int\limits^{+\infty}_{0} dr
    \displaystyle\int d\Omega \:
    r^{2} \psi^{*}_{f}(\mathbf{r})
    \biggl( \displaystyle\frac{\partial}{\partial \mathbf{r}}
            \psi_{i}(\mathbf{r}) \biggr)
    \mathbf{A}_{l_{ph} \mu}^{*} (\mathbf{r}, E) = \\

  & = &
  \displaystyle\sum\limits_{\mu = \pm 1}
    h_{\mu}
    \displaystyle\int\limits^{+\infty}_{0} dr
    \displaystyle\int d\Omega \:
    r^{2} \varphi^{*}_{f}(r) Y_{l_{f}m}^{*}({\mathbf n}_{r}^{f})
    \biggl(
      -\displaystyle\frac{d\varphi_{i}(r)}{dr}
      \mathbf{T}_{01,0}(\mathbf{n}^{i}_{r})
    \biggr) \times \\
    & & \times \;
    \Biggl\{
      \sqrt{\displaystyle\frac{l_{ph}+1}{2l_{ph}+1}}
      j_{l_{ph}-1}(kr) \: \mathbf{T}_{l_{ph}l_{ph}-1,\mu}^{*}({\mathbf n}_{ph}) -
      \sqrt{\displaystyle\frac{l_{ph}}{2l_{ph}+1}}
      j_{l_{ph}+1}(kr) \: \mathbf{T}_{l_{ph}l_{ph}+1,\mu}^{*}({\mathbf n}_{ph})
    \Biggr\} = \\

  & = &
  -\sqrt{\displaystyle\frac{l_{ph}+1}{2l_{ph}+1}}
    \displaystyle\int\limits^{+\infty}_{0}
    \varphi^{*}_{f}(r)
    \displaystyle\frac{d\varphi_{i}(r)}{dr}
    j_{l_{ph}-1}(kr) \:
    r^{2} dr

    \displaystyle\sum\limits_{\mu = \pm 1}
    h_{\mu}
    \displaystyle\int
    Y_{l_{f}m}^{*}({\mathbf n}_{r}^{f})
    \mathbf{T}_{01,0}(\mathbf{n}^{i}_{r})
    \: \mathbf{T}_{l_{ph}l_{ph}-1,\mu}^{*}({\mathbf n}_{ph}) \: d\Omega - \\

  & & + \;
  \sqrt{\displaystyle\frac{l_{ph}}{2l_{ph}+1}}
    \displaystyle\int\limits^{+\infty}_{0}
    \varphi^{*}_{f}(r)
    \displaystyle\frac{d\varphi_{i}(r)}{dr} j_{l_{ph}+1}(kr) \:
    r^{2} dr

    \displaystyle\sum\limits_{\mu = \pm 1}
    h_{\mu}
    \displaystyle\int
    Y_{l_{f}m}^{*}({\mathbf n}_{r}^{f})
    \mathbf{T}_{01,0}(\mathbf{n}^{i}_{r})
    \: \mathbf{T}_{l_{ph}l_{ph}+1,\mu}^{*}({\mathbf n}_{ph}) \: d\Omega.
\end{array}
\]
So, the components $p_{l_{ph}}^{M}$ and $p_{l_{ph}}^{E}$ have the form:
\begin{equation}
\begin{array}{lcl}
  p_{l_{ph}}^{M} & = &
  -\displaystyle\int\limits^{+\infty}_{0}
    \varphi^{*}_{f}(r)
    \displaystyle\frac{d\varphi_{i}(r)}{dr}
    j_{l_{ph}}(kr) \:
    r^{2} dr

    \displaystyle\sum\limits_{\mu = \pm 1}
    \mu h_{\mu}
    \displaystyle\int
    Y_{l_{f}m}^{*}({\mathbf n}_{r}^{f}) \:
    \mathbf{T}_{01,0}(\mathbf{n}^{i}_{r}) \:
    \mathbf{T}_{l_{ph}l_{ph},\mu}^{*} ({\mathbf n}_{ph}) \: d\Omega, \\

  p_{l_{ph}}^{E} & = &
  -\sqrt{\displaystyle\frac{l_{ph}+1}{2l_{ph}+1}}
    \displaystyle\int\limits^{+\infty}_{0}
    \varphi^{*}_{f}(r)
    \displaystyle\frac{d\varphi_{i}(r)}{dr}
    j_{l_{ph}-1}(kr) \:
    r^{2} dr

    \displaystyle\sum\limits_{\mu = \pm 1}
    h_{\mu}
    \displaystyle\int
    Y_{l_{f}m}^{*}({\mathbf n}_{r}^{f})
    \mathbf{T}_{01,0}(\mathbf{n}^{i}_{r})
    \: \mathbf{T}_{l_{ph}l_{ph}-1,\mu}^{*}({\mathbf n}_{ph}) \: d\Omega - \\

  & & + \;
  \sqrt{\displaystyle\frac{l_{ph}}{2l_{ph}+1}}
    \displaystyle\int\limits^{+\infty}_{0}
    \varphi^{*}_{f}(r)
    \displaystyle\frac{d\varphi_{i}(r)}{dr} j_{l_{ph}+1}(kr) \:
    r^{2} dr

    \displaystyle\sum\limits_{\mu = \pm 1}
    h_{\mu}
    \displaystyle\int
    Y_{l_{f}m}^{*}({\mathbf n}_{r}^{f})
    \mathbf{T}_{01,0}(\mathbf{n}^{i}_{r})
    \: \mathbf{T}_{l_{ph}l_{ph}+1,\mu}^{*}({\mathbf n}_{ph}) \: d\Omega.
\end{array}
\label{eq.3.4.8}
\end{equation}
Let's introduce the following symbols:
\begin{equation}
\begin{array}{lcl}
  J(l_{f},n) & = &
  \displaystyle\int\limits^{+\infty}_{0}
    \varphi^{*}_{f}(l,r) \displaystyle\frac{d\varphi_{i}(r)}{dr}
    j_{n}(kr) \: r^{2} dr, \\

  I_{M}(l_{f}, l_{ph}, n) & = &
    \displaystyle\sum\limits_{\mu = \pm 1}
    \mu h_{\mu}
    \displaystyle\int
    Y_{l_{f}m}^{*}({\mathbf n}_{r}^{f}) \:
    \mathbf{T}_{01,0}(\mathbf{n}^{i}_{r})
    \: \mathbf{T}_{l_{ph} n,\mu}^{*}({\mathbf n}_{ph}) \: d\Omega, \\

  I_{E}(l_{f}, l_{ph}, n) & = &
    \displaystyle\sum\limits_{\mu = \pm 1}
    h_{\mu}
    \displaystyle\int
    Y_{l_{f}m}^{*}({\mathbf n}_{r}^{f}) \:
    \mathbf{T}_{01,0}(\mathbf{n}^{i}_{r})
    \: \mathbf{T}_{l_{ph} n,\mu}^{*}({\mathbf n}_{ph}) \: d\Omega.
\end{array}
\label{eq.3.4.9}
\end{equation}
Then expressions (\ref{eq.3.4.8}) are written so:
\begin{equation}
\begin{array}{lcl}
  p_{l_{ph}}^{M} & = & - I_{M}(l_{f},l_{ph}, l_{ph}) \cdot J(l,l), \\
  p_{l_{ph}}^{E} & = &
    -\sqrt{\displaystyle\frac{l_{ph}+1}{2l_{ph}+1}} I_{E}(l_{f},l_{ph},l_{ph}-1) \cdot J(l_{f},l_{ph}-1) +
    \sqrt{\displaystyle\frac{l_{ph}}{2l_{ph}+1}} I_{E}(l_{f},l_{ph},l_{ph}+1) \cdot J(l_{f},l_{ph}+1).
\end{array}
\label{eq.3.4.10}
\end{equation}

Using the following value of the Clebsh-Gordon coefficient (see Appendix~\ref{app.3}; also~\cite{Eisenberg.1973}, Table~1 in p.~317):
\begin{equation}
  (110|1, -1, 0) = (110|-1, 1, 0) = \sqrt{\displaystyle\frac{1}{3}},
\label{eq.3.4.11}
\end{equation}
from (\ref{eq.3.2.3}) and (\ref{eq.3.4.7}) we find:
\begin{equation}
\begin{array}{c}
  \mathbf{T}_{01,0}(\mathbf{n}^{i}_{r}) =
    \displaystyle\sum\limits_{\mu = \pm 1} (110|-\mu\mu 0) \:
      Y_{1,-\mu}(\mathbf{n}^{i}_{r}) \: \mathbf{\xi}_{\mu} =
    \sqrt{\displaystyle\frac{1}{3}}
      \displaystyle\sum\limits_{\mu = \pm 1}
      Y_{1,-\mu}(\mathbf{n}^{i}_{r}) \: \mathbf{\xi}_{\mu}, \\
  \displaystyle\frac{\partial}{\partial \mathbf{r}}
    \psi_{i}(\mathbf{r}) =
  -\sqrt{\displaystyle\frac{1}{3}}
    \displaystyle\frac{d\varphi_{i}(r)}{dr}
    \displaystyle\sum\limits_{\mu = -1, 1}
    Y_{1,-\mu}(\mathbf{n}^{i}_{r}) \: \mathbf{\xi}_{\mu}
\end{array}
\label{eq.3.4.12}
\end{equation}
and from (\ref{eq.3.4.9}) we obtain:
\begin{equation}
\begin{array}{lcl}
  I_{M}(l_{f}, l_{ph}, n) & = &
  \sqrt{\displaystyle\frac{1}{3}}
    \displaystyle\sum\limits_{\mu = \pm 1}
      \mu h_{\mu}
      \displaystyle\int \:
      Y_{l_{f} m}^{*}({\mathbf n}_{r}^{f})
      \sum\limits_{\mu' = \pm 1}
      Y_{1,-\mu'}(\mathbf{n}^{i}_{r}) \: \mathbf{\xi}_{\mu'}
      \mathbf{T}_{l_{ph}n,\mu}^{*} ({\mathbf n}_{ph}) \: d\Omega, \\
  I_{E}(l_{f}, l_{ph}, n) & = &
  \sqrt{\displaystyle\frac{1}{3}}
    \displaystyle\sum\limits_{\mu = \pm 1}
      h_{\mu}
      \displaystyle\int \:
      Y_{l_{f}m}^{*}({\mathbf n}_{r}^{f})
      \sum\limits_{\mu' = \pm 1}
      Y_{1,-\mu'}(\mathbf{n}^{i}_{r}) \: \mathbf{\xi}_{\mu'}
      \mathbf{T}_{l_{ph}n,\mu}^{*} ({\mathbf n}_{ph}) \: d\Omega.
\end{array}
\label{eq.3.4.13}
\end{equation}

If to study photon emission into all final states with different values of $m$ and the same number $l$, then as the wave function of the final state amidst $\psi_{f}(\mathbf{r})$ in (\ref{eq.3.4.1}) we shall use superposition at all states with all possible $m$:
\begin{equation}
  \psi_{f} (\mathbf{r}) = \displaystyle\sum\limits_{m} \varphi_{f} (r) \: Y_{l_{f}m}({\mathbf n}_{r}^{f}).
\label{eq.3.4.14}
\end{equation}
Let's assume that the radial component of wave function $\varphi_{f} (r)$ does not depend on $m$ for selected $l_{f}$. We rewrite (\ref{eq.3.4.14}) so:
\begin{equation}
  \psi_{f} (\mathbf{r}) = \varphi_{f} (r) \: \displaystyle\sum\limits_{m} Y_{l_{f}m}({\mathbf n}_{r}^{f}).
\label{eq.3.4.15}
\end{equation}
The transition matrix element into the superposition of all states with different $m$ at the same $l_{f}$ is defined by (\ref{eq.3.4.10}) where it needs to change expressions (\ref{eq.3.4.13}) for the angular integrals $I_{M}(l_{f}, l_{ph}, n)$ and $I_{M}(l_{f},l_{ph},n)$ into the following:
\begin{equation}
\begin{array}{lcl}
  \tilde{I}_{M}(l_{f},l_{ph},n) & = &
  \sqrt{\displaystyle\frac{1}{3}} \;
    \displaystyle\sum\limits_{\mu = \pm 1}
    \displaystyle\sum\limits_{m}
      \mu h_{\mu}
      \displaystyle\int \:
      Y_{l_{f}m}^{*}({\mathbf n}_{r}^{f})
      \sum\limits_{\mu' = \pm 1}
      Y_{1,-\mu'}(\mathbf{n}^{i}_{r}) \: \mathbf{\xi}_{\mu'}
      \mathbf{T}_{l_{ph}n,\mu}^{*} ({\mathbf n}_{ph}) \: d\Omega, \\
  \tilde{I}_{E}(l_{f},l_{ph},n) & = &
  \sqrt{\displaystyle\frac{1}{3}} \;
    \displaystyle\sum\limits_{\mu = \pm 1}
    \displaystyle\sum\limits_{m}
      h_{\mu}
      \displaystyle\int \:
      Y_{l_{f}m}^{*}({\mathbf n}_{r}^{f})
      \sum\limits_{\mu' = \pm 1}
      Y_{1,-\mu'}(\mathbf{n}^{i}_{r}) \: \mathbf{\xi}_{\mu'}
      \mathbf{T}_{l_{ph}n,\mu}^{*} ({\mathbf n}_{ph}) \: d\Omega.
\end{array}
\label{eq.3.4.16}
\end{equation}
%-----------------------------------------------------------------------------------------------------------------------

%-----------------------------------------------------------------------------------------------------------------------
\subsection{Vectors $\mathbf{n}^{i}_{r}$, $\mathbf{n}^{f}_{r}$, $\mathbf{n}_{ph}$ and calculations of the angular integrals
\label{sec.3.5}}

Let's analyze a physical sense of vectors $\mathbf{n}^{i}_{r}$, $\mathbf{n}^{f}_{r}$ and $\mathbf{n}_{ph}$. At first, we consider the vector $\mathbf{n}^{i}_{r}$. According to definition of $\psi_{i} (\mathbf{r})$, it determines orientation of radius-vector $\mathbf{r}$ from the center of frame system to point where this wave function describes the particle before the emission of photon. Such description of the particle has a probabilistic sense and is fulfilled over whole space. Here, angles $\theta_{i}$ and $\varphi_{i}$ of $\mathbf{n}^{i}_{r}$ characterize orientation of this radius-vector concerning axis $Oz$ of the frame system.

The vector $\mathbf{n}^{f}_{r}$ determines orientation of radius-vector $\mathbf{r}$ from the center of the frame system to point where $\psi_{f} (\mathbf{r})$ describes the particle after the emission of photon. Such a description of the particle is fulfilled over whole space and has the probabilistic sense also. Here, change of direction of motion (or tunneling) of the particle in result of the photon emission can be characterized by change of quantum numbers $l$ and $m$ in the angular wave function: $Y_{00}(\mathbf{n}^{i}_{r}) \to Y_{lm}(\mathbf{n}^{f}_{r})$ (which changes the probability of appearance of this particle in different directions, and angular asymmetry is appeared). Angles $\theta_{f}$ and $\varphi_{f}$ of $\mathbf{n}^{f}_{r}$ characterize orientation of radius-vector $\mathbf{r}$ relatively axis $Oz$ of the frame system.
I.~e. two vectors $\mathbf{n}^{i}_{r}$ and $\mathbf{n}^{f}_{r}$ have similar sense and coincide in the angular integrals (\ref{eq.3.4.13}):
\begin{equation}
  \mathbf{n}^{i}_{r} = \mathbf{n}^{f}_{r}.
\label{eq.3.5.1}
\end{equation}

The vector $\mathbf{n}_{ph}$ determines orientation of radius-vector $\mathbf{r}$ from the center of the frame system to point where wave function of photon describes its ``appearance''. Using such a logic, we write:
\begin{equation}
  \mathbf{n}_{ph} = \mathbf{n}^{i}_{r} = \mathbf{n}^{f}_{r} = \mathbf{n}_{r}.
\label{eq.3.5.2}
\end{equation}
We use such frame system where axis $z$ is parallel to vector $\mathbf{k}$ of the photon emission. Than dependent on $\mathbf{r}$ integrant function in the matrix element represents amplitude (its square is probability) of appearance of the particle at point $\mathbf{r}$ after emission of photon, if this photon has emitted along axis $z$. Then angle $\theta$ (of vector $\mathbf{n}_{\mathbf{r}}$) is the angle between direction of the particle motion (with possible tunneling) and direction of the photon emission.

On the basis of (\ref{eq.3.5.2}) we rewrite the angular integrals (\ref{eq.3.4.13}) so:
\begin{equation}
\begin{array}{lcl}
  I_{M}(l_{f}, l_{ph}, n) & = &
  \sqrt{\displaystyle\frac{1}{3}}
    \displaystyle\sum\limits_{\mu = \pm 1}
      \mu h_{\mu}
      \displaystyle\int \:
      Y_{l_{f}m}^{*}({\mathbf n}_{r})
      \sum\limits_{\mu' = \pm 1}
      Y_{1,-\mu'}(\mathbf{n}_{r}) \: \mathbf{\xi}_{\mu'}
      \mathbf{T}_{l_{ph}n,\mu}^{*} ({\mathbf n}_{r}) \: d\Omega, \\
  I_{E}(l_{f}, l_{ph}, n) & = &
  \sqrt{\displaystyle\frac{1}{3}}
    \displaystyle\sum\limits_{\mu = \pm 1}
      h_{\mu}
      \displaystyle\int \:
      Y_{l_{f}m}^{*}({\mathbf n}_{r})
      \sum\limits_{\mu' = \pm 1}
      Y_{1,-\mu'}(\mathbf{n}_{r}) \: \mathbf{\xi}_{\mu'}
      \mathbf{T}_{l_{ph}n,\mu}^{*} ({\mathbf n}_{r}) \: d\Omega.
\end{array}
\label{eq.3.5.3}
\end{equation}
Writing from (\ref{eq.3.2.3}) vector spherical harmonic:
\begin{equation}
  \mathbf{T}_{l_{ph}n, \mu} (\mathbf{n_{r}}) =
    \sum\limits_{\mu^{\prime\prime} = \pm 1} (n, 1, l_{ph} \big| \mu-\mu^{\prime\prime}, \mu^{\prime\prime}, \mu) \;
    Y_{n, \mu-\mu^{\prime\prime}}(\mathbf{n_{r}}) \;
    \mathbf{\xi}_{\mu^{\prime\prime}}
\label{eq.3.5.4}
\end{equation}
and taking into account orthogonality of vectors $\mathbf{\xi}_{\pm 1}$, we obtain:
\begin{equation}
\begin{array}{lcl}
  I_{M}(l_{f},l_{ph},n) & = &
    \sqrt{\displaystyle\frac{1}{3}}
    \displaystyle\sum\limits_{\mu = \pm 1} \mu h_{\mu}
    \sum\limits_{\mu^{\prime} = \pm 1} (n, 1, l_{ph} \big| \mu-\mu^{\prime}, \mu^{\prime}, \mu) \;
    \displaystyle\int \:
      Y_{l_{f}m}^{*}({\mathbf n}_{r}) \,
      Y_{1,-\mu^{\prime}}(\mathbf{n}_{r}) \,
      Y_{n, \mu-\mu^{\prime}}^{*}(\mathbf{n_{r}}) \;
      d\Omega, \\

  I_{E}(l_{f},l_{ph},n) & = &
    \sqrt{\displaystyle\frac{1}{3}}
    \displaystyle\sum\limits_{\mu = \pm 1} h_{\mu}
    \sum\limits_{\mu^{\prime} = \pm 1} (n, 1, l_{ph} \big| \mu-\mu^{\prime}, \mu^{\prime}, \mu) \;
    \displaystyle\int \:
      Y_{l_{f}m}^{*}({\mathbf n}_{r}) \,
      Y_{1,-\mu^{\prime}}(\mathbf{n}_{r}) \,
      Y_{n, \mu-\mu^{\prime}}^{*}(\mathbf{n_{r}}) \;
      d\Omega.
\end{array}
\label{eq.3.5.5}
\end{equation}

We use the following definition of the spherical functions $Y_{lm}(\theta, \varphi)$
(according to \cite{Landau.v3.1989}, p.~119, (28,7)--(28,8); also Appendix~\ref{app.2}):
\begin{equation}
\begin{array}{lcl}
  Y_{lm}(\theta,\varphi) & = &
    (-1)^{\frac{m+|m|}{2}} \; i^{l} \;
    \sqrt{\displaystyle\frac{2l+1}{4\pi} \displaystyle\frac{(l-|m|)!}{(l+|m|)!} }  \; P_{l}^{|m|} (\cos{\theta}) \cdot e^{im\varphi}
\end{array}
\label{eq.3.5.6}
\end{equation}
where $P_{l}^{m}(\cos{\theta})$ is the Legendre's polynomial (see Appendix~\ref{app.1}).
Rewrite the angular integral in (\ref{eq.3.5.5}) so:
\begin{equation}
\begin{array}{l}
  \displaystyle\int \:
    Y_{lm}^{*}({\mathbf n}_{r}) \,
    Y_{1,-\mu^{\prime}}(\mathbf{n}_{r}) \,
    Y_{n, \mu-\mu^{\prime}}^{*}(\mathbf{n_{r}}) \;
    d\Omega = \\

  = (-1)^{\frac{m+|m| - \mu^{\prime}+1 + \mu-\mu^{\prime}+|\mu-\mu^{\prime}|}{2}} \; (-1)^{l+n} \; i^{l+n+1} \;
    \sqrt{\displaystyle\frac{2l+1}{4\pi} \displaystyle\frac{(l-|m|)!}{(l+|m|)!} } \;
    \sqrt{\displaystyle\frac{3}{8\pi}} \;
    \sqrt{\displaystyle\frac{2n+1}{4\pi} \displaystyle\frac{(n-|\mu-\mu^{\prime}|)!}{(n+|\mu-\mu^{\prime}|)!}}
    \times \\
  \;\times
    \displaystyle\int\limits_{0}^{2\pi} \, e^{i(-m -\mu^{\prime} -\mu+\mu^{\prime})\varphi} \: d\varphi \cdot
    \displaystyle\int\limits_{0}^{\pi} \:
      P_{l}^{|m|}(\cos{\theta}) \; P_{1}^{1}(\cos{\theta}) \; P_{n}^{|\mu-\mu^{\prime}|} (\cos{\theta}) \cdot
      \sin{\theta} \, d\theta \,d\varphi.
\end{array}
\label{eq.3.5.7}
\end{equation}
The integral over $\varphi$ is nonzero only in the case:
\[
  m = -\mu.
\]
Taking into account $\mu = \pm 1$, we obtain restrictions on possible values of $m$ and $l_{f}$:
\begin{equation}
\begin{array}{cc}
  m = -\mu = \pm 1,  &  l_{f} \ge 1,
\end{array}
\label{eq.3.5.8}
\end{equation}
and also
\begin{equation}
  n \ge |\mu - \mu^{\prime}| = |m + \mu^{\prime}|.
\label{eq.3.5.9}
\end{equation}
We obtain:
\begin{equation}
\begin{array}{l}
  \displaystyle\int \:
    Y_{lm}^{*}({\mathbf n}_{r}) \,
    Y_{1,-\mu^{\prime}}(\mathbf{n}_{r}) \,
    Y_{n, \mu-\mu^{\prime}}^{*}(\mathbf{n_{r}}) \;
    d\Omega = \\

  = (-1)^{l+n-\mu^{\prime}+1 + \frac{|m+\mu^{\prime}|}{2}} \; i^{l+n+1} \;
    \sqrt{\displaystyle\frac{3\,(2l+1)\,(2n+1)}{32\pi}\;
          \displaystyle\frac{(l-1)!}{(l+1)!} \;
          \displaystyle\frac{(n-|m+\mu^{\prime}|)!}{(n+|m+\mu^{\prime}|)!}} \;
    \times \\
  \;\times
    \displaystyle\int\:
      P_{l}^{1}(\cos{\theta}) \; P_{1}^{1}(\cos{\theta}) \; P_{n}^{|m+\mu^{\prime}|} (\cos{\theta}) \cdot
      \sin{\theta} \, d\theta \,d\varphi.
\end{array}
\label{eq.3.5.10}
\end{equation}
Let's introduce the following coefficient $C_{l_{f} l_{ph} n}^{m \mu^{\prime}}$:
\begin{equation}
  C_{l_{f} l_{ph} n}^{m \mu^{\prime}} =
    (-1)^{l_{f}+n+1 - \mu^{\prime} + \frac{|m+\mu^{\prime}|}{2}} \;
    (n, 1, l_{ph} \big| -m-\mu^{\prime}, \mu^{\prime}, -m) \;
    \sqrt{\displaystyle\frac{(2l_{f}+1)\,(2n+1)}{32\pi}\;
          \displaystyle\frac{(l_{f}-1)!}{(l_{f}+1)!} \;
          \displaystyle\frac{(n-|m+\mu^{\prime}|)!}{(n+|m+\mu^{\prime}|)!}}
\label{eq.3.5.11}
\end{equation}
and function $f_{l_{f}n}^{m \mu^{\prime}}(\theta)$:
\begin{equation}
  f_{l_{f} n}^{m \mu^{\prime}}(\theta) =
    P_{l_{f}}^{1}(\cos{\theta}) \; P_{1}^{1}(\cos{\theta}) \; P_{n}^{|m+\mu^{\prime}|} (\cos{\theta}).
\label{eq.3.5.12}
\end{equation}
Then we obtain the total angular integrals $I_{M}(l_{f},l_{ph},n)$ and $I_{E}(l_{f},l_{ph},n)$, after integrating over $\varphi$:
\begin{equation}
\begin{array}{lcl}
  I_{M}(l_{f}; l_{ph}, n) & = &
    -m \,h_{-m} \;
    i^{l_{f}+n+1} \;
    \displaystyle\sum\limits_{\mu^{\prime} = \pm 1}
    C_{l_{f} l_{ph} n}^{m \mu^{\prime}}
    \displaystyle\int\limits_{0}^{\pi} \:
      f_{l_{f}n}^{m \mu^{\prime}}(\theta) \; \sin{\theta}\,d\theta, \\

  I_{E}(l_{f}; l_{ph},n) & = &
    h_{-m} \;
    i^{l_{f}+n+1} \;
    \displaystyle\sum\limits_{\mu^{\prime} = \pm 1}
    C_{l_{f}l_{ph} n}^{m \mu^{\prime}}
    \displaystyle\int\limits_{0}^{\pi} \:
      f_{l_{f}n}^{m \mu^{\prime}}(\theta) \; \sin{\theta}\,d\theta.
\end{array}
\label{eq.3.5.13}
\end{equation}

Now let's introduce the following differential expressions of these angular integrals:
\begin{equation}
\begin{array}{lcl}
  \displaystyle\frac{d\, I_{M}(l_{f}; l_{ph}, n)}{\sin{\theta}\,d\theta} & = &
    -m \,h_{-m} \;
    i^{l_{f}+n+1} \;
    \displaystyle\sum\limits_{\mu^{\prime} = \pm 1}
    C_{l_{f} l_{ph} n}^{m \mu^{\prime}} f_{l_{f}n}^{m \mu^{\prime}}(\theta), \\

  \displaystyle\frac{d\, I_{E}(l_{f}; l_{ph},n)}{\sin{\theta}\,d\theta} & = &
    h_{-m} \;
    i^{l_{f}+n+1} \;
    \displaystyle\sum\limits_{\mu^{\prime} = \pm 1}
    C_{l_{f}l_{ph} n}^{m \mu^{\prime}} f_{l_{f}n}^{m \mu^{\prime}}(\theta)
\end{array}
\label{eq.3.5.14}
\end{equation}
and define the differential matrix elements $dp_{l}^{M}$ and $dp_{l}^{E}$ dependent on the angle $\theta$:
\begin{equation}
\begin{array}{lcl}
  \displaystyle\frac{d \,p_{l}^{M}}{\sin{\theta}\,d\theta} & = &
    - \displaystyle\frac{d\, I_{M}(l_{f},l_{ph},l_{ph})}{\sin{\theta}\,d\theta} \cdot J(l_{f},l_{ph}) =
    m \,h_{-m} \;
    i^{l_{f}+l_{ph}+1} \;
    J(l_{f},l_{ph})
    \displaystyle\sum\limits_{\mu^{\prime} = \pm 1}
    C_{l_{f}l_{ph}l_{ph}}^{m \mu^{\prime}} f_{l_{f}l_{ph}}^{m \mu^{\prime}}(\theta), \\

  \displaystyle\frac{d \,p_{l}^{E}}{\sin{\theta}\,d\theta} & = &
    -\sqrt{\displaystyle\frac{l_{ph}+1}{2l_{ph}+1}} \cdot
       \displaystyle\frac{d \,I_{E}(l_{f},l_{ph},l_{ph}-1)}{\sin{\theta}\,d\theta} \cdot J(l_{f},l_{ph}-1) +
    \sqrt{\displaystyle\frac{l_{ph}}{2l_{ph}+1}} \cdot
      \displaystyle\frac{d\,I_{E}(l_{f},l_{ph},l_{ph}+1)}{\sin{\theta}\,d\theta} \cdot J(l_{f},l_{ph}+1) = \\

  & = &
    - h_{-m} \;
    i^{l_{f}+l_{ph}} \;
    \sqrt{\displaystyle\frac{l_{ph}+1}{2l_{ph}+1}} \, J(l_{f},l_{ph}-1)
    \displaystyle\sum\limits_{\mu^{\prime} = \pm 1}
      C_{l_{f},l_{ph},l_{ph}-1}^{m \mu^{\prime}} \: f_{l_{f},l_{ph}-1}^{m \mu^{\prime}}(\theta) \: + \\
  & + &
    h_{-m}
    i^{l_{f}+l_{ph}+2} \;
    \sqrt{\displaystyle\frac{l_{ph}}{2l_{ph}+1}} \, J(l_{f},l_{ph}+1)
    \displaystyle\sum\limits_{\mu^{\prime} = \pm 1}
      C_{l_{f},l_{ph},l_{ph}+1}^{m \mu^{\prime}} \: f_{l_{f},l_{ph}+1}^{m \mu^{\prime}}(\theta).
\end{array}
\label{eq.3.5.15}
\end{equation}
One can see that integration of such functions by angle $\theta$ with limits from 0 to $\pi$ gives the total matrix elements $p_{l}^{M}$ and $p_{l}^{E}$ exactly.

For transition into superposition of all possible final states with different $m$ at the same $l_{f}$ instead of (\ref{eq.3.5.15}) we obtain:
\begin{equation}
\begin{array}{lcl}
  \displaystyle\frac{d \,\tilde{p}_{l}^{M}}{\sin{\theta}\,d\theta} & = &
    i^{l_{f}+l_{ph}+1} \;
    J(l_{f},l_{ph})
    \displaystyle\sum\limits_{m = \pm 1}
    m \,h_{-m} \;
    \displaystyle\sum\limits_{\mu^{\prime} = \pm 1}
    C_{l_{f}l_{ph}l_{ph}}^{m \mu^{\prime}} f_{l_{f}l_{ph}}^{m \mu^{\prime}}(\theta), \\

  \displaystyle\frac{d \,\tilde{p}_{l}^{E}}{\sin{\theta}\,d\theta} & = &
    -i^{l_{f}+l_{ph}} \;
    \sqrt{\displaystyle\frac{l_{ph}+1}{2l_{ph}+1}} \, J(l_{f},l_{ph}-1)
    \displaystyle\sum\limits_{m = \pm 1}
    h_{-m} \;
    \displaystyle\sum\limits_{\mu^{\prime} = \pm 1}
      C_{l_{f},l_{ph},l_{ph}-1}^{m \mu^{\prime}} \: f_{l_{f},l_{ph}-1}^{m \mu^{\prime}}(\theta) \: - \\
  & - &
    i^{l_{f}+l_{ph}} \;
    \sqrt{\displaystyle\frac{l_{ph}}{2l_{ph}+1}} \, J(l_{f},l_{ph}+1)
    \displaystyle\sum\limits_{m = \pm 1}
    h_{-m} \;
    \displaystyle\sum\limits_{\mu^{\prime} = \pm 1}
      C_{l_{f},l_{ph},l_{ph}+1}^{m \mu^{\prime}} \: f_{l_{f},l_{ph}+1}^{m \mu^{\prime}}(\theta).
\end{array}
\label{eq.3.5.16}
\end{equation}
%-----------------------------------------------------------------------------------------------------------------------

%-----------------------------------------------------------------------------------------------------------------------
% \newpage
\subsection{The angular and integral matrix elements at the first values of $l_{f}$, $l_{ph}$
\label{sec.3.6}}

We shall find the matrix element at the first values of $l_{f}$ and $l_{ph}$. We use
\begin{equation}
\begin{array}{cc}
  l_{f} = 1, &
  l_{ph} = 1.
\end{array}
\label{eq.3.6.1}
\end{equation}
From (\ref{eq.3.5.16}) we write:
\begin{equation}
\begin{array}{lcl}
  \displaystyle\frac{d \,\tilde{p}_{1}^{M}}{\sin{\theta}\,d\theta} & = &
    - \: i\,
    J(1,1) \cdot
    \displaystyle\sum\limits_{m = \pm 1}
    m \,h_{-m} \;
    \displaystyle\sum\limits_{\mu^{\prime} = \pm 1}
    C_{111}^{m \mu^{\prime}} f_{11}^{m \mu^{\prime}}(\theta), \\

  \displaystyle\frac{d \,\tilde{p}_{1}^{E}}{\sin{\theta}\,d\theta} & = &
    \sqrt{\displaystyle\frac{2}{3}} \, J(1,0)
    \displaystyle\sum\limits_{m = \pm 1}
    h_{-m}
    \displaystyle\sum\limits_{\mu^{\prime} = \pm 1}
      C_{110}^{m \mu^{\prime}} \: f_{10}^{m \mu^{\prime}}(\theta) \: +
    \sqrt{\displaystyle\frac{1}{3}} \, J(1,2)
    \displaystyle\sum\limits_{m = \pm 1}
    h_{-m}
    \displaystyle\sum\limits_{\mu^{\prime} = \pm 1}
      C_{112}^{m \mu^{\prime}} \: f_{12}^{m \mu^{\prime}}(\theta).
\end{array}
\label{eq.3.6.2}
\end{equation}
Calculating coefficients $C_{11 n}^{m \mu^{\prime}}$ and functions $f_{1 n}^{m \mu^{\prime}}(\theta)$ (see Appendix~\ref{app.4}, \ref{app.5}), we obtain:
\[
\begin{array}{lcl}
  \displaystyle\frac{d \,\tilde{p}_{1}^{M}}{\sin{\theta}\,d\theta} & = &
    - \: i\, J(1,1) \cdot
    \biggl( - h_{1} \; C_{111}^{-1, 1} f_{11}^{-1, 1}(\theta) +
      h_{-1} \; C_{111}^{1, -1} f_{11}^{1, -1}(\theta)\biggr) = \\

  & = &
    - \: i\, J(1,1) \cdot
    \biggl( h_{1} \cdot
           \displaystyle\frac{3}{8} \: \sqrt{\displaystyle\frac{1}{2\,\pi}} \cdot
           \sin^{2}{\theta} \cos{\theta} +
           h_{-1} \cdot
           \displaystyle\frac{3}{8} \: \sqrt{\displaystyle\frac{1}{2\,\pi}} \cdot
           \sin^{2}{\theta} \cos{\theta}
    \biggr) = \\

  & = &
    - \: i\, J(1,1) \cdot
    \displaystyle\frac{3}{8} \: \sqrt{\displaystyle\frac{1}{2\,\pi}} \cdot
    \Bigl( h_{1} + h_{-1} \Bigr) \cdot
    \sin^{2}{\theta} \cos{\theta} = \\

  & = &
    - \displaystyle\frac{3}{8} \: \sqrt{\displaystyle\frac{1}{\pi}} \cdot
    J(1,1) \cdot
    \sin^{2}{\theta} \cos{\theta},
\end{array}
\]
\[
\begin{array}{lcl}
  \displaystyle\frac{d \,\tilde{p}_{1}^{E}}{\sin{\theta}\,d\theta} & = &

    \sqrt{\displaystyle\frac{2}{3}} \, J(1,0)
    \biggl( h_{1} C_{110}^{-1, 1} \: f_{10}^{-1, 1}(\theta) +
      h_{-1} C_{110}^{1, -1} \: f_{10}^{1, -1}(\theta)
    \biggr) + \\
\vspace{2mm}
  & + &
    \sqrt{\displaystyle\frac{1}{3}} \, J(1,2)
    \biggl(
      h_{1} C_{112}^{-1, -1} \: f_{12}^{-1, -1}(\theta) +
      h_{1} C_{112}^{-1, 1}  \: f_{12}^{-1, 1}(\theta) +
      h_{-1} C_{112}^{1, -1} \: f_{12}^{1, -1}(\theta) +
      h_{-1} C_{112}^{1, 1}  \: f_{12}^{1, 1}(\theta)
    \biggr) = \\

  & = &
    \sqrt{\displaystyle\frac{2}{3}} \, J(1,0)
    \biggl( - h_{1} \cdot
            \displaystyle\frac{1}{8} \: \sqrt{\displaystyle\frac{3}{2\,\pi}} \cdot
            \sin^{2}{\theta} -
            h_{-1} \cdot
            \displaystyle\frac{1}{8} \: \sqrt{\displaystyle\frac{3}{2\,\pi}} \cdot
            \sin^{2}{\theta}
    \biggr) + \\
  & + &
    \sqrt{\displaystyle\frac{1}{3}} \, J(1,2)
    \biggl( h_{1} \cdot
            \displaystyle\frac{1}{16} \: \sqrt{\displaystyle\frac{3}{2\,\pi}} \cdot
            3 \sin^{4}{\theta} -
            h_{1} \cdot
            \displaystyle\frac{1}{8} \: \sqrt{\displaystyle\frac{3}{2\,\pi}} \cdot
            \frac{1}{2} \sin^{2}{\theta} \:(3\cos^{2}{\theta} - 1) - \\
\vspace{2mm}
  & - &
            h_{-1} \cdot
            \displaystyle\frac{1}{8} \: \sqrt{\displaystyle\frac{3}{2\,\pi}} \cdot
            \frac{1}{2} \sin^{2}{\theta} \:(3\cos^{2}{\theta} - 1) +
            h_{-1} \cdot
            \displaystyle\frac{1}{16} \: \sqrt{\displaystyle\frac{3}{2\,\pi}} \cdot
            3 \sin^{4}{\theta}
    \biggr) = \\

  & = &
    - \sqrt{\displaystyle\frac{2}{3}} \, J(1,0) \cdot
    \displaystyle\frac{1}{8} \: \sqrt{\displaystyle\frac{3}{2\,\pi}} \cdot
    \Bigl( h_{1} + h_{-1} \Bigr) \cdot
    \sin^{2}{\theta} \: + \\
  \vspace{3mm}
  & + &
    \sqrt{\displaystyle\frac{1}{3}} \, J(1,2) \cdot
    \Biggl[
      \displaystyle\frac{1}{16} \: \sqrt{\displaystyle\frac{3}{2\,\pi}} \cdot
      \Bigl( h_{1} + h_{-1} \Bigr) \cdot
      3 \sin^{4}{\theta} -
      \displaystyle\frac{1}{8} \: \sqrt{\displaystyle\frac{3}{2\,\pi}} \cdot
      \Bigl( h_{1} + h_{-1} \Bigr)
      \frac{1}{2} \sin^{2}{\theta} \:(3\cos^{2}{\theta} - 1)
    \Biggr] = \\

  & = &
    i \: \displaystyle\frac{1}{8} \: \sqrt{\displaystyle\frac{2}{\pi}} \cdot  J(1,0) \cdot  \sin^{2}{\theta} \: + \:
    i \: \displaystyle\frac{1}{8} \: \sqrt{\displaystyle\frac{1}{\pi}} \cdot  J(1,2) \cdot
    \sin^{2}{\theta} \: \Bigl( 1 - 3 \sin^{2}{\theta} \Bigr) \\
\end{array}
\]
or
\vspace{3mm}
\begin{equation}
\begin{array}{lcl}
  \displaystyle\frac{d \,\tilde{p}_{1}^{M}}{\sin{\theta}\,d\theta} & = &
    - \displaystyle\frac{3}{8} \: \sqrt{\displaystyle\frac{1}{\pi}} \cdot
    J(1,1) \cdot
    \sin^{2}{\theta} \cos{\theta}, \\
  \displaystyle\frac{d \,\tilde{p}_{1}^{E}}{\sin{\theta}\,d\theta} & = &
    i \: \displaystyle\frac{1}{8} \: \sqrt{\displaystyle\frac{2}{\pi}} \cdot  J(1,0) \cdot  \sin^{2}{\theta} \: + \:
    i \: \displaystyle\frac{1}{8} \: \sqrt{\displaystyle\frac{1}{\pi}} \cdot  J(1,2) \cdot
    \sin^{2}{\theta} \: \Bigl( 1 - 3 \sin^{2}{\theta} \Bigr).
\end{array}
\label{eq.3.6.3}
\end{equation}

Also we find the integral matrix elements. Integrating (\ref{eq.3.6.3}) by angle $\theta$, we obtain:
\[
\begin{array}{lcl}
  \tilde{p}_{1}^{M} & = &
    -\displaystyle\frac{3}{8} \: \sqrt{\displaystyle\frac{1}{\pi}} \; J(1,1) \cdot
      \displaystyle\int\limits_{0}^{\pi}
      \sin^{2}{\theta} \cos{\theta} \cdot \sin{\theta}\; d\theta =
    -\displaystyle\frac{3}{8} \: \sqrt{\displaystyle\frac{1}{\pi}} \; J(1,1) \cdot
      \displaystyle\int\limits_{0}^{\pi} \sin^{3}{\theta} \; d\sin{\theta} = 0,
\end{array}
\]
\[
\begin{array}{lcl}
  \tilde{p}_{1}^{E} & = &
    i \: \displaystyle\frac{1}{8} \: \sqrt{\displaystyle\frac{2}{\pi}} \;  J(1,0) \cdot
      \displaystyle\int\limits_{0}^{\pi}
      \sin^{2}{\theta} \cdot \sin{\theta}\;d\theta + \:
    i \: \displaystyle\frac{1}{8} \: \sqrt{\displaystyle\frac{1}{\pi}} \;  J(1,2) \cdot
      \displaystyle\int\limits_{0}^{\pi}
      \sin^{2}{\theta} \: \Bigl( 1 - 3 \sin^{2}{\theta} \Bigr) \cdot \sin{\theta}\;d\theta = \\

  & = &
    -i \: \displaystyle\frac{1}{8} \: \sqrt{\displaystyle\frac{2}{\pi}} \;  J(1,0) \cdot
      \displaystyle\int\limits_{0}^{\pi}
      \bigl(1 - \cos^{2}{\theta} \bigr) \;d \cos{\theta} - \:
    i \: \displaystyle\frac{1}{8} \: \sqrt{\displaystyle\frac{1}{\pi}} \;  J(1,2) \cdot
      \displaystyle\int\limits_{0}^{\pi}
      \Bigl(-2 + 5 \cos^{2}{\theta} - 3 \cos^{4}{\theta} \Bigr) \;d\cos{\theta} = \\

\vspace{2mm}
  & = &
    -i \: \displaystyle\frac{1}{8} \: \sqrt{\displaystyle\frac{2}{\pi}} \;  J(1,0) \cdot
      \biggl\{\cos{\theta} - \displaystyle\frac{\cos^{3}{\theta}}{3} \biggr\} \bigg|_{0}^{\pi} - \:
    i \: \displaystyle\frac{1}{8} \: \sqrt{\displaystyle\frac{1}{\pi}} \;  J(1,2) \cdot
      \biggl\{-2\cos{\theta} + \displaystyle\frac{5 \cos^{3}{\theta}}{3} -
            \displaystyle\frac{3 \cos^{5}{\theta}}{5}
      \biggr\} \bigg|_{0}^{\pi} = \\

\vspace{2mm}
  & = &
    i \: \displaystyle\frac{1}{4} \: \sqrt{\displaystyle\frac{2}{\pi}} \;  J(1,0) \cdot
      \biggl\{ 1 - \displaystyle\frac{1}{3} \biggr\} + \:
    i \: \displaystyle\frac{1}{4} \: \sqrt{\displaystyle\frac{1}{\pi}} \;  J(1,2) \cdot
      \biggl\{-2 + \displaystyle\frac{5}{3} - \displaystyle\frac{3}{5} \biggr\} =

    i \: \displaystyle\frac{1}{6} \, \sqrt{\displaystyle\frac{2}{\pi}} \cdot
    \biggl\{ J(1,0) - \displaystyle\frac{7}{5} \, \sqrt{\displaystyle\frac{1}{2}} \cdot J(1,2) \biggr\}
\end{array}
\]
or
\begin{equation}
\begin{array}{lcl}
  \tilde{p}_{1}^{M} & = & 0, \\
  \tilde{p}_{1}^{E} & = &
    i \: \displaystyle\frac{1}{6} \, \sqrt{\displaystyle\frac{2}{\pi}} \cdot
    \Bigl\{ J(1,0) - \displaystyle\frac{7}{10} \, \sqrt{2} \cdot J(1,2) \Bigr\}.
\end{array}
\label{eq.3.6.4}
\end{equation}
% *******************************************************************************************************************

% *******************************************************************************************************************
% \newpage
\section{Probability of photon emission with impulse $\mathbf{k}$ and polarization $\mathbf{e}^{(\alpha)}$
\label{sec.4}}

Now we define probability of photon emission when before emission (i.e. in the initial $i$-state) there is the given flux of particles with quantum numbers $l_{i}=m_{i}=0$, and it needs to find average number of particles after photon emission (i.e. in the final $f$-state), scattering with the interesting angle relatively direction of photon emission and having given quantum numbers $l_{f}$, $m_{f}$. Further, we shall consider two approaches for definition of such probability.
%-----------------------------------------------------------------------------------------------------------------------

%-----------------------------------------------------------------------------------------------------------------------
\subsection{Angular probability of photon emission in dependence on direction of impulse of the $\alpha$-particle
\label{sec.4.1}}

In formulation of the first definition of angular probability we assume that direction of particle motion (with taking into account its tunneling) before and after photon emission, direction of photon emission are defined by \underline{impulse} $\mathbf{p}_{i}$ of the particle before emission, by the \underline{impulse} $\mathbf{p}_{f}$ of the particle after photon emission and by the \underline{impulse} $\mathbf{k}$ of photon emitted, correspondingly.

Probability of transition of the system in time unit from the initial $i$-state into final $f$-states with parameters, being in the given interval $d \nu_{f}$, with photon emission with possible impulses, in the given interval $d \nu_{ph}$, we define on the basis of matrix element $a_{fi}$ in the form (\ref{eq.2.5.9}) so:
\begin{equation}
\begin{array}{ll}
  d W = \displaystyle\frac{|a_{fi}|^{2}}{T} \cdot d\nu, &
  \hspace{7mm}
  d \nu = d\nu_{f} \cdot d\nu_{ph}.
\end{array}
\label{eq.4.1.1}
\end{equation}
Taking into account Exp.~(\ref{eq.2.5.11}) for square of the matrix element, we obtain (see \cite{Landau.v3.1989}, (42,5) sec.~42, p.~189; \cite{Berestetsky.1989}, sec.~44, p.~191):
\begin{equation}
  d W = 2\pi \:|F_{fi}|^{2} \: \delta (w_{f} - w_{i} + w) \cdot d\nu
\label{eq.4.1.2}
\end{equation}
where $d \nu$ is a set of parameters, characterized photon and particle in the final $f$-state.
In a general case, $d \nu_{ph}$ denotes a set of parameters, described photon and having continuous region of values. If the emission of photon with given impulse $\mathbf{k}$ is considered, then:
\begin{equation}
  d \nu_{ph} = \displaystyle\frac{d k^{3}}{(2\pi)^{3}} =
               \displaystyle\frac{w^{2} \, dw \,d\Omega_{ph}}{(2\pi c)^{3}}
\label{eq.4.1.3}
\end{equation}
where $d\Omega_{ph} = d\,\cos{\theta_{ph}} = \sin{\theta_{ph}} \,d\theta_{ph} \,d\varphi_{ph}$, $k_{ph}=w/c$. Here, one assume that wave function of photon (plane wave) is normalized on the one photon inside unite space volume $V=1$. Then $d \nu_{ph}$ is number of states inside unit phase volume $V\,dk^{3}$.

Now we shall be interesting in such transition of the system, when the particle after photon emission propagates with impulse inside the given interval $\mathbf{d p}_{f}$. Let's define interval of states $d \nu_{f}$, like (\ref{eq.4.1.3}) (also \cite{Landau.v3.1989} p.~599; in contrast with \cite{Bogoliubov.1980} p.~171--175 --- without devision on $(2\pi\hbar)^{3}$):
\begin{equation}
  d \nu_{f} = \displaystyle\frac{\mathbf{d p}_{f}}{(2\pi\hbar)^{3}} =
  \displaystyle\frac{dp_{x}\, dp_{y}\, dp_{z}}{(2\pi\hbar)^{3}} =
  \displaystyle\frac{p_{f}^{2} \, dp_{f}} {(2\pi\hbar)^{3}} \cdot d\Omega_{\mathbf{p}f}
\label{eq.4.1.4}
\end{equation}
where $d\Omega_{f} = d\,\cos{\theta_{f}} = \sin{\theta_{f}} \,d\theta_{f} \,d\varphi_{f}$, $p_{f}=|\mathbf{p}_{f}|$. Substituting definitions for $d \nu_{f}$ and $d \nu_{ph}$ into (\ref{eq.4.1.1}), integrating (\ref{eq.4.1.2}) at $dw$, we obtain:
\begin{equation}
\begin{array}{cc}
  d W    = \displaystyle\frac{w_{fi}^{2}\; p_{f}^{2}\; |F_{fi}|^{2}}{(2\pi)^{5}\, \hbar^{3}\, c^{3}} \;
           dp_{f}\; d \Omega_{ph}\, d \Omega_{f}, &
  w_{fi} = w_{i} - w_{f} = \displaystyle\frac{E_{i} - E_{f}}{\hbar}.
\end{array}
\label{eq.4.1.5}
\end{equation}
One can note, that in result of such one-dimensional integration by $dw$ we obtain law of conservation of energy, which was not taken into account previously (like, as in quantum field theory after 4-dimensional integration one can obtain law of conservation of energy-impulse; for example, see \cite{Bogoliubov.1980} p.~171--175). Substituting Exp.~(\ref{eq.2.5.4}) for $F_{fi}$ into (\ref{eq.4.1.5}), we obtain:
\begin{equation}
  d W = \displaystyle\frac{Z_{eff}^{2} \,e^{2}}{m^{2}}\:
            \displaystyle\frac{w_{fi}\: p_{f}^{2}}{(2\pi)^{4}\, \hbar^{2}\, c^{3}}\;
            \Bigl|p(k_{i}, k_{f})\Bigr|^{2}\; dp_{f}\; d \Omega_{ph} \, d \Omega_{f} =
        \displaystyle\frac{Z_{eff}^{2} \,e^{2}}{(2\pi)^{4}\, \hbar^{2}\, c^{3}}\:
            \displaystyle\frac{w_{fi}\: p_{f}^{2}}{m^{2}} \;
            \Biggl|\displaystyle\sum\limits_{\alpha =1,2}
            \mathbf{e}^{(\alpha), *} \; \mathbf{p} (k_{i}, k_{f}) \Biggr|^{2} \;
            dp_{f}\; d \Omega_{ph} \, d \Omega_{f}.
\label{eq.4.1.6}
\end{equation}

In consideration of impulse interval $dp_{f}$ it needs to take into account, that modul of impulse $p_{f}$ in the final $f$-state is not arbitrary but is defined by law of energy conservation (\ref{eq.4.1.5}). Using (as in \cite{Amusia.1990}, see p.~33):
\begin{equation}
  \displaystyle\frac{p_{f}^{2}}{2m} = \displaystyle\frac{p_{i}^{2}}{2m} - \hbar w_{ph},
\label{eq.4.1.7}
\end{equation}
we obtain
\begin{equation}
  p_{f}\: dp_{f} = -\hbar\,m\: dw_{ph}.
\label{eq.4.1.8}
\end{equation}
Taking into account this expression (its absolute values), from (\ref{eq.4.1.6}) we define \emph{differential angular probability of photon emission} so:
\begin{equation}
\begin{array}{ccccl}
  \displaystyle\frac{d W(\varphi_{f}, \theta_{f})} {dw_{ph}\: d\,\Omega_{ph}\: d\,\Omega_{f}} & = &
  \displaystyle\frac{d W(\varphi_{f}, \theta_{f})}
    {dw_{ph}\: d\,\Omega_{ph} \: \sin{\theta_{f}}\, d\,\theta_{f}\, d\varphi_{f}} & = &
      \displaystyle\frac{Z_{eff}^{2} \,e^{2}} {(2\pi)^{4}\, \hbar\,c^{3}}\:
      \displaystyle\frac{w_{fi}\:p_{f}}{m} \;
      \Biggl|\displaystyle\sum\limits_{\alpha =1,2}
      \mathbf{e}^{(\alpha), *} \; \mathbf{p} (k_{i}, k_{f}) \Biggr|^{2}, \\
  \displaystyle\frac{d W(\varphi_{f}, \theta_{f})} {dw_{ph}\: d\,\Omega_{ph}\: d\cos{\theta_{f}}} & = &
  \displaystyle\frac{d W(\varphi_{f}, \theta_{f})} {dw_{ph}\: d\,\Omega_{ph}\: \sin{\theta_{f}}\, d\,\theta_{f}} & = &
      \displaystyle\frac{Z_{eff}^{2} \,e^{2}} {(2\pi)^{3}\, \hbar\, c^{3}}\:
      \displaystyle\frac{w_{fi}\:p_{f}}{m} \;
      \Biggl|\displaystyle\sum\limits_{\alpha =1,2}
      \mathbf{e}^{(\alpha), *} \; \mathbf{p} (k_{i}, k_{f}) \Biggr|^{2}.
\end{array}
\label{eq.4.1.9}
\end{equation}
Taking into account, that the wave functions of the particle before and after photon emission have not the impulses $\mathbf{p}_{i,f}$, but they have their wave vectors $k_{i,f}$:
\begin{equation}
  k_{i,f} = \displaystyle\frac{p_{i,f}}{\hbar},
\label{eq.4.1.10}
\end{equation}
we rewrite:
\begin{equation}
\begin{array}{ccccl}
  \displaystyle\frac{d W(\varphi_{f}, \theta_{f})} {dw_{ph}\: d\,\Omega_{ph}\: d\,\Omega_{f}} & = &
  \displaystyle\frac{d W(\varphi_{f}, \theta_{f})}
    {dw_{ph}\: d\,\Omega_{ph} \: \sin{\theta_{f}}\, d\,\theta_{f}\, d\varphi_{f}} & = &
      \displaystyle\frac{Z_{eff}^{2}\, e^{2}} {(2\pi)^{4}\, c^{3}}\:
      \displaystyle\frac{w_{fi}\:k_{f}}{m} \;
      \Biggl|\displaystyle\sum\limits_{\alpha =1,2}
      \mathbf{e}^{(\alpha), *} \; \mathbf{p} (k_{i}, k_{f}) \Biggr|^{2}, \\

  \displaystyle\frac{d W(\varphi_{f}, \theta_{f})} {dw_{ph}\: d\,\Omega_{ph}\: d\cos{\theta_{f}}} & = &
  \displaystyle\frac{d W(\varphi_{f}, \theta_{f})} {dw_{ph}\: d\,\Omega_{ph}\: \sin{\theta_{f}}\, d\,\theta_{f}} & = &
      \displaystyle\frac{Z_{eff}^{2}\, e^{2}} {(2\pi\, c)^{3}}\:
      \displaystyle\frac{w_{fi}\:k_{f}}{m} \;
      \Biggl|\displaystyle\sum\limits_{\alpha =1,2}
      \mathbf{e}^{(\alpha), *} \; \mathbf{p} (k_{i}, k_{f}) \Biggr|^{2}.
\end{array}
\label{eq.4.1.11}
\end{equation}
These expressions define probability of emission of photon with impulse $\mathbf{k}$ (with averaging by polarizations $\mathbf{e}^{(\alpha)}$), where impulse of the particle after emission lays inside interval for the angle $d\,\Omega_{f}$ or angle $d\,\theta_{f}$. Note, that in definition of $F_{fi}$ integration by whole space at $\mathbf{r}$ is fulfilled (i.~e. averaging of wave functions of the particle before and after photon emission, wave function of photon is realized by whole space).

The probability of photon emission is inversely proportional to normalized volume $V$, which can be used arbitrary. With a purpose to obtain the characteristics, which characterizes the process of emission and does not depend on $V$, it needs to divide the differential probability of emission $d\, W$ on the flux $j$ of outgoing $\alpha$-particles in $\alpha$-decay, which is inversely proportional to this volume $V$ also. Write the differential probability so:
\begin{equation}
  d W(\varphi_{f}, \theta_{f}) = n_{i}\, v(\mathbf{p}_{i}) \cdot dP,
\label{eq.4.1.12}
\end{equation}
where $n_{i}$ is average number of particles in time unit before photon emission (in used by us normalization for wave function in the initial $i$-state we have $n_{i}=1$), $v(\mathbf{p}_{i})$ is module of velocity of outgoing particle in the frame system, where colliding center is not moved (which coincides with laboratory frame, where the second particle is not moved). Factor $P$ is proportional to the element of the angle of the particle after its scattering in result of photon emission and we shall name it as \emph{differential absolute probability} (while value $dW$ we shall name as the \emph{relative probability}). Such a definition coincides with definition (3.4.12) for the differential cross-section in \cite{Ahiezer.1981} (see sec.~3.4.4, p.~162; also sec.~4.3, p.~238--242).

According to (9.8) from \cite{Landau.v2.1988} (see sec.~9, p.~46), for particles with finite mass we have:
\begin{equation}
  \mathbf{p}_{i} = \displaystyle\frac{E_{i}\,\mathbf{v}_{i}} {c^{2}}.
\label{eq.4.1.13}
\end{equation}
From here we obtain:
\begin{equation}
\begin{array}{cc}
  \mathbf{v}_{i} = \displaystyle\frac{c^{2}\,\mathbf{p}_{i}} {E_{i}}, & \hspace{10mm}
  v_{i} = |\mathbf{v}_{i}| = \displaystyle\frac{c^{2}\,p_{i}} {E_{i}},
\end{array}
\label{eq.4.1.14}
\end{equation}
that agrees with $v(\mathbf{p}) = |\mathbf{p}| / p_{0}$ from \cite{Bogoliubov.1980} (at $c=1$, see sec.~21.4, p.~174).
Taking into account, that wave function of the particle before and after emission have not impulse $\mathbf{p}_{i,f}$ of this particle, but have the wave vector $k_{i,f}$, on the basis of (\ref{eq.4.1.10}) we rewrite expression for velocity so:
\begin{equation}
  v_{i} = \displaystyle\frac{\hbar\,c^{2}\,k_{i}} {E_{i}}.
\label{eq.4.1.15}
\end{equation}
From here we obtain equation of connection between differential relative and absolute probabilities:
\begin{equation}
  d\,P (\varphi_{f}, \theta_{f}) =
    \displaystyle\frac{d W(\varphi_{f}, \theta_{f})} {n_{i}\, v(\mathbf{k})} =
    d W(\varphi_{f}, \theta_{f}) \cdot \displaystyle\frac{E_{i}} {\hbar\, c^{2}\, k_{i}}
\label{eq.4.1.16}
\end{equation}
and from here we find:
\[
\begin{array}{ccl}
  d\,P (\varphi_{f}, \theta_{f}) & = &
    \displaystyle\frac{Z_{eff}^{2} \,e^{2}}{(2\pi)^{4}\, c^{3}}\:
      \displaystyle\frac{w_{fi}\: k_{f}}{m} \; \Bigl|p(k_{i}, k_{f})\Bigr|^{2}
      dw_{ph}\; d \Omega_{ph} \, d \Omega_{f} \cdot
      \displaystyle\frac{E_{i}} {\hbar\,c^{2}\,k_{i}} = \\
  & = &
    \displaystyle\frac{Z_{eff}^{2} \,e^{2}}{(2\pi)^{4}\, \hbar\, c^{5}}\:
      \displaystyle\frac{w_{ph}\,E_{i}\: k_{f}}{m\,k_{i}} \; \Bigl|p(k_{i}, k_{f})\Bigr|^{2} \cdot
      dw_{ph}\; d \Omega_{ph} \, d \Omega_{f}
\end{array}
\]
or
\begin{equation}
\begin{array}{ccl}
  \displaystyle\frac{d\,P (\varphi_{f}, \theta_{f})}{dw_{ph}\; d\Omega_{ph}\, d\Omega_{f}} & = &
    \displaystyle\frac{Z_{eff}^{2} \,e^{2}}{(2\pi)^{4}\, \hbar\, c^{5}}\:
      \displaystyle\frac{w_{ph}\,E_{i}\: k_{f}}{m\,k_{i}} \; \Bigl|p(k_{i}, k_{f})\Bigr|^{2}; \\
  \displaystyle\frac{d\,P (\varphi_{f}, \theta_{f})}{dw_{ph}\; d\Omega_{ph}\, d\cos{\theta_{f}}} & = &
    \displaystyle\frac{Z_{eff}^{2} \,e^{2}}{(2\pi)^{3}\,\hbar\, c^{5}}\:
    \displaystyle\frac{w_{ph}\,E_{i}\: k_{f}}{m\,k_{i}} \; \Bigl|p(k_{i}, k_{f})\Bigr|^{2}.
\end{array}
\label{eq.4.1.17}
\end{equation}

\vspace{5mm} We also define \emph{intensity of photon emission in dependence on angles} by multiplication of corresponding probabilities on $\hbar w$ (as in sec.45 \cite{Berestetsky.1989}):
\begin{equation}
\begin{array}{ccl}
  \displaystyle\frac{d I(\varphi_{f}, \theta_{f})} {dw_{ph}\: d\,\Omega_{ph}\: d\,\Omega_{f}} & = &
      \displaystyle\frac{Z_{eff}^{2} \,e^{2}\, \hbar} {(2\pi)^{4}\, c^{3}}\:
      \displaystyle\frac{w_{ph}^{2}\:k_{f}}{m} \;
      \Biggl|\displaystyle\sum\limits_{\alpha =1,2}
      \mathbf{e}^{(\alpha), *} \; \mathbf{p} (k_{i}, k_{f}) \Biggr|^{2}, \\

  \displaystyle\frac{d I(\varphi_{f}, \theta_{f})} {dw_{ph}\: d\,\Omega_{ph}\: d\cos{\theta_{f}}} & = &
      \displaystyle\frac{Z_{eff}^{2} \,e^{2}\, \hbar} {(2\pi\, c)^{3}}\:
      \displaystyle\frac{w_{ph}^{2}\:k_{f}}{m} \;
      \Biggl|\displaystyle\sum\limits_{\alpha =1,2}
      \mathbf{e}^{(\alpha), *} \; \mathbf{p} (k_{i}, k_{f}) \Biggr|^{2}.
\end{array}
\label{eq.4.1.18}
\end{equation}
We see, that so defined angular relative probabilities (\ref{eq.4.1.11}), absolute probabilities (\ref{eq.4.1.16}) and intensities (\ref{eq.4.1.17}) have real values.
%-----------------------------------------------------------------------------------------------------------------------

%-----------------------------------------------------------------------------------------------------------------------
% \newpage
\subsection{Angular probability of photon emission in dependence on direction of motion of the $\alpha$-particle
\label{sec.4.2}}

Now we consider another way of introduction of angular probability of bremsstrahlung.
The probability of transition of the system (during time unit) from the initial $i$-state into the final $f$-state, being in the given interval $d \nu_{f}$, with emission of photon with possible impulses inside the given interval $d \nu_{ph}$, we define so (see \cite{Landau.v3.1989}, (42,5) sec.~42, p.~189; \cite{Berestetsky.1989}, sec.~44, p.~191):
\begin{equation}
\begin{array}{ll}
  d W = \displaystyle\frac{|a_{fi}|^{2}}{T} \cdot d\nu =
    2\pi \:|F_{fi}|^{2} \: \delta (w_{f} - w_{i} + w) \cdot d\nu, &
  \hspace{7mm}
  d \nu = d\nu_{f} \cdot d\nu_{ph}
\end{array}
\label{eq.4.2.1}
\end{equation}
where $d \nu$ is values characterizing photon and particle in the final $f$-state. If the emission of photon wiht impulse $\mathbf{k}$ is considered then
\begin{equation}
  d \nu_{ph} = \displaystyle\frac{d^{3} k}{(2\pi)^{3}} =
          \displaystyle\frac{w^{2} \, dw \,d\Omega_{ph}}{(2\pi c)^{3}},
\label{eq.4.2.2}
\end{equation}
where $d\Omega_{ph} = d\,\cos{\theta_{ph}} = \sin{\theta_{ph}} \,d\theta_{ph} \,d\varphi_{ph}$, $k_{ph}=w/c$.
Then integrating (\ref{eq.4.2.1}) by $dw$, we obtain:
\begin{equation}
\begin{array}{cc}
  d W    = \displaystyle\frac{w_{fi}^{2}\; |F_{fi}|^{2}}{(2\pi)^{2} \:c^{3}} \; d \Omega_{ph}\, d\nu_{f}, &
  w_{fi} = w_{i} - w_{f} = \displaystyle\frac{E_{i} - E_{f}}{\hbar}.
\end{array}
\label{eq.4.2.3}
\end{equation}
Now we note concerning interval $d\,\nu_{f}$. In definition (\ref{eq.4.2.1}) we use matrix element $F_{fi}$ which we define as integral over space with possible summation by some quantum numbers of the system in the final $f$-state. One can consider such procedure as averaging by these characteristics and then $F_{fi}$ does not depend on them. Therefore, we shall suppose that interval $d\,\nu_{f}$ in definition (\ref{eq.4.2.1}) takes into account only such additional characteristics and quantum numbers of the system in the final $f$-state, by which integration or summation was not fulfilled in definition of $F_{fi}$.

\vspace{5mm}
Substituting Exp. (\ref{eq.2.5.4}) of $F_{fi}$ into (\ref{eq.4.2.3}), we obtain:
\begin{equation}
\begin{array}{lcl}
  d W & = & \displaystyle\frac{Z_{eff}^{2} \,e^{2}}{m^{2}}\:
            \displaystyle\frac{\hbar\, w_{fi}}{2\pi \,c^{3}} \; \Bigl|p(k_{i}, k_{f})\Bigr|^{2} \;
            d \Omega_{ph} \, d\nu_{f} =
      \displaystyle\frac{Z_{eff}^{2} \hbar\, \,e^{2}}{2\pi \,c^{3}}\:
            \displaystyle\frac{w_{fi}}{m^{2}} \;
            \Biggl|\displaystyle\sum\limits_{\alpha =1,2}
            \mathbf{e}^{(\alpha), *} \; \mathbf{p} (k_{i}, k_{f}) \Biggr|^{2} \; d \Omega_{ph} \, d \nu_{f}.
\end{array}
\label{eq.4.2.4}
\end{equation}
This expression represents probability of the photon emission with impulse $\mathbf{k}$ (and with averaging by polarization $\mathbf{e}^{(\alpha)}$) where the integration over angles of the particle motion after the photon emission has already fulfilled. Such probability is averaged over all possible directions of the particle motion after emission and therefore does not depend on them.

For description of the photon emission with impulse $\mathbf{k}$ with taking into account direction $\mathbf{n}_{\mathbf{r}}^{f}$ of motion (or tunneling) of the particle after emission we introduce the probability of emission so:

\vspace{3mm}
\noindent
\emph{
We shall define \underline{differential angular probability} concerning angle $\theta$ (and \underline{differential angular} \underline{probability} \underline{concerning space angle} $\Omega$) such a function, definite integral of which by the angle $\theta$ with limits from 0 to $\pi$ (definite space integral over angles $\theta$ and $\varphi$) corresponds exactly to the total probability of photon emission (\ref{eq.4.2.5}).}

\vspace{3mm}
\noindent
Let's consider two function:
\begin{equation}
\begin{array}{ccl}
  \displaystyle\frac{d W(\varphi_{f}, \theta_{f})} {d\,\Omega_{ph} \: d\,\Omega_{f}} =
  \displaystyle\frac{d W(\varphi_{f}, \theta_{f})}
    {d\,\Omega_{ph} \: \sin{\theta_{f}}\, d\,\theta_{f}\, d\varphi_{f}} & = &
      \displaystyle\frac{Z_{eff}^{2}\, \hbar\, e^{2}}{2\pi\, c^{3}}\:
      \displaystyle\frac{w_{fi}}{m^{2}} \;
    \displaystyle\frac{d}{d\,\Omega_{f}}\,
      \Bigl| p (k_{i}, k_{f}) \Bigr|^{2}, \\

  \displaystyle\frac{d W(\theta_{f})} {d\,\Omega_{ph} \: d\cos{\theta_{f}}} =
  \displaystyle\frac{d W(\theta_{f})} {d\,\Omega_{ph} \: \sin{\theta_{f}}\, d\,\theta_{f}} & = &
      \displaystyle\frac{Z_{eff}^{2}\, \hbar\, e^{2}}{2\pi\, c^{3}}\:
      \displaystyle\frac{w_{fi}}{m^{2}} \;
    \displaystyle\frac{d}{\sin{\theta_{f}} \,d \theta_{f}}\,
      \Bigl| p (k_{i},k_{f}) \Bigr|^{2}.
\end{array}
\label{eq.4.2.5}
\end{equation}
Introducing differential matrix elements dependent on angle $\theta$ and space angle $\Omega$ (like definition (\ref{eq.3.5.16})), we obtain:
\begin{equation}
\begin{array}{ccl}
\vspace{2mm}
  \displaystyle\frac{d W(\varphi_{f}, \theta_{f})} {d\,\Omega_{ph} \: d\,\Omega_{f}} & = &
  \displaystyle\frac{Z_{eff}^{2}\, \hbar\, e^{2}}{2\pi\, c^{3}}\: \displaystyle\frac{w_{fi}}{m^{2}} \;
    \biggl\{p\,(k_{i},k_{f}) \displaystyle\frac{d\, p^{*}(k_{i},k_{f}, \Omega_{f})}{d\,\Omega_{f}} + {\rm h. e.} \biggr\}, \\

  \displaystyle\frac{d W(\theta_{f})} {d\,\Omega_{ph} \: d\cos{\theta_{f}}} & = &
  \displaystyle\frac{Z_{eff}^{2}\, \hbar\, e^{2}}{2\pi\, c^{3}}\: \displaystyle\frac{w_{fi}}{m^{2}} \;
    \biggl\{p\,(k_{i},k_{f}) \displaystyle\frac{d\, p^{*}(k_{i},k_{f}, \theta_{f})}{d\cos{\theta_{f}}} + {\rm h. e.} \biggr\}.
\end{array}
\label{eq.4.2.6}
\end{equation}
One can see that so constructed functions satisfies exactly for such definition and, therefore, they can be used as definitions of the angular probabilities.

\emph{The total (integrated over angles) probability of the photon emission} is:
\begin{equation}
\begin{array}{ccl}
\vspace{2mm}
  W & = &
    \displaystyle\frac{Z_{eff}^{2}\, \hbar\, e^{2}}{2\pi\, c^{3}}\: \displaystyle\frac{w_{fi}}{m^{2}} \;
    \Bigl| p\,(k_{i},k_{f}) \Bigr|^{2}.
\end{array}
\label{eq.4.2.7}
\end{equation}

\emph{Intensity of emission in dependence on angles} we define by multiplication of corresponding probabilities on $\hbar w$ (like sec.~45 in \cite{Berestetsky.1989}):
\begin{equation}
\begin{array}{ccl}
  d I(\varphi_{f}, \theta_{f}, \,\Omega_{ph}) & = &
    \displaystyle\frac{Z_{eff}^{2}\, \hbar^{2}\, e^{2}}{2\pi\,c^{3}}\: \displaystyle\frac{w_{fi}^{2}}{m^{2}} \;
    \biggl\{p\,(k_{i},k_{f}) \displaystyle\frac{d\, p^{*}(k_{i},k_{f},\Omega_{f})}{d\,\Omega_{f}} + {\rm h. e.} \biggr\}, \\

  d I(\theta_{f}, \,\Omega_{ph}) & = &
    \displaystyle\frac{Z_{eff}^{2}\, \hbar^{2}\, e^{2}}{2\pi\,c^{3}}\: \displaystyle\frac{w_{fi}^{2}}{m^{2}} \;
    \biggl\{p\,(k_{i},k_{f}) \displaystyle\frac{d\, p^{*}(k_{i},k_{f},\theta_{f})}{d\cos{\theta_{f}}} + {\rm h. e.} \biggr\}.
\end{array}
\label{eq.4.2.8}
\end{equation}
From (\ref{eq.4.2.7}) and (\ref{eq.4.2.8}) we see that so defined angular probabilities and intensities are real.

If the probability $W$ has dimension of mass and coincides with width $\Gamma$ then one can define inverse value $\tau$ to it:
\begin{equation}
\begin{array}{cc}
  \tau = \displaystyle\frac{\hbar}{\Gamma}, &
  \Gamma = W.
\end{array}
\label{eq.4.2.9}
\end{equation}
According to \cite{Bogoliubov.1980} (see p.~175), in considering of transition of the system from the initial state to the final one the value $\tau$ represents half-life of this system in the initial state, i.~e. before photon emission.

\emph{Differential absolute probability} we define on the basis of the found differential relative probabilities, like this has done in the previous section. Using (\ref{eq.4.1.16}), we obtain ($n_{i}=1$):
\[
\begin{array}{ccl}
  d\,P (\varphi_{f}, \theta_{f}) & = &
    \displaystyle\frac{d W(\varphi_{f}, \theta_{f})} {n_{i}\, v(\mathbf{k})} =
    d W(\varphi_{f}, \theta_{f}) \cdot \displaystyle\frac{E_{i}} {\hbar\, c^{2}\, k_{i}} = \\
  & = &
    \displaystyle\frac{Z_{eff}^{2}\, \hbar\, e^{2}}{2\pi\, c^{3}}\:
      \displaystyle\frac{w_{fi}}{m^{2}} \;
      \biggl\{p\,(k_{i},k_{f}) \displaystyle\frac{d\, p^{*}(k_{i},k_{f}, \Omega_{f})}{d\,\Omega_{f}} + {\rm h. e.} \biggr\}\;
      d \Omega_{ph} \, d \Omega_{f} \cdot
      \displaystyle\frac{E_{i}} {\hbar\,c^{2}\,k_{i}} = \\
  & = &
    \displaystyle\frac{Z_{eff}^{2} \,e^{2}}{2\pi\, c^{5}}\:
      \displaystyle\frac{w_{ph}\,E_{i}}{m^{2}\,k_{i}} \;
      \biggl\{p\,(k_{i},k_{f}) \displaystyle\frac{d\, p^{*}(k_{i},k_{f}, \Omega_{f})}{d\,\Omega_{f}} + {\rm h. e.} \biggr\} \cdot
      d \Omega_{ph} \, d \Omega_{f}
\end{array}
\]
or
\begin{equation}
\begin{array}{ccl}
  \displaystyle\frac{d\,P (\varphi_{f}, \theta_{f})}{d\Omega_{ph}\, d\Omega_{f}} & = &
    \displaystyle\frac{Z_{eff}^{2} \,e^{2}}{2\pi\, c^{5}}\:
      \displaystyle\frac{w_{ph}\,E_{i}}{m^{2}\,k_{i}} \;
      \biggl\{p\,(k_{i},k_{f}) \displaystyle\frac{d\, p^{*}(k_{i},k_{f}, \Omega_{f})}{d\,\Omega_{f}} + {\rm h. e.} \biggr\}; \\
  \displaystyle\frac{d\,P (\varphi_{f}, \theta_{f})}{d\Omega_{ph}\, d\cos{\theta_{f}}} & = &
    \displaystyle\frac{Z_{eff}^{2} \,e^{2}}{2\pi\,c^{5}}\:
      \displaystyle\frac{w_{ph}\,E_{i}}{m^{2}\,k_{i}} \;
      \biggl\{p\,(k_{i},k_{f}) \displaystyle\frac{d\, p^{*}(k_{i},k_{f}, \Omega_{f})}{d\,\cos{\theta_{f}}} + {\rm h. e.} \biggr\}.
\end{array}
\label{eq.4.2.10}
\end{equation}
%-----------------------------------------------------------------------------------------------------------------------

%-----------------------------------------------------------------------------------------------------------------------
% \newpage
\subsection{Multipolar approach
\label{sec.4.3}}

Let's find the angular probability (\ref{eq.4.2.7}) in dependence on the angle $\theta$ for the first values $l_{f}=1$ and $l_{ph}=1$. According to (\ref{eq.3.4.4}), for the matrix element $p_{1}\, (k_{i},k_{f})$ we have:
\begin{equation}
\begin{array}{lcl}
\vspace{2mm}
  \tilde{p}_{1}\, (k_{i},k_{f}) & = &
    \sqrt{2\pi} \cdot
      \biggl\{
        (-i)^{l} \sqrt{2l+1} \:
        \Bigl[ \tilde{p}_{l}^{M} - i\, \tilde{p}_{l}^{E} \Bigr] \biggr\} \Biggr|_{l=1} =
    -\sqrt{6\pi}\, \Bigl( i\,\tilde{p}_{1}^{M} + \tilde{p}_{1}^{E} \Bigr), \\

  \displaystyle\frac{d \, \tilde{p}_{1}\, (k_{i},k_{f})}{\sin{\theta}\,d\theta} & = &
    \sqrt{2\pi} \cdot
      \biggl\{
      (-i)^{l} \sqrt{2l+1} \:
      \Bigl[
        \displaystyle\frac{d\, \tilde{p}_{l}^{M}}{\sin{\theta}\,d\theta} -
        i\, \displaystyle\frac{d\, \tilde{p}_{l}^{E}}{\sin{\theta}\,d\theta}
      \Bigr]
      \biggr\} \Biggr|_{l=1} =
    -\sqrt{6\pi} \,
      \biggl\{
        i\, \displaystyle\frac{d\, \tilde{p}_{1}^{M}}{\sin{\theta}\,d\theta} +
        \displaystyle\frac{d\, \tilde{p}_{1}^{E}}{\sin{\theta}\,d\theta}
      \biggr\}.
\end{array}
\label{eq.4.3.1}
\end{equation}
Using the found angular electrical and magnetic components of the matrix element (\ref{eq.3.6.3}):
\[
\begin{array}{lcl}
  \displaystyle\frac{d \,\tilde{p}_{1}^{M}}{\sin{\theta}\,d\theta} & = &
    - \displaystyle\frac{3}{8} \: \sqrt{\displaystyle\frac{1}{\pi}} \cdot
    J(1,1) \cdot
    \sin^{2}{\theta} \cos{\theta}, \\
  \displaystyle\frac{d \,\tilde{p}_{1}^{E}}{\sin{\theta}\,d\theta} & = &
    i \: \displaystyle\frac{1}{8} \: \sqrt{\displaystyle\frac{2}{\pi}} \cdot  J(1,0) \cdot  \sin^{2}{\theta} \: + \:
    i \: \displaystyle\frac{1}{8} \: \sqrt{\displaystyle\frac{1}{\pi}} \cdot  J(1,2) \cdot
    \sin^{2}{\theta} \: \Bigl( 1 - 3 \sin^{2}{\theta} \Bigr)
\end{array}
\]
and these integral components (\ref{eq.3.6.4}):
\[
\begin{array}{lcl}
  \tilde{p}_{1}^{M} & = & 0, \\
  \tilde{p}_{1}^{E} & = &
    i \: \displaystyle\frac{1}{6} \, \sqrt{\displaystyle\frac{2}{\pi}} \cdot
    \Bigl\{ J(1,0) - \displaystyle\frac{7}{10} \, \sqrt{2} \cdot J(1,2) \Bigr\},
\end{array}
\]
from (\ref{eq.4.3.1}) we obtain:
\begin{equation}
\begin{array}{lcl}
\vspace{2mm}
  \tilde{p}_{1}\, (k_{i},k_{f}) & = &
    - i \, \sqrt{\displaystyle\frac{1}{3}} \cdot
    \Bigl\{ J(1,0) - \displaystyle\frac{7}{10} \, \sqrt{2} \cdot J(1,2) \Bigr\}, \\

  \displaystyle\frac{d \, \tilde{p}_{1}\, (k_{i},k_{f})}{\sin{\theta}\,d\theta} & = &
    i\; \displaystyle\frac{\sqrt{6}}{8} \: \cdot
    \biggl\{
      3\,J(1,1) \cdot \cos{\theta} - \sqrt{2}\, J(1,0) - J(1,2) \cdot \Bigl( 1 - 3 \sin^{2}{\theta} \Bigr)
    \biggr\} \cdot \sin^{2}{\theta}.
\end{array}
\label{eq.4.3.2}
\end{equation}

Now we find the \emph{relative} angular probability from (\ref{eq.4.2.6}):
\[
\begin{array}{ccl}
\vspace{3mm}
  \displaystyle\frac{d W^{E1+M1}_{1}(\theta_{f})} {d\,\Omega_{ph} \: d\cos{\theta_{f}}} & = &
  \displaystyle\frac{Z_{eff}^{2}\, \hbar\, e^{2}}{2\pi\, c^{3}}\: \displaystyle\frac{w_{fi}}{m^{2}}\;
    \biggl\{p\,(k_{i},k_{f}) \displaystyle\frac{d\, p^{*}(k_{i},k_{f})}{d\cos{\theta_{f}}} + {\rm h. e.} \biggr\} = \\

  & = &
    \displaystyle\frac{Z_{eff}^{2}\, \hbar\, e^{2}}{2\pi\, c^{3}}\: \displaystyle\frac{w_{fi}}{m^{2}} \;
    \biggl\{
      i\, \sqrt{\displaystyle\frac{1}{3}} \cdot
      \Bigl\{ J(1,0) - \displaystyle\frac{7}{10} \, \sqrt{2} \cdot J(1,2) \Bigr\} \times \\
\vspace{4mm}
  & \times &
      i\; \displaystyle\frac{\sqrt{6}}{8} \cdot
      \biggl\{ 3\,J^{*}(1,1) \cdot \cos{\theta} - \sqrt{2}\,J^{*}(1,0) -
        J^{*}(1,2) \cdot \Bigl( 1 - 3 \sin^{2}{\theta} \Bigr)
      \biggr\} \cdot \sin^{2}{\theta}
    + {\rm h. e.} \biggr\} = \\

  & = &
    \displaystyle\frac{Z_{eff}^{2}\, \hbar\, e^{2}}{2\pi\, c^{3}}\: \displaystyle\frac{w_{fi}}{m^{2}} \;
    \biggl\{
      \displaystyle\frac{\sqrt{2}}{8} \cdot
      \Bigl[ J(1,0) - \displaystyle\frac{7}{10} \, \sqrt{2} \cdot J(1,2) \Bigr] \times \\
\vspace{4mm}
  & \times &
      \Bigl[
        \sqrt{2}\,J^{*}(1,0) + J^{*}(1,2) \cdot \Bigl( 1 - 3 \sin^{2}{\theta} \Bigr)
        - 3\,J^{*}(1,1) \cdot \cos{\theta}
      \Bigr]
    + {\rm h. e.} \biggr\} \cdot \sin^{2}{\theta} = \\

  & = &
    \displaystyle\frac{Z_{eff}^{2}\, \hbar\, e^{2}}{8\,\pi\, c^{3}}\: \displaystyle\frac{w_{fi}}{m^{2}} \;
    \biggl\{ \Bigl[ J(1,0) - \displaystyle\frac{7}{10} \, \sqrt{2} \cdot J(1,2) \Bigr] \times \\
  & \times &
      \Bigl[
        J^{*}(1,0) +
        \displaystyle\frac{1}{\sqrt{2}} J^{*}(1,2) \cdot \Bigl( 1 - 3 \sin^{2}{\theta} \Bigr)
        - \displaystyle\frac{3}{\sqrt{2}} \,J^{*}(1,1) \cdot \cos{\theta}
      \Bigr]
    + {\rm h. e.} \biggr\} \cdot \sin^{2}{\theta}
\end{array}
\]
or
\begin{equation}
\begin{array}{ccl}
\vspace{3mm}
  \displaystyle\frac{d W^{E1+M1}_{1}(\theta_{f})} {d\,\Omega_{ph} \: d\cos{\theta_{f}}} & = &
    \displaystyle\frac{Z_{eff}^{2}\, \hbar\, e^{2}}{8\,\pi\, c^{3}}\: \displaystyle\frac{w_{fi}}{m^{2}} \;
    \biggl\{ \Bigl[ J(1,0) - \displaystyle\frac{7}{10} \, \sqrt{2} \cdot J(1,2) \Bigr] \times \\
  & \times &
      \Bigl[
        J^{*}(1,0) +
        \displaystyle\frac{1}{\sqrt{2}} J^{*}(1,2) \cdot \Bigl( 1 - 3 \sin^{2}{\theta} \Bigr)
        - \displaystyle\frac{3}{\sqrt{2}} \,J^{*}(1,1) \cdot \cos{\theta}
      \Bigr]
    + {\rm h. e.} \biggr\} \cdot \sin^{2}{\theta}
\end{array}
\label{eq.4.3.3}
\end{equation}
and the \emph{absolute} angular probability from (\ref{eq.4.2.10}):
\begin{equation}
\begin{array}{ccl}
\vspace{3mm}
  \displaystyle\frac{d P^{E1+M1}_{1}(\theta_{f})} {d\,\Omega_{ph} \: d\cos{\theta_{f}}} & = &
    \displaystyle\frac{Z_{eff}^{2}\, e^{2}}{8\,\pi\, c^{5}}\:
    \displaystyle\frac{w_{fi}}{m^{2}}\,
    \displaystyle\frac{E_{i}}{k_{i}}\;
    \biggl\{ \Bigl[ J(1,0) - \displaystyle\frac{7}{10} \, \sqrt{2} \cdot J(1,2) \Bigr] \times \\
  & \times &
      \Bigl[
        J^{*}(1,0) +
        \displaystyle\frac{1}{\sqrt{2}} J^{*}(1,2) \cdot \Bigl( 1 - 3 \sin^{2}{\theta} \Bigr)
        - \displaystyle\frac{3}{\sqrt{2}} \,J^{*}(1,1) \cdot \cos{\theta}
      \Bigr]
    + {\rm h. e.} \biggr\} \cdot \sin^{2}{\theta}.
\end{array}
\label{eq.4.3.4}
\end{equation}
One can see that such effect takes place:  \emph{while magnetic component $p^{M}_{1}$ equals to zero, its differential part introduces nonzero contribution into the total probability of the photon emission}. One can rewrite this expression so:
\begin{equation}
\begin{array}{ccl}
  \displaystyle\frac{d W^{E1+M1}_{1}(\theta_{f})} {d\,\Omega_{ph} \: d\cos{\theta_{f}}} & = &
    \displaystyle\frac{d W^{E1}_{1}(\theta_{f})} {d\,\Omega_{ph} \: d\cos{\theta_{f}}} +
    \Delta\, \displaystyle\frac{d\, W^{M1}_{1}(\theta_{f})} {d\,\Omega_{ph} \: d\cos{\theta_{f}}},
\end{array}
\label{eq.4.3.5}
\end{equation}
where the first item has a form:
\begin{equation}
\begin{array}{ccl}
  \displaystyle\frac{d W^{E1}_{1}(\theta_{f})} {d\,\Omega_{ph} \: d\cos{\theta_{f}}} & = &
    \displaystyle\frac{Z_{eff}^{2}\, \hbar\, e^{2}}{8\,\pi\, c^{3}}\: \displaystyle\frac{w_{fi}}{m^{2}} \;
    \biggl\{ \Bigl[ J(1,0) - \displaystyle\frac{7}{10} \, \sqrt{2} \cdot J(1,2) \Bigr] \times \\
  & \times &
    \Bigl[
      J^{*}(1,0) + \displaystyle\frac{1}{\sqrt{2}} J^{*}(1,2) \cdot \Bigl( 1 - 3 \sin^{2}{\theta} \Bigr)
    \Bigr] + {\rm h. e.} \biggr\} \cdot \sin^{2}{\theta},
\end{array}
\label{eq.4.3.6}
\end{equation}
and determines probability of photon emission only on the basis of electric multipole E1. The second item in (\ref{eq.4.3.5}) has a form:
\begin{equation}
\begin{array}{ccl}
  \Delta\, \displaystyle\frac{d\, W^{E1+M1}_{1}(\theta_{f})} {d\,\Omega_{ph} \: d\cos{\theta_{f}}} & = &
    -\displaystyle\frac{3\,Z_{eff}^{2}\, \hbar\, e^{2}}{8\,\sqrt{2}\: \pi\, c^{3}}\: \displaystyle\frac{w_{fi}}{m^{2}} \;
    \biggl\{
      \Bigl[ J(1,0) - \displaystyle\frac{7}{10} \, \sqrt{2} \cdot J(1,2) \Bigr] \cdot J^{*}(1,1) + {\rm h. e.}
    \biggr\} \cdot \sin^{2}{\theta}\, \cos{\theta}
\end{array}
\label{eq.4.3.7}
\end{equation}
and determines correction to the probability of the photon emission in result of inclusion of the magnetic multipole M1.
In (\ref{eq.4.3.7}) one can see, that the found correction can be separated into radial and angular components. \emph{Therefore, in dependence on the angle it takes influence on emission equally for different energies of the photons emitted}. From here, one can see such value of the angle, when influence of the magnetic component on the photon emission will be minimal or maximal:
\begin{equation}
\begin{array}{ccl}
  f(\theta) & = & \sin^{2}{\theta}\, \cos{\theta}; \\
  \displaystyle\frac{d\,f(\theta)}{d\,\theta} & = &
    \displaystyle\frac{d}{d\,\theta} \sin^{2}{\theta}\,\cos{\theta} =
    2 \sin{\theta}\,\cos^{2}{\theta} - \sin^{3}{\theta} =
    \sin{\theta} \cdot \Bigl( 2\,\cos^{2}{\theta} - \sin^{2}{\theta} \Bigr) = \\
    & = &
    \sin{\theta} \cdot \Bigl( 2 - 3 \sin^{2}{\theta} \Bigr) = 0.
\end{array}
\label{eq.4.3.8}
\end{equation}
In result, we find (at $0 \le \theta \le \pi$):
\begin{equation}
\begin{array}{cll}
  1) & \theta_{1} = 0 & \mbox{ --- influence is absent}, \\
  2) & \theta_{2} = \arcsin{\sqrt{\displaystyle\frac{2}{3}}} & \mbox{ --- influence is maximal}.
\end{array}
\label{eq.4.3.9}
\end{equation}
% *******************************************************************************************************************

% *******************************************************************************************************************
\section{Nucleus--$\alpha$-particle potential
\label{sec.5}}

To describe the interaction between the $\alpha$-particle and daughter nucleus we use the potential given in~\cite{Denisov.2005.PHRVA} (see relations~(6)--(10) in the cited paper, also \cite{Maydanyuk.2008.EPJA,Maydanyuk.2008.MPLA}) in the general form
\begin{equation}
  V (r, \theta, l, Q) = v_{C} (r, \theta) + v_{N} (r, \theta, Q) + v_{l} (r)
\label{eq.5.1.1}
\end{equation}
where the Coulomb $v_{C} (r, \theta)$, nuclear $v_{N} (r, \theta, Q)$ and centrifugal $v_{l} (r)$ components are
\begin{equation}
  v_{C} (r, \theta) =
  \left\{
  \begin{array}{ll}
    \displaystyle\frac{2 Z e^{2}} {r}\;
      \biggl(1 + \displaystyle\frac{3 R^{2}} {5 r^{2}}\; \beta_{2} Y_{20}(\theta) \biggr), &
      \mbox{for  } r \ge r_{m}, \\
    \displaystyle\frac{2 Z e^{2}} {r_{m}}\;
    \biggl\{
      \displaystyle\frac{3}{2} -
      \displaystyle\frac{r^{2}}{2r_{m}^{2}} +
      \displaystyle\frac{3 R^{2}} {5 r_{m}^{2}}
      \Bigl(2 - \displaystyle\frac{r^{3}}{r_{m}^{3}} \Bigr)\;
      \beta_{2}\, Y_{20}(\theta)
    \biggr\}, &
    \mbox{for  } r < r_{m},
  \end{array}
  \right.
\label{eq.5.1.2}
\end{equation}

\begin{equation}
\begin{array}{l}
  \vspace{0mm}
  v_{N} (r, \theta, Q) = \displaystyle\frac{V(A,Z,Q)} {1 + \exp{\displaystyle\frac{r-r_{m}(\theta)} {d}}}, \\
\end{array}
\label{eq.5.1.3}
\end{equation}

\begin{equation}
\begin{array}{l}
  v_{l} (r) = \displaystyle\frac{l\,(l+1)} {2mr^{2}}.
\end{array}
\label{eq.5.1.4}
\end{equation}
We define the parameters of the Coulomb and nuclear components as (see relations ~(14), (16)--(19) in~\cite{Denisov.2005.PHRVA}):
\begin{equation}
\begin{array}{rcl}
  V(A,Z,Q) & = & -(30.275 - 0.45838 \, Z/A^{1/3} + 58.270\,I - 0.24244 \, Q),
\end{array}
\label{eq.5.1.5}
\end{equation}
\begin{equation}
\begin{array}{rcl}
  R & = & R_{p}\:(1 + 3.0909/R_{p}^{2}) + 0.1243\,t, \\
\end{array}
\label{eq.5.1.6}
\end{equation}
\begin{equation}
\begin{array}{rcl}
  R_{p} & = & 1.24 \,A^{1/3} \: (1 + 1.646/A - 0.191\,I), \\
\end{array}
\label{eq.5.1.7}
\end{equation}
\begin{equation}
\begin{array}{rcl}
  t & = & I - 0.4 \, A/(A+200), \\
\end{array}
\label{eq.5.1.8}
\end{equation}
\begin{equation}
\begin{array}{rcl}
  d & = & 0.49290, \\
\end{array}
\label{eq.5.1.9}
\end{equation}
\begin{equation}
\begin{array}{rcl}
  I & = & (A-2Z) / A.
\end{array}
\label{eq.5.1.10}
\end{equation}
According to relations (21)--(22) in~\cite{Denisov.2005.PHRVA}, we also use:
\begin{equation}
\begin{array}{cclccl}
  r_{m}(\theta) & = & 1.5268 + R (\theta), &
  \hspace{3mm}
  R(\theta)     & = & R \: (1 + \beta_{2} Y_{20}(\theta) ).
\end{array}
\label{eq.5.1.11}
\end{equation}
Here, $A$ and $Z$ are the nucleon and proton numbers of the daughter nucleus, respectively; $Q$ is the $Q_{\alpha}$-value, for the $\alpha$-decay, $R$ is the radius of the daughter nucleus, $V(A,Z,Q,\theta)$ is the strength of the nuclear component; $r_{m}$ is the effective radius of the nuclear component, $d$ is the parameter of the diffuseness;
$Y_{20}(\theta)$ is the spherical harmonic function of the second order, $\theta$ is the angle between the direction of the leaving $\alpha$-particle and the axis of the axial symmetry of the daughter nucleus;
$\beta_{2}$ is the parameter of the quadruple deformation of the daughter nucleus.
%-----------------------------------------------------------------------------------------------------------------------

%-----------------------------------------------------------------------------------------------------------------------
\subsection{Spherically symmetric $\alpha$-decay
\label{sec.5.2}}

According to \cite{Muntyan.2003}, the deformation parameter $\beta_{2}$ for the decaying  $^{214}\mbox{\rm Po}$ nucleus is sufficiently small that allows us to apply the spherically symmetric approximation for the nucleus--$\alpha$-particle potential (\ref{eq.5.1.1})--(\ref{eq.5.1.11}) and to use formulas (\ref{eq.5.1.1})--(\ref{eq.5.1.4}) for the calculation of the bremsstrahlung spectrum during the $\alpha$-decay of such a nucleus.

In order to obtain the spectrum, we have to know WFs in the initial and
final states. In the spherically symmetric approximation one can
rewrite the total WFs by separating the radial and angular components:
\begin{equation}
\begin{array}{cclcl}
  \varphi_{i}(r,\theta,\phi) & = &
    R_{i}(r) \: Y_{l_{i} m_{i}} (\theta, \phi) & = &
    \displaystyle\frac{\chi_{i}(r)}{r} \: Y_{l_{i} m_{i}} (\theta, \phi), \\
  \varphi_{f}(r,\theta,\phi) & = &
    R_{f}(r) \: Y_{l_{f} m_{f}} (\theta, \phi) & = &
    \displaystyle\frac{\chi_{f}(r)}{r} \: Y_{l_{f} m_{f}} (\theta, \phi).
\end{array}
\label{eq.5.2.1}
\end{equation}
We find the radial components $\chi_{i,f}(r)$ numerically on the base of the given nucleus--$\alpha$-particle potential. Here, we use the following boundary conditions: the $i$-state of the system before the photon emission is a pure decaying state, and therefore for its description we use WF for the $\alpha$-decay; after the photon emission the state of the system is changed and it is more convenient to use WF as the scattering of the $\alpha$-particle by the daughter nucleus for the description of the $f$-state. So, we impose the following boundary conditions on the radial components $\chi_{i,f}(r)$:
\begin {equation}
\begin {array} {ll}
  \mbox {initial $i$-state:} & \chi_{i}(r \to +\infty) \to G(r)+iF(r), \\
  \mbox {final $f$-state:} & \chi _ {f} (r=0) = 0,
\end {array}
\label{eq.5.2.2}
\end {equation}
where $F$ and $G$ are the Coulomb functions.
%-----------------------------------------------------------------------------------------------------------------------

%-----------------------------------------------------------------------------------------------------------------------
\section{Calculations and analysis
\label{sec.6}}

\subsection{Bremsstrahlung spectra for $^{210}\mbox{\rm Po}$, $^{214}\mbox{\rm Po}$, $^{226}\mbox{\rm Ra}$ and $^{244}\mbox{\rm Cm}$: comparison theory and experiments
\label{sec.6.1}}

To estimate efficiency of the definition of the angular absolute (normalized) probability of the photon emission, we shall calculate the absolute spectra for the $^{210}\mbox{\rm Po}$, $^{214}\mbox{\rm Po}$ and $^{226}\mbox{\rm Ra}$ nuclei in such definition and the proposed approach where experimental data exist.

\begin{figure}[htbp]
\centerline{\includegraphics[width=130mm]{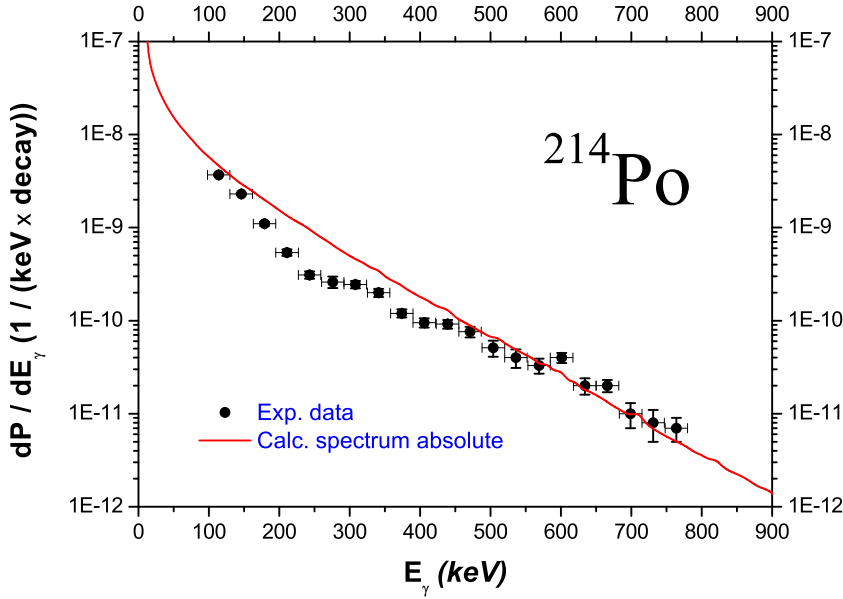}}
\vspace{-8mm}
\caption{\small
The calculated absolute spectrum and experimental data \cite{Maydanyuk.2008.EPJA} of the bremsstrahlung probability in $\alpha$-decay of the $^{214}{\rm Po}$ nucleus
(in calculations, we have used:
$R_{max} = 2000$~MeV)
\label{fig.6.1}}
\end{figure}

The best result in agreement between theory and experiment we have obtained for the $^{214}\mbox{\rm Po}$ nucleus. In Fig.~\ref{fig.6.1} the calculated absolute probability of bremsstrahlung emission for the $^{214}\mbox{\rm Po}$ nucleus and the newest experimental data in~\cite{Maydanyuk.2008.EPJA} for this nucleus are presented (here, there is no any normalization of the calculated curve relatively experimental data).
In calculations, we use the second definition of the angular absolute probability based on direction of $\alpha$-particle motion. The probability we calculate by (\ref{eq.4.3.4}), using approximation of $l=0$ in the calculation of the matrix element $p\,(w, \vartheta)$ (because according to our estimations the next value $l$ does not give a noticeable deformation of the found bremsstrahlung spectrum). The angle $\vartheta$ between the directions of the $\alpha$-particle motion (with possible tunneling) and the photon emission is used equal to $90^{\circ}$. The nucleus--$\alpha$-particle potential is defined in (\ref{eq.5.1.1})--(\ref{eq.5.1.4}), its parameters are defined in (\ref{eq.5.1.5})--(\ref{eq.5.1.11}). $Q_{\alpha}$-value is 7,865 keV according to~\cite{Buck.1993.ADNDT} (see p.~63).
Radial components of wave functions of the decaying system in states before and after the photon emission are calculated concerning such nucleus--$\alpha$-particle potential in spherically symmetric approximation (at $\beta_{2}=0$). The boundary conditions are used in form (\ref{eq.5.2.2}).
For this nucleus we also have \cite{Buck.1993.ADNDT} (see p.~63):
$b_{\alpha}^{\rm abs} = 100$ percents (brenching ratio for $\alpha$-decays that populate the given favored daughter state, given as a percentage of all decays),
$T^{\rm exp}_{1/2,\, \alpha} = 1,6 \cdot 10^{-8}$~sec.
In this figure one can see that the calculated spectrum for $^{214}{\rm Po}$ by the proposed approach is in enough good agreement with the experimental data for this nucleus inside the region from 100 keV up to 750 keV.

In the next Fig.~\ref{fig.6.2} the calculated absolute probabilities of the bremsstrahlung in $\alpha$-decay of the $^{210}\mbox{\rm Po}$ and $^{226}\mbox{\rm Ra}$ nuclei and experimental data in \cite{Boie.2007.PRL} and \cite{Maydanyuk.2008.MPLA} for these nuclei are presented.
\begin{figure}[htbp]
\centerline{\includegraphics[width=88mm]{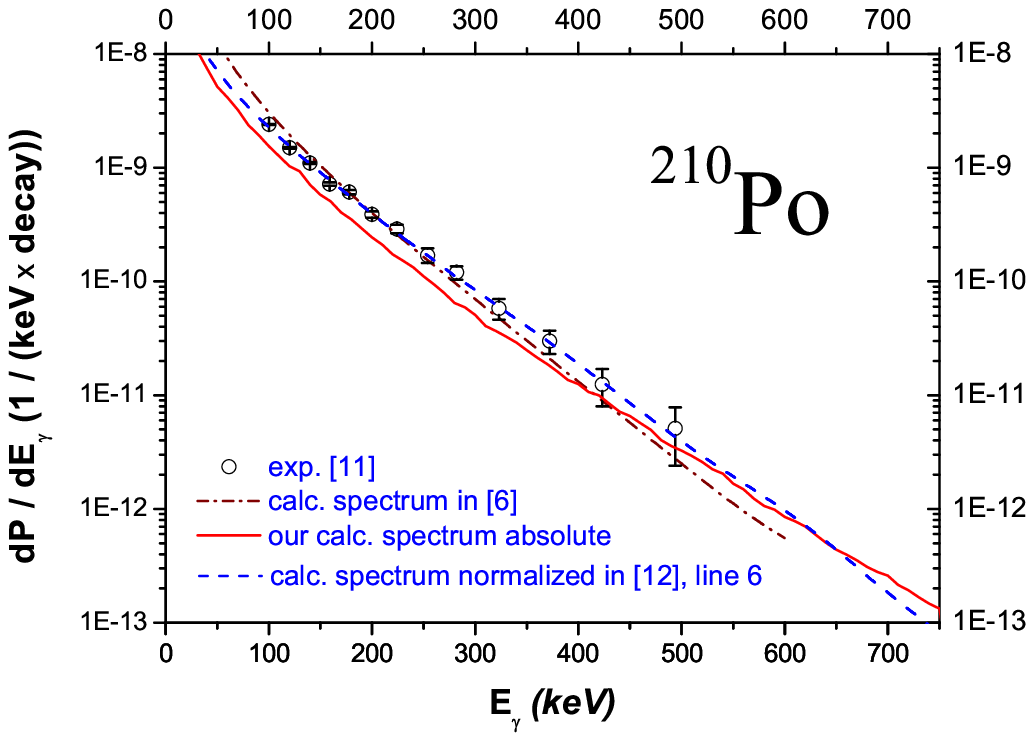}
\includegraphics[width=88mm]{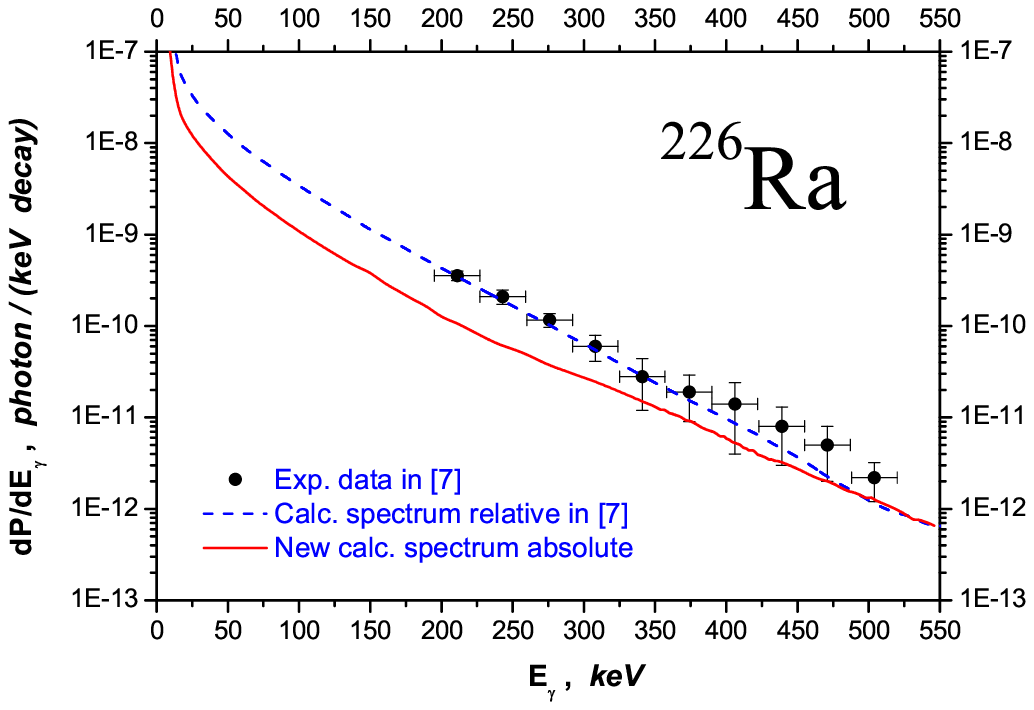}}
\vspace{-5mm}
\caption{\small
The absolute bremsstrahlung probabilities in $\alpha$-decay of the spherical $^{210}{\rm Po}$ and deformed $^{226}{\rm Ra}$ nuclei and experimental data in \cite{Boie.2007.PRL} and \cite{Maydanyuk.2008.MPLA} for these nuclei
\label{fig.6.2}}
\end{figure}
In calculations, the angular absolute probability, the nucleus--$\alpha$-particle potential, its parameters, algorithms of calculations of wave functions and their boundary conditions are defined and calculated by the same approach as for the nucleus $^{214}{\rm Po}$.
Here, we use:
$Q_{\alpha}$-value is 5,439 keV for $^{210}{\rm Po}$ and 4,904 keV for $^{226}{\rm Ra}$, according to~\cite{Buck.1993.ADNDT};
the angle $\vartheta$ between the directions of the $\alpha$-particle motion and the photon emission is $90^{\circ}$.
In figures one can see that for both nuclei for low energies of the photons emitted the calculated spectra are located below experimental data, but for energies from 350 keV and higher we have obtained good agreement between theory and experiment.
One note that for both these nuclei there is less convergence in calculations of the spectra in a comparison with calculation of the spectrum for $^{214}{\rm Po}$ that can be explained by larger tunneling regions for such two nuclei.
From here one can suppose: \emph{the tunneling region for studied nucleus is larger, the convergence in calculations of the bremsstrahlung spectra is obtained with larger difficulty, and study of bremsstrahlung from the tunneling region is more difficult}.
In figures one can see a little difference (tendency) between the calculated absolute spectrum and the relative spectrum in \cite{Maydanyuk.2006.EPJA,Maydanyuk.2008.MPLA} that cam be explained by different combination of integrals in the total matrix element in result of different expansions of wave function of the photon emitted.

We also add the calculated absolute spectrum for $^{244}{\rm Cm}$, comparing it with the high limit of errors of experimental data in~\cite{Kasagi.1997.JPHGB,Kasagi.1997.PRLTA}.
In calculations we use:
$Q_{\alpha}$-value is 5,940 keV \cite{Buck.1993.ADNDT} and $\vartheta=90^{\circ}$.
\begin{figure}[htbp]
\centerline{\includegraphics[width=88mm]{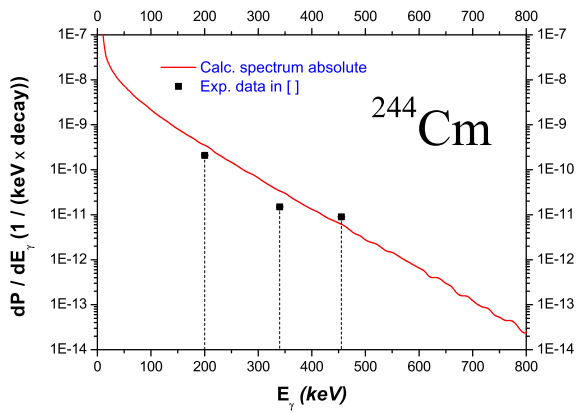}}
\vspace{-5mm}
\caption{\small
The absolute bremsstrahlung probabilities in $\alpha$-decay of the $^{244}{\rm Cm}$ nucleus and experimental data \cite{Kasagi.1997.JPHGB} for this nucleus
\label{fig.6.3}}
\end{figure}
From this figure we see that our calculated curve is located close to the high limit of error of experimental data. One can conclude that the agreement between theory and experiment is not bad and this nucleus has also sufficient experimental and theoretical basis for further study of bremsstrahlung processes in $\alpha$-decay.
%
% The bremsstrahlung spectra obtained in the first definition of the probability, based on direction of photon impulse $\mathbf{k}$, are located far essentially from the experimental data for these nuclei.
This indicates to effectiveness of the proposed definition of the angular absolute probability of the photons emitted based on the direction of $\alpha$-particle motion, and confirms effectiveness of the developed method of the calculations of the absolute spectra.

%-----------------------------------------------------------------------------------------------------------------------

%-----------------------------------------------------------------------------------------------------------------------
\subsection{Bremsstrahlung dependence on $Q_{\alpha}$ and predictions of the bremsstrahlung probability during $\alpha$-decay of isotopes of ${\rm Th}$
\label{sec.6.2}}

\begin{figure}[htbp]
\centerline{%\includegraphics[width=88mm]{Abs_spectrum_228Th.eps}
\includegraphics[width=88mm]{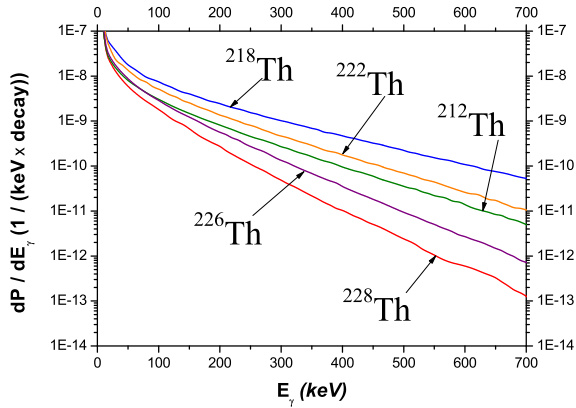}}
\vspace{-5mm}
\caption{\small
The predicted absolute bremsstrahlung probabilities in $\alpha$-decay of the $^{228}{\rm Th}$ nucleus and its isotopes
\label{fig.6.4}}
\end{figure}

In \cite{Kasagi.1997.JPHGB} it was noted about current investigations of bremsstrahlung accompanying the $\alpha$-decay of the $^{228}{\rm Th}$ nucleus. It can be interesting on the basis of the proposed approach to estimate the absolute bremsstrahlung probability for this nucleus. Results of such calculations are presented in Fig.~\ref{fig.6.4}.
In calculations we use:
the angle $\vartheta$ between the directions of the $\alpha$-particle motion (with possible tunneling) and the photon emission is $90^{\circ}$, 
$Q_{\alpha}$-value is 5.555 keV according to~\cite{Buck.1993.ADNDT} (see p.~63).

In~\cite{Maydanyuk.2008.MPLA} we explained the difference between the photon emission probabilities (both experimental and theoretical results) in the $\alpha$-decay of $^{226}{\rm Ra}$ and $^{214}{\rm Po}$ (at first, dependence of the bremsstrahlung probability on the $\alpha$-particle energy was analyzed in~\cite{So_Kim.2000.JKPS}):
\emph{``The difference between the two sets of data can be attributed to the different structure of the two nuclei, which affects the motion of the $\alpha$-particle inside the barrier. The ratio between the two sets of data of the photon emission probability $dP / dE_{\gamma}$ is strongly characterized by the different $\alpha$-decay energy for $^{214}{\rm Po}$ (E$_{\alpha}$=7.7 MeV)  and $^{226}{\rm Ra}$ (E$_{\alpha}$=4.8 MeV) concerning the shapes of the alpha-nucleus barriers for these nuclei.''} 
The difference between the $\alpha$-particle energies for the decaying $^{214}{\rm Po}$ and $^{226}{\rm Ra}$ nuclei is directly connected with different tunneling regions for these nuclei, which is directly connected with different contributions of the photons emission from tunneling and external regions, interference terms into the total spectra. And we obtained the property: \emph{The tunneling region is larger, the bremsstrahlung spectrum is smaller.}
The smaller values of the calculated total emission probability for $^{226}{\rm Ra}$ than the one for $^{214}{\rm Po}$ can be explained by a consequence of the fact that outside the barrier the Coulomb field (and its derivative respect to $r$) that acts on the $\alpha$-particle in the case of $^{226}{\rm Ra}$ is smaller than in the case of  $^{214}{\rm Po}$ because the external wide region results for the $^{214}{\rm Po}$ nucleus larger than for $^{226}{\rm Ra}$ and therefore the $\gamma$-emission probability for the $^{214}{\rm Po}$ nucleus is bigger.
In~Fig.\ref{fig.6.4} we have seen the demonstration of this property for isotopes of ${\rm Th}$.
In the Tabl.~\ref{table.6.1} one can see in detail how the bremsstrahlung probability depends on $Q_{\alpha}$-value of the nucleus for different energies of the photons emitted.
\begin{table}
\begin{center}
\begin{tabular}{|c|c|c|c|c|c|c|c|c|} \hline
 \multicolumn{4}{|c|}{$\alpha$-decay data} &
 \multicolumn{5}{|c|}{Bremsstrahlung probability, 1 / keV / decay}
 \\ \cline{1-9}
  $A_{p}$ &
  $Q_{\alpha}$, MeV &
  $b_{\alpha}^{\rm abs}$, \% &
  $T_{1/2,\, \alpha}^{\rm exp}$, sec &

  100 keV &
  200 keV &
  300 keV &
  400 keV &
  500 keV \\ \hline
  212 & 7.987 & 100.0 & 3.0\,E-2   & 3.0\,E-9 & 8.1\,E-10 & 2.7\,E-10 & 9.5\,E-11 & 3.5\,E-11 \\
  218 & 9.881 & 100.0 & 1.1\,E-7   & 7.5\,E-9 & 2.5\,E-9  & 1.0\,E-9  & 4.7\,E-10 & 2.2\,E-10 \\
  222 & 8.164 & 100.0 & 2.8\,E-3   & 5.2\,E-9 & 1.3\,E-9  & 4.6\,E-10 & 1.7\,E-10 & 7.0\,E-11 \\
  226 & 6.487 & 75.5  & 2.5\,E+3   & 2.9\,E-9 & 5.6\,E-10 & 1.3\,E-10 & 3.5\,E-11 & 9.4\,E-12 \\
  228 & 5.555 & 72.7  & 8.3\,E+7   & 1.8\,E-9 & 2.8\,E-10 & 4.9\,E-11 & 1.0\,E-11 & 1.9\,E-12 \\
  \hline
\end{tabular}
\end{center}
\caption{Estimated values of the bremsstrahlung probability during $\alpha$-decay of the $^{228}{\rm Th}$ nucleus and its isotopes
\label{table.6.1}}
\end{table}
%-----------------------------------------------------------------------------------------------------------------------

%-----------------------------------------------------------------------------------------------------------------------
\subsection{Bremsstrahlung dependence on effective charge and bremsstrahlung during proton emission from nucleus \label{sec.6.3}}

As we have seen above, $Q_{\alpha}$-value of the $\alpha$-decay has strong influence on the bremsstrahlung spectra. Now we put a question: \emph{which are other parameters having essential influence on the bremsstrahlung spectrum?}

Let's consider Fig.~4 and Fig.~7 in \cite{Ploeg.1995.PRC} where the calculated $\gamma$-ray emission probabilities for the spontaneous fission of $^{252}{\rm Cf}$ are presented. One can see that the emission probability is changed essentially in dependence on mass split. From here one can suppose that the other characteristic which takes influence on the emission probability essentially should depend on \underline{relative} numbers of mass and charge of the daughter nucleus and the emitting charged particle. The idea proposed in \cite{So_Kim.2000.JKPS} about influence of the electromagnetic charge of the daughter nucleus on the bremsstrahlung probability reflects this property only partially, and we see that \emph{effective charge of the decaying system is more directly connected with such property!} Now if to consider the formula (\ref{eq.4.3.4}) of the bremsstrahlung probability then one can find its direct dependence on square of the effective charge, i.~e. we have obtained the real basis for such supposition.

One can find that the effective charge for the $\alpha$-decay is smaller (and, perhaps, essentially) in comparison with many other types of decays (and it is else smaller for heavier nuclei; square of effective charge equals to 0.18482 for $\alpha$-decay of $^{214}{\rm Po}$, and to 0.16 for $\alpha$-decay of $^{210}{\rm Po}$)! The first such example of decay, which has larger effective charge, we find from literature: \emph{this is proton emission from nucleus}. Let's estimate the bremsstrahlung in this decay. To describe the interaction between the proton and daughter nucleus we use the proton-nucleus potential in standard optical model form given in a famous paper~\cite{Becchetti.1969.PR} (see formulas (5), (8)).
Performing preliminary calculations, we neglect spin-orbit and imaginary components, add centrifugal component and obtain:
\begin{equation}
  V (r, l) = v_{C} (r) + v_{N} (r, R_{R}, d_{R}) + v_{l} (r)
\label{eq.6.3.1}
\end{equation}
where the Coulomb $v_{C} (r)$, nuclear $v_{N} (r, R_{R}, d_{R})$ and centrifugal $v_{l} (r)$ components are
\begin{equation}
  v_{C} (r) =
  \left\{
  \begin{array}{ll}
    \displaystyle\frac{Z\, e^{2}} {r}, &
      \mbox{for  } r \ge R_{c}, \\
    \displaystyle\frac{Z\, e^{2}} {2\,R_{c}}\;
      \biggl\{ 3 - \displaystyle\frac{r^{2}}{R_{c}^{2}} \biggr\}, &
      \mbox{for  } r < R_{c},
  \end{array}
  \right.
\label{eq.6.3.2}
\end{equation}
\begin{equation}
  v_{N} (r, R_{R}, d_{R}) =
  \displaystyle\frac{V_{R}(A,Z,E)} {1 + \exp{\displaystyle\frac{r-R_{R}} {d_{R}}}},
\label{eq.6.3.3}
\end{equation}
\begin{equation}
  v_{l} (r) = \displaystyle\frac{l\,(l+1)} {2mr^{2}}.
\label{eq.6.3.4}
\end{equation}
We take radii in the form:
\begin{equation}
  R_{R} = r_{R}\, A^{1/3}
\label{eq.6.3.5}
\end{equation}
and define the optimum proton-nucleus standard OM parameters so:
\begin{equation}
\begin{array}{rcl}
  V_{R}(A,Z,E) & = & -(54.0 - 0.32\,E + 0.4\,Z/A^{1/3} + 24.0\,I), \\
  I & = & (N-Z)/A, \\
  r_{R} & = & 1.17, \\
  d_{R} & = & 0.75
\end{array}
\label{eq.6.3.5}
\end{equation}
where $A$ and $Z$ are mass and proton numbers of the daughter nucleus, $E$ is incident lab energy.
In search of the convenient nuclei for analysis, we use Table~2 in~\cite{Aberg_Nazarewicz.1997.PRC} and select only 4 proton emitters which have decay from state $2s_{1/2}$: $^{157}{\rm Ta}$, $^{161}{\rm Re}$, $^{167}{\rm Ir}$ and $^{185}{\rm Bi}$.

\begin{figure}[htbp]
\centerline{\includegraphics[width=88mm]{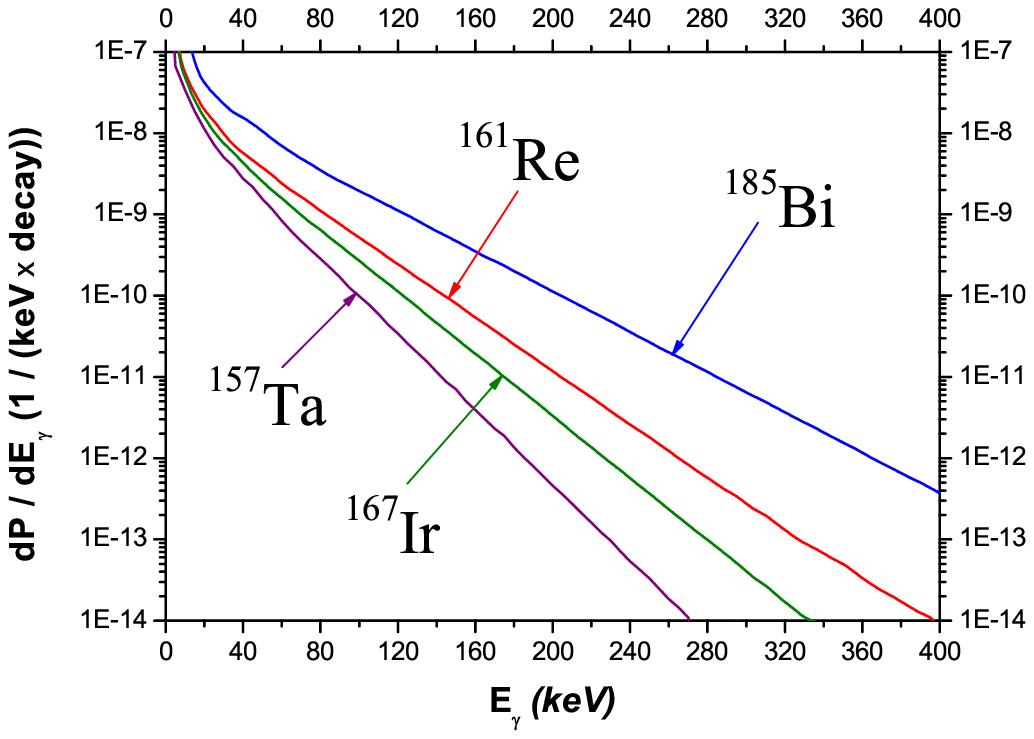}
\includegraphics[width=88mm]{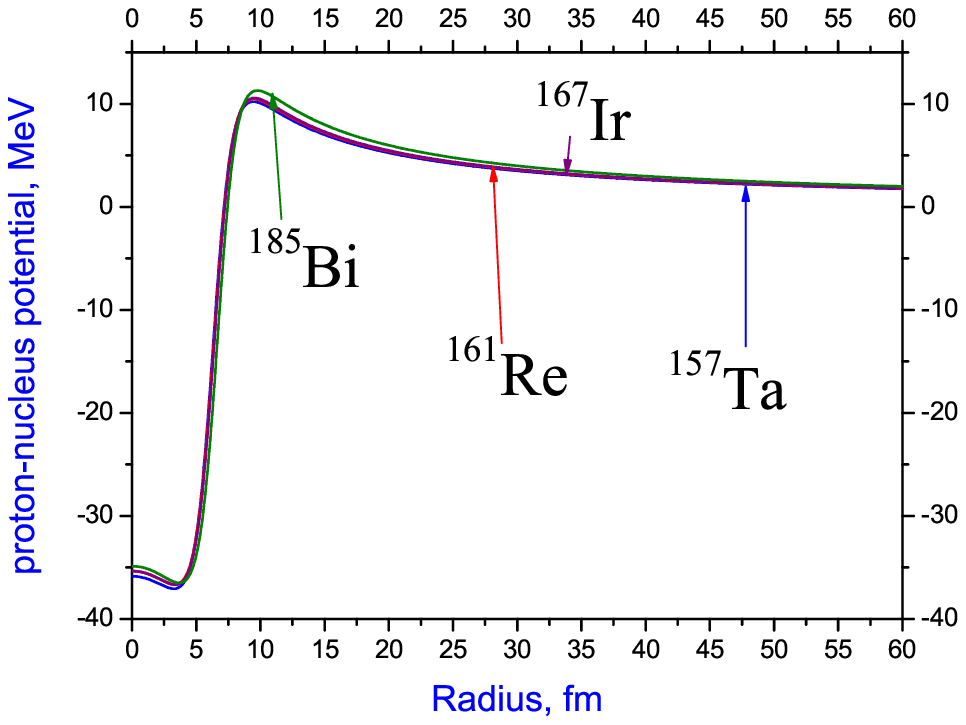}}
\vspace{-8mm}
\caption{\small
The bremsstrahlung during proton decay of the $^{157}{\rm Ta}$, $^{161}{\rm Re}$, $^{167}{\rm Ir}$ and $^{185}{\rm Bi}$ nuclei
(in calculations, we have used: $R_{max} = 2000$~MeV):
(a) --- the absolute bremsstrahlung probabilities;
(b) --- the proton-nucleus potentials for studied proton emitters
\label{fig.6.5}}
\end{figure}

The result of calculations of bremsstrahlung probabilities during proton decay are presented in Fig.~\ref{fig.6.5}. In calculations we use angle between photon emission and proton motion equaled to $90^{\circ}$.
In Tabl.~\ref{table.6.2} one can see values of some parameters of the proton-nucleus potential for the studied nuclei. From here one can find that the different proton emitters have practically similar effective charges, but different essentially tunneling regions.

\begin{table}
\begin{center}
\begin{tabular}{|c|c|c|c|c|c|c|c|c|c|} \hline
 \multicolumn{4}{|c|}{Proton decay data} &
 \multicolumn{2}{|c|}{Parameters} &
 \multicolumn{2}{|c|}{Turning points} &
 Tunneling &
 Eff. charge,
 \\ \cline{1-8}
  Nucleus &
  $Q_{p}$, keV &
  Orbit &
  $T_{1/2,\, p}^{\rm WKB}$, sec &
  $R_{R}$, fm & $V_{R}$, MeV &
  1-st, fm & 2-nd, fm &
  region, fm &
  $E_{\rm eff}^{2}$
  \\ \hline
  $^{157}_{73}{\rm Ta}_{83}$ & 947  & $2s_{1/2}$ & 210 $ms$
      & 6.29 & -60.89 & 7.25 & 110.96 & 103.71 & 0.286259 \\
  $^{161}_{75}{\rm Re}_{86}$ & 1214 & $2s_{1/2}$ & 180 $\mu s$
      & 6.3517 & -60.8638 & 7.32 & 88.96 & 81.64 & 0.285328 \\
  $^{167}_{77}{\rm Ir}_{90}$ & 1086 & $2s_{1/2}$ &  35 $ms$
      & 6.3912 & -61.0585 & 7.32 & 100.79 & 93.46 & 0.287924 \\
  $^{185}_{83}{\rm Bi}_{98}$ & 1611 & $2s_{1/2}$ & 3.1 $\mu s$
      & 6.6546 & -61.8599 & 7.56 & 74.28 & 66.72 & 0.303988 \\
  \hline
\end{tabular}
\end{center}
\caption{Parameters of the proton decay of some proton emitters
($E_{\rm eff}^{2}$ is square of effective charge)
\label{table.6.2}}
\end{table}

We conclude:
\emph{the bremsstrahlung probabilities in proton decay (from state $2s_{1/2}$) have similar order of values in a comparison with the bremsstrahlung probabilities in $\alpha$-decay}.

%-----------------------------------------------------------------------------------------------------------------------

%-----------------------------------------------------------------------------------------------------------------------
\subsection{Predictions of the bremsstrahlung spectra during ternary fission
\label{sec.6.5}}

Nuclear fission accompanied by light charged particle emission which is often called as \emph{ternary fission} has been widely studied (see \cite{Daniel.2004.PRC} and references cited therein). Study of $\gamma$-emission during such process has been causing increased interest.
Let's estimate the absolute probability of the photons emission during emission of $\alpha$-particle from the $^{252}{\rm Cf}$ nucleus (which is the popular nucleus used in study of such problem).
In calculations we use:
parameters of the nucleus--$\alpha$-particle potential are used according the proposed approach above,
$Q_{\alpha}$-value is 6.257 keV,
$R = 7.684$~fm;
$b_{\alpha}^{\rm abs} = 81.6$ percents (brenching ratio for $\alpha$-decays that populate the given favored daughter state, given as a percentage of all decays),
$T^{\rm exp}_{1/2,\, \alpha} = 1.0 \cdot 10^{+8}$~sec.
In the next Fig.~\ref{fig.6.6} the absolute bremsstrahlung probability during emission of $\alpha$-decay from $^{252}{\rm Cf}$ is presented.

\begin{figure}[htbp]
\centerline{\includegraphics[width=88mm]{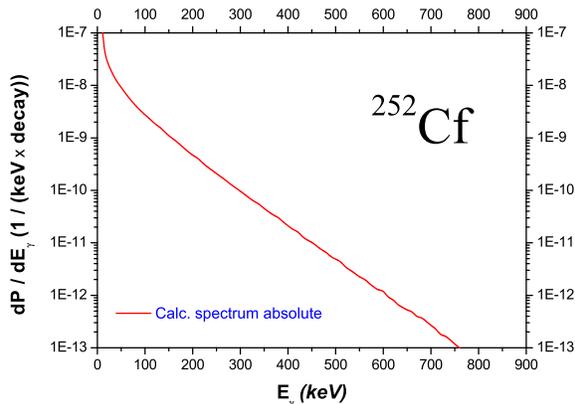}}
\vspace{-8mm}
\caption{\small
The absolute bremsstrahlung probabilities in $\alpha$-decay of the $^{252}{\rm Cf}$ nucleus
(in calculations, we have used:
$R_{max} = 5000$~MeV)
\label{fig.6.6}}
\end{figure}
%-----------------------------------------------------------------------------------------------------------------------

%-----------------------------------------------------------------------------------------------------------------------
% \newpage
% \subsection{Predictions of the bremsstrahlung spectra during $\alpha$-decay of light nuclei
% \label{sec.6.3}}

% It can be interesting to estimate on the basis of proposed approach of calculations of absolute probabilities, which spectra of bremsstrahlung other nuclei have.
% In fig.~4 absolute bremsstrahlung probability in $\alpha$-decay of $^{241}{\rm Am}$ is presented.

%-----------------------------------------------------------------------------------------------------------------------

%-----------------------------------------------------------------------------------------------------------------------
% \subsection{Bremsstrahlung spectra in $\alpha$-decay of strongly deformed nuclei
% \label{sec.6.4}}

% Let's consider the nucleus which has $\beta_{2}= $ and analyze how tunneling region is changes between maximum and minimum of deformation of its shape.

%-----------------------------------------------------------------------------------------------------------------------

%-----------------------------------------------------------------------------------------------------------------------
\section{Conclusion}

In this paper the improved multipolar model of bremsstrahlung accompanied the $\alpha$-decay is presented. The angular formalism of calculations of the matrix elements is stated in details. A new definition of the angular (differential) absolute probability of the photon emission (i.~e. without normalization on experimental data) in the $\alpha$-decay is proposed where direction of motion of the $\alpha$-particle outside (with its tunneling inside barrier) is defined on the basis of angular distribution of its spacial wave function. Effectiveness of the proposed definition, the developed formalism of the model and accuracy of the calculations of the bremsstrahlung spectra are analyzed in their comparison with experimental data for the $^{210}{\rm Po}$, $^{214}{\rm Po}$, $^{226}{\rm Ra}$ and $^{244}{\rm Cm}$ nuclei.
Here, note the following.
\begin{itemize}
\item
The best result have been obtained in agreement between the calculated absolute probability of the bremsstrahlung emission for the $^{214}\mbox{\rm Po}$ nucleus and the newest experimental data in~\cite{Maydanyuk.2008.EPJA} for this nucleus inside the region of photons energies from 100 keV up to 750 keV (see Fig.~\ref{fig.6.1}, $Q_{\alpha} = 7.865$~keV, the angle $\vartheta$ between the directions of the $\alpha$-particle motion and the photon emission is used $90^{\circ}$).

\item
The calculated absolute probabilities of the bremsstrahlung emission in $\alpha$-decay of the $^{210}\mbox{\rm Po}$ and $^{226}\mbox{\rm Ra}$ nuclei for low energies of the photons emitted are located below experimental data \cite{Boie.2007.PRL} and \cite{Maydanyuk.2008.MPLA}, but for energies from 350 keV and higher we have obtained good agreement between our model and experiment (see Fig.~\ref{fig.6.2}, $Q_{\alpha} = 5.439$~keV for $^{210}\mbox{\rm Po}$ and $Q_{\alpha} = 4.904$~keV for $^{226}\mbox{\rm Ra}$, $\vartheta=90^{\circ}$).

\item
The calculated absolute probability of the bremsstrahlung emission for the $^{244}{\rm Cm}$ nucleus is located close to the high limit of error of experimental data~\cite{Kasagi.1997.JPHGB,Kasagi.1997.PRLTA}
(see Fig.~\ref{fig.6.3}, $Q_{\alpha} = 5.940$~keV and $\vartheta=90^{\circ}$).
\end{itemize}

Analyzing the formalism of the model, we establish:
\begin{itemize}
\item
A presence of oscillations of the bremsstrahlung probability in $\alpha$-decay defined on the basis of non-stationary quantum approach (theoretically, at first time in fully quantum approach). In such context, period of these oscillations has information about duration of presence of the $\alpha$-particle inside the region of electromagnetic forces of the daughter nucleus (from the $\alpha$-particle formation inside the nucleus up to its passing through electrons shells).

\item
A property: for any studied nucleus the tunneling region is larger, the convergence in calculations of the bremsstrahlung spectrum is weaker, and study of bremsstrahlung from the tunneling region is more difficult.

\item
A principal difference between two experiments \cite{Kasagi.1997.JPHGB,Kasagi.1997.PRLTA} and \cite{D'Arrigo.1994.PHLTA} (opened at first time):
in experimental data~\cite{Kasagi.1997.JPHGB,Kasagi.1997.PRLTA} for the $^{210}{\rm Po}$ nucleus the contribution of the magnetic component M1 into the total spectrum is close to maximal while in the experimental data~\cite{D'Arrigo.1994.PHLTA} for the $^{214}{\rm Po}$ and $^{226}{\rm Ra}$ nuclei we have obtained zero contribution of this magnetic component.

\item
The non small dependence of the bremsstrahlung probability on the effective charge of the decaying system.
\end{itemize}

On the basis of such model the bremsstrahlung probabilities in the $\alpha$-decay of $^{228}{\rm Th}$ and some its isotopes, the bremsstrahlung probabilities during proton emission from the $^{157}{\rm Ta}$, $^{161}{\rm Re}$, $^{167}{\rm Ir}$ and $^{185}{\rm Bi}$ nuclei are predicted (in absolute scale, at first time in the fully quantum approach). According to analysis, the bremsstrahlung probabilities in the proton decay (from $s_{1/2}$ state) have similar order of values in comparison with the bremsstrahlung probabilities in the $\alpha$-decay.
On such a basis one can hope that experimental study of bremsstrahlung in the proton decay can be interesting.

\section*{Acknowledgements
\label{sec.acknowledgements}}

The author is appreciated to 
Dr. Alexander~K.~Zaichenko for his assistance in computer realization of numerical methods in calculations of wave functions,
Prof.~Vladislav~S.~Olkhovsky for useful discussions concerning realizations of multiple internal reflections in the problem of $\alpha$-decay and comments of definition of phase times,
Prof.~Giorgio~Giardina for useful discussions concerning main formalism of the model, dependence of the bremsstrahlung spectra on $Q_{\alpha}$-value of $\alpha$-decay, aspects to investigate deformed nuclei in this problem,
Prof.~Volodimir~M.~Kolomietz for useful discussions and critical comments concerning the general formalism of the presented model,
Dr.~Sergei~N.~Fedotkin for useful comments concerning definitions of absolute and normalized probabilities of the photons emitted during the $\alpha$-decay,
Dr.~Alexander~G.~Magner for useful comments concerning determination of wave function of the $\alpha$-decaying system,
Dr.~Vladislav~Kobychev for interesting discussions concerning behavior of the bremsstrahlung spectra for photon energies close to zero.

% *******************************************************************************************************************

% *******************************************************************************************************************
% \newpage
% \appendix
\section{Appendix
\label{app}}

\subsection{Legandre's polynomials
\label{app.1}}

\emph{Legandre's polynomials} $P_{l}\,(\cos{\theta})$ and \emph{associated Legandre's polynomials} $P_{l}^{m} (\cos{\theta})$ are defined as in \cite{Landau.v3.1989} (see p.~752--754, (c,1)--(c,4); \cite{Eisenberg.1973} (2.6), p.~34):
\begin{equation}
\begin{array}{lcl}
  P_{l} \,(\cos{\theta}) & = &
    \displaystyle\frac{1}{2^{l}\,l!} \:
    \displaystyle\frac{d^{l}}{(d\,\cos{\theta})^{l}} \: (\cos^{2}{\theta} - 1)^{l}, \\
  P_{l}^{m} (\cos{\theta}) & = &
    \sin^{m}{\theta} \displaystyle\frac{d^{m}\, P_{l}(\cos{\theta})}{(d\,\cos{\theta})^{m}} =
    \displaystyle\frac{1}{2^{l}\, l!}
    \sin^{m}{\theta}
    \displaystyle\frac{d^{l+m}}{(d\,\cos{\theta})^{l+m}} (\cos^{2}{\theta} - 1)^{l} = \\
    & = &
    (-1)^{m} \: \displaystyle\frac{(l+m)!}{(l-m)! \: 2^{l} \, l!} \sin^{-m}{\theta}
      \displaystyle\frac{d^{l-m}}{(d\,\cos{\theta})^{l-m}} (\cos^{2}{\theta} - 1)^{l}.
\end{array}
\label{eq.app.1.1}
\end{equation}
where $m = 0 \ldots l$. For these polynomials the normalization condition is fulfilled подчиняются ($\mu = \cos{\theta}$; see~\cite{Landau.v3.1989}, (c,6)--(c,9) p.~753--754):
\begin{equation}
\begin{array}{lcllcl}
  \displaystyle\int\limits_{-1}^{1} \bigl[ P_{l} \,(\mu) \bigr]^{2} \; d\mu = \displaystyle\frac{2}{2l + 1}, &
  \hspace{10mm}
  \displaystyle\int\limits_{-1}^{1} P_{l} \,(\mu) \: P_{l^{\prime}} \,(\mu) \; d\mu & = & 0, \\
  \displaystyle\int\limits_{-1}^{1} \bigl[ P_{l}^{m} \,(\mu) \bigr]^{2} \; d\mu =
    \displaystyle\frac{2}{2l + 1} \: \displaystyle\frac{(l+m)!}{(l-m)!}, &
  \hspace{10mm}
  \displaystyle\int\limits_{-1}^{1} P_{l}^{m} \,(\mu) \: P_{l^{\prime}}^{m} \,(\mu) \; d\mu & = & 0.
\end{array}
\label{eq.app.1.2}
\end{equation}
At the first coefficients $l$ and $m$ polynomials are:
\begin{equation}
\begin{array}{lcl}
  \vspace{3mm} P_{0}^{0} (\cos{\theta}) & = & 1, \\

  \vspace{3mm} P_{1}^{0} (\cos{\theta}) & = & \cos{\theta}, \\
  \vspace{3mm} P_{1}^{1} (\cos{\theta}) & = & \sin{\theta}, \\

  \vspace{3mm} P_{2}^{0} (\cos{\theta})  & = & \frac{1}{2} \:(3\cos^{2}{\theta} - 1), \\
  \vspace{3mm} P_{2}^{1} (\cos{\theta})  & = & 3 \,\sin{\theta} \cos{\theta}, \\
  \vspace{3mm} P_{2}^{2} (\cos{\theta})  & = & 3 \,\sin^{2}{\theta}.
\end{array}
\label{eq.app.1.3}
\end{equation}
%-----------------------------------------------------------------------------------------------------------------------

%-----------------------------------------------------------------------------------------------------------------------
% \newpage
% \vspace{5mm}
\subsection{Spherical functions $Y_{lm}$
\label{app.2}}

On the score of different definitions of the spherical functions in literature, we present the definition of them used in this paper. We define the \emph{spherical functions} $Y_{lm}(\theta,\varphi)$ according to \cite{Landau.v3.1989} (see p.~119, (28,7)--(28,8)):
\begin{equation}
\begin{array}{lcl}
  Y_{lm}(\theta,\varphi) & = &
    (-1)^{\frac{m+|m|}{2}} \; i^{l} \;
    \sqrt{\displaystyle\frac{2l+1}{4\pi} \displaystyle\frac{(l-|m|)!}{(l+|m|)!} }  \; P_{l}^{|m|} (\cos{\theta}) \cdot e^{im\varphi},
\end{array}
\label{eq.app.2.1}
\end{equation}
where $P_{l}^{m}(\cos{\theta})$ are \emph{associated Legandre's polynomial} (see Appendix~\ref{app.1}).
For the functions $Y_{lm}(\theta,\varphi)$ the following condition of normalization is fulfilled (see~\cite{Landau.v3.1989}, (28,3), p.~118):
\begin{equation}
  \displaystyle\int\limits_{0}^{2\pi}
  \displaystyle\int\limits_{0}^{\pi}
    Y_{l^{\prime} m^{\prime}}^{*}(\theta,\varphi) \: Y_{l m}(\theta,\varphi) \: \sin{\theta} \;
    d\theta \: d\varphi =
    \delta_{l l^{\prime}} \delta_{m m^{\prime}}.
\label{eq.app.2.2}
\end{equation}
The functions $Y_{lm}(\theta,\varphi)$, differed by sign $m$, are connected by relation (see (28,9), p. 119 in \cite{Landau.v3.1989}):
\begin{equation}
  Y_{lm}^{*} (\theta,\varphi) = (-1)^{l-m} \: Y_{l -m}(\theta,\varphi).
\label{eq.app.2.3}
\end{equation}
Write expressions for some first normalized spherical functions $Y_{lm}(\theta,\varphi)$ (see~\cite{Landau.v3.1989}, p.~754--755):
\begin{equation}
\begin{array}{lcllcl}
  Y_{00} & = & \displaystyle\frac{1}{\sqrt{4\pi}}, &
  Y_{10} & = & i \; \sqrt{\displaystyle\frac{3}{4\pi}} \; \cos{\theta}, \\

  Y_{1, \:\pm 1} & = & \mp \,i \; \sqrt{\displaystyle\frac{3}{8\pi}} \; \sin{\theta} \cdot e^{\pm i\varphi}, &
  Y_{20} & = & \sqrt{\displaystyle\frac{5}{16\pi}} \; (1 - 3\cos^{2}{\theta}), \\

  Y_{2, \:\pm 1} & = &
    \pm \,\sqrt{\displaystyle\frac{15}{8\pi}} \; \cos{\theta} \sin{\theta} \cdot e^{\pm i\varphi}, &
  Y_{2, \:\pm 2} & = & - \,\sqrt{\displaystyle\frac{15}{32\pi}} \; \sin^{2}{\theta} \cdot e^{\pm 2i\varphi},
\end{array}
\label{eq.app.2.4}
\end{equation}
\begin{equation}
\begin{array}{lcl}
  Y_{30} & = & -\,i \; \sqrt{\displaystyle\frac{7}{16\pi}} \; \cos{\theta} \,(5 \cos^{2}{\theta} - 3), \\
  Y_{3, \:\pm 1} & = & \pm \:i \; \sqrt{\displaystyle\frac{21}{64\pi}} \; \sin{\theta} \,(5 \cos^{2}{\theta} - 1)
                       \cdot e^{\pm i\varphi}, \\

  Y_{3, \:\pm 2} & = & - \,i \; \sqrt{\displaystyle\frac{105}{32\pi}} \; \cos{\theta} \sin^{2}{\theta}
                       \cdot e^{\pm 2i\varphi}, \\
  Y_{3, \:\pm 3} & = & \pm \:i \; \sqrt{\displaystyle\frac{35}{64\pi}} \; \sin^{3}{\theta}
                       \cdot e^{\pm 3i\varphi}.
\end{array}
\label{eq.app.2.5}
\end{equation}
%-----------------------------------------------------------------------------------------------------------------------

%-----------------------------------------------------------------------------------------------------------------------
% \newpage
\subsection{Clebsch-Gordan coefficients
\label{app.3}}

We define Clebsch-Gordan coefficients, according to Table ПА.1 in \cite{Eisenberg.1973} (see~p.317), rewriting it as Table 1.
\begin{table}
% \hspace{-20mm}
\begin{center}
\begin{tabular}{|c|c|c|} \hline
  & \multicolumn{2}{|c|}{$( j_{a} 1 j \big| m_{a} m_{b} m)$} \\ \cline{2-3}
        &
    $m_{b}=1$ &
    $m_{b}=-1$ \\ \hline
    $j=j_{a}+1$ &
    $\Biggl(\displaystyle\frac{(j_{a}+m)\,(j_{a}+m+1)} {(2j_{a}+2)\,(2j_{a}+2)} \Biggr)^{1/2}$  &
    $\Biggl(\displaystyle\frac{(j_{a}-m)\,(j_{a}-m+1)} {(2j_{a}+1)\,(2j_{a}+2)} \Biggr)^{1/2}$  \\
    $j=j_{a}$ &
    $-\Biggl(\displaystyle\frac{(j_{a}+m)\,(j_{a}-m+1)} {2j_{a}\,(j_{a}+1)} \Biggr)^{1/2}$  &
    $\Biggl(\displaystyle\frac{(j_{a}-m)\,(j_{a}+m+1)} {2j_{a}\,(j_{a}+1)} \Biggr)^{1/2}$  \\
    $j=j_{a}-1$ &
    $\Biggl(\displaystyle\frac{(j_{a}-m)\,(j_{a}-m+1)} {2j_{a}\,(2j_{a}+1)} \Biggr)^{1/2}$  &
    $\Biggl(\displaystyle\frac{(j_{a}+m+1)\,(j_{a}+m)} {2j_{a}\,(2j_{a}+1)} \Biggr)^{1/2}$  \\ \hline
\end{tabular}
\end{center}
\caption{\small Clebsch-Gordan coefficients.
\label{table.3.1}}
\end{table}
Using the table~\ref{table.3.1}, we find:
\begin{equation}
\begin{array}{lclclcc}
  (0 1 1 \big| \,2,  -1, \,1) & = &
    \sqrt{\displaystyle\frac{(j_{a}-m)\,(j_{a}-m+1)} {(2j_{a}+1)\,(2j_{a}+2)} } & = &
    \sqrt{\displaystyle\frac{(0-1)\,(0-1+1)} {(2 \cdot 0+1)\,(2 \cdot 0 +2)} } & = & 0, \\
  (0 1 1 \big| \,0, \,1, \,1) & = &
    \sqrt{\displaystyle\frac{(j_{a}+m)\,(j_{a}+m+1)} {(2j_{a}+2)\,(2j_{a}+2)} } & = &
    \sqrt{\displaystyle\frac{(0+1)\,(0+1+1)} {(2 \cdot 0+2)\,(2 \cdot 0 +2)} } & = &
    \sqrt{\displaystyle\frac{1}{2}}, \\
  (0 1 1 \big| \,0,  -1, -1) & = &
    \sqrt{\displaystyle\frac{(j_{a}-m)\,(j_{a}-m+1)} {(2j_{a}+1)\,(2j_{a}+2)} } & = &
    \sqrt{\displaystyle\frac{(0+1)\,(0+1+1)} {(2 \cdot 0+1)\,(2 \cdot 0+2)} } & = &
    \sqrt{\displaystyle\frac{1}{2}}, \\
  (0 1 1 \big|  -2, \,1, -1) & = &
    \sqrt{\displaystyle\frac{(j_{a}+m)\,(j_{a}+m+1)} {(2j_{a}+2)\,(2j_{a}+2)} } & = &
    \sqrt{\displaystyle\frac{(0-1)\,(0-1+1)} {(2 \cdot 0+2)\,(2 \cdot 0 +2)} } & = & 0;
\end{array}
\label{eq.app.3.1}
\end{equation}

\begin{equation}
\begin{array}{lclclcc}
  (1 1 1 \big| \,2,  -1, \,1) & = &
    \sqrt{\displaystyle\frac{(j_{a}-m)\,(j_{a}+m+1)} {2j_{a}\,(j_{a}+1)} } & = &
    \sqrt{\displaystyle\frac{(1-1)\,(1+1+1)} {2 \cdot 1 \cdot(1+1)}} & = & 0,\\
  (1 1 1 \big| \,0, \,1, \,1) & = &
    -\sqrt{\displaystyle\frac{(j_{a}+m)\,(j_{a}-m+1)} {2j_{a}\,(j_{a}+1)} } & = &
    -\sqrt{\displaystyle\frac{(1+1)\,(1-1+1)} {2 \cdot 1 \cdot(1+1)}} & = &
    -\sqrt{\displaystyle\frac{1}{2}}, \\
  (1 1 1 \big| \,0,  -1, -1) & = &
    \sqrt{\displaystyle\frac{(j_{a}-m)\,(j_{a}+m+1)} {2j_{a}\,(j_{a}+1)} } & = &
    \sqrt{\displaystyle\frac{(1+1)\,(1-1+1)} {2 \cdot 1 \cdot (1+1)}} & = &
    \sqrt{\displaystyle\frac{1}{2}}, \\
  (1 1 1 \big|  -2, \,1, -1) & = &
    -\sqrt{\displaystyle\frac{(j_{a}+m)\,(j_{a}-m+1)} {2j_{a}\,(j_{a}+1)} } & = &
    -\sqrt{\displaystyle\frac{(1-1)\,(1+1+1)} {2 \cdot 1 \cdot (1+1)}} & = & 0;
\end{array}
\label{eq.app.3.2}
\end{equation}

\begin{equation}
\begin{array}{lclclcc}
  (2 1 1 \big| \,2,  -1, \,1) & = &
    \sqrt{\displaystyle\frac{(j_{a}+m+1)\,(j_{a}+m)} {2j_{a}\,(2j_{a}+1)} } & = &
    \sqrt{\displaystyle\frac{(2+1+1)\,(2+1)} {2 \cdot 2 \cdot (2 \cdot 2 + 1)} } & = &
    \sqrt{\displaystyle\frac{3}{5}}, \\
  (2 1 1 \big| \,0, \,1, \,1) & = &
    \sqrt{\displaystyle\frac{(j_{a}-m)\,(j_{a}-m+1)} {2j_{a}\,(2j_{a}+1)} } & = &
    \sqrt{\displaystyle\frac{(2-1)\,(2-1+1)}  {2 \cdot 2 \cdot (2 \cdot 2 + 1)} } & = &
    \sqrt{\displaystyle\frac{1}{10}}, \\
  (2 1 1 \big| \,0,  -1, -1) & = &
    \sqrt{\displaystyle\frac{(j_{a}+m+1)\,(j_{a}+m)} {2j_{a}\,(2j_{a}+1)} } & = &
    \sqrt{\displaystyle\frac{(2-1+1)\,(2-1)}  {2 \cdot 2 \cdot (2 \cdot 2 + 1)} } & = &
    \sqrt{\displaystyle\frac{1}{10}}, \\
  (2 1 1 \big|  -2, \,1, -1) & = &
    \sqrt{\displaystyle\frac{(j_{a}-m)\,(j_{a}-m+1)} {2j_{a}\,(2j_{a}+1)} } & = &
    \sqrt{\displaystyle\frac{(2+1)\,(2+1+1)}  {2 \cdot 2 \cdot (2 \cdot 2 + 1)} } & = &
    \sqrt{\displaystyle\frac{3}{5}}.
\end{array}
\label{eq.app.3.3}
\end{equation}
%-----------------------------------------------------------------------------------------------------------------------

%-----------------------------------------------------------------------------------------------------------------------
\subsection{Coefficients $C_{l_{f} l_{ph} n}^{m \mu^{\prime}}$
\label{app.4}}

We define coefficients $C_{l_{f} l_{ph} n}^{m \mu^{\prime}}$ so:
\begin{equation}
  C_{l_{f} l_{ph} n}^{m \mu^{\prime}} =
    (-1)^{l_{f}+n+1 - \mu^{\prime} + \frac{|m+\mu^{\prime}|}{2}} \;
    (n, 1, l_{ph} \big| -m-\mu^{\prime}, \mu^{\prime}, -m) \;
    \sqrt{\displaystyle\frac{(2l_{f}+1)\,(2n+1)}{32\pi}\;
          \displaystyle\frac{(l_{f}-1)!}{(l_{f}+1)!} \;
          \displaystyle\frac{(n-|m+\mu^{\prime}|)!}{(n+|m+\mu^{\prime}|)!}}
\label{eq.app.4.1}
\end{equation}

Let's consider case of $l_{f}=1$, $l_{ph}=1$ and $n=0$. From(\ref{eq.3.5.9}) we obtain:
\begin{equation}
  m = - \mu^{\prime} = \pm 1.
\label{eq.app.4.2}
\end{equation}
The coefficient $C_{l_{f} l_{ph} n}^{m \mu^{\prime}}$ is:
\begin{equation}
\begin{array}{lcl}
  C_{110}^{m \mu^{\prime}} & = &
    (-1)^{1 + 0 + 1 - \mu^{\prime} + 0} \;
    (0 1 1 \big| \;0, \mu^{\prime}, -m) \;
    \sqrt{\displaystyle\frac{(2 \cdot 1 +1)\,(2\cdot 0+1)}{32\pi}\;
          \displaystyle\frac{(1-1)!}{(1+1)!} \;
          \displaystyle\frac{(0-0)!}{(0+0)!}} = \\
  & = &
    - \sqrt{\displaystyle\frac{3}{64\pi}} \cdot (0 1 1 \big| \; 0, \mu^{\prime}, -m).
\end{array}
\label{eq.app.4.3}
\end{equation}
Taking into account values (\ref{eq.app.3.1}) for the following Clebsch-Gordan coefficients:
\[
\begin{array}{lclcl}
  (0 1 1 \big| \;0, 1, 1)  & = & (0 1 1 \big| \;0, -1, -1) & = & \sqrt{\displaystyle\frac{1}{2}},
\end{array}
\]
we obtain:
\begin{equation}
\begin{array}{lclcccl}
  C_{110}^{-1 -1} & = & 0, & & & & \\

  C_{110}^{-1 1} & = &
    -\sqrt{\displaystyle\frac{3}{64\pi}} \cdot (0 1 1 \big| \;0, 1, 1) & = &
    -\sqrt{\displaystyle\frac{3}{128\,\pi}} & = &
    -\displaystyle\frac{1}{8} \cdot \sqrt{\displaystyle\frac{3}{2\,\pi}}, \\

  C_{110}^{1 -1} & = &
    -\sqrt{\displaystyle\frac{3}{64\pi}} \cdot (0 1 1 \big| \;0, -1, -1) & = &
    -\sqrt{\displaystyle\frac{3}{128\,\pi}} & = &
    -\displaystyle\frac{1}{8} \cdot \sqrt{\displaystyle\frac{3}{2\,\pi}}, \\

  C_{110}^{11} & = & 0. & & & &
\end{array}
\label{eq.app.4.4}
\end{equation}

At $l_{f}=1$, $l_{ph}=1$ and $n=1$ property (\ref{eq.app.4.2}) is fulfilled also. The coefficients $C_{l_{f} l_{ph} n}^{m \mu^{\prime}}$ obtain the form:
\begin{equation}
\begin{array}{lcl}
  C_{111}^{m \mu^{\prime}} & = &
    (-1)^{1 + 1 + 1 - \mu^{\prime} + 0} \;
    (1 1 1 \big| \;0, \mu^{\prime}, -m) \;
    \sqrt{\displaystyle\frac{(2 \cdot 1 +1)\,(2\cdot 1+1)}{32\pi}\;
          \displaystyle\frac{(1-1)!}{(1+1)!} \;
          \displaystyle\frac{(1-0)!}{(1+0)!}} = \\
  & = &
    \sqrt{\displaystyle\frac{9}{64\pi}} \cdot (1 1 1 \big| \; 0, \mu^{\prime}, -m).
\end{array}
\label{eq.app.4.5}
\end{equation}
Taking into account values in (\ref{eq.app.3.2}) for the following Clebsch-Gordan coefficients:
\[
\begin{array}{cc}
  (1 1 1 \big| \;0, 1, 1)  = -\sqrt{\displaystyle\frac{1}{2}}, &
  (1 1 1 \big| \;0, -1, -1) = \sqrt{\displaystyle\frac{1}{2}},
\end{array}
\]
we find:
\begin{equation}
\begin{array}{lclcl}
  C_{111}^{-1 -1} & = & 0, & & \\

  C_{111}^{-1 1} & = &
    \sqrt{\displaystyle\frac{9}{64\pi}} \cdot (1 1 1 \big| \;0, 1, 1) & = &
    -\displaystyle\frac{3}{8} \cdot \sqrt{\displaystyle\frac{1}{2\,\pi}}, \\

  C_{111}^{1 -1} & = &
    \sqrt{\displaystyle\frac{9}{64\pi}} \cdot (1 1 1 \big| \:0, -1, -1) & = &
    \displaystyle\frac{3}{8} \cdot \sqrt{\displaystyle\frac{1}{2\,\pi}}, \\

  C_{111}^{11} & = & 0. & &
\end{array}
\label{eq.app.4.6}
\end{equation}

At $l_{f}=1$, $l_{ph}=1$ and $n=2$ the property (\ref{eq.app.4.2}) is not fulfilled. We have:
\begin{equation}
\begin{array}{lcl}
  C_{112}^{m \mu^{\prime}} & = &
    (-1)^{1 + 2 + 1 - \mu^{\prime} + \frac{|m+\mu^{\prime}|}{2}} \;
    (2 1 1 \big| -m-\mu^{\prime}, \mu^{\prime}, -m) \;
    \sqrt{\displaystyle\frac{(2\cdot 1+1)\,(2\cdot 2+1)}{32\,\pi}\;
          \displaystyle\frac{(1-1)!}{(1+1)!} \;
          \displaystyle\frac{(2-|m+\mu^{\prime}|)!}{(2+|m+\mu^{\prime}|)!}} = \\
  & = &
    (-1)^{- \mu^{\prime} + \frac{|m+\mu^{\prime}|}{2}} \;
    \sqrt{\displaystyle\frac{15}{64\,\pi}\;
          \displaystyle\frac{(2-|m+\mu^{\prime}|)!}{(2+|m+\mu^{\prime}|)!}} \cdot
    (2 1 1 \big| -m-\mu^{\prime}, \mu^{\prime}, -m).
\end{array}
\label{eq.app.4.7}
\end{equation}
Rewrite at different $m = \pm 1$ and $\mu^{\prime} = \pm 1$:
\begin{equation}
\begin{array}{lcl}
  C_{112}^{-1 -1} & = &
    (-1)^{1 + \frac{|-1-1|}{2}} \;
    \sqrt{\displaystyle\frac{15}{64\,\pi}\;
          \displaystyle\frac{(2-|-1-1|)!}{(2+|-1-1|)!}} \cdot
    (2 1 1 \big| \;1+1, -1, 1) = \\
  & = &
    \sqrt{\displaystyle\frac{15}{64\,\pi}\; \displaystyle\frac{1}{4!}} \cdot (2 1 1 \big| \;2, -1, 1) =
    \displaystyle\frac{1}{16} \: \sqrt{\displaystyle\frac{5}{2\,\pi}} \cdot (2 1 1 \big| \;2, -1, 1), \\

  C_{112}^{-1 1} & = &
    (-1)^{-1 + \frac{|-1+1|}{2}} \;
    \sqrt{\displaystyle\frac{15}{64\,\pi}\; \displaystyle\frac{(2-|-1+1|)!}{(2+|-1+1|)!}} \cdot
    (2 1 1 \big| \;1-1, 1, 1) = \\
  & = &
    - \sqrt{\displaystyle\frac{15}{64\,\pi}} \cdot (2 1 1 \big| \;0 1 1) =
    - \displaystyle\frac{1}{8} \sqrt{\displaystyle\frac{15}{\pi}} \cdot (2 1 1 \big| \;0 1 1), \\

  C_{112}^{1 -1} & = &
    (-1)^{1 + \frac{|1-1|}{2}} \;
    \sqrt{\displaystyle\frac{15}{64\,\pi}\; \displaystyle\frac{(2-|1-1|)!}{(2+|1-1|)!}} \cdot
    (2 1 1 \big| -1+1, -1, -1) = \\
  & = &
    - \sqrt{\displaystyle\frac{15}{64\,\pi}} \cdot (2 1 1 \big| \;0, -1, -1) =
    - \displaystyle\frac{1}{8} \sqrt{\displaystyle\frac{15}{\pi}} \cdot (2 1 1 \big| \;0, -1, -1), \\

  C_{112}^{11} & = &
    (-1)^{-1 + \frac{|1+1|}{2}} \;
    \sqrt{\displaystyle\frac{15}{64\,\pi}\; \displaystyle\frac{(2-|1+1|)!}{(2+|1+1|)!}} \cdot
    (2 1 1 \big| -1-1, 1, -1) = \\
  & = &
    \sqrt{\displaystyle\frac{15}{64\,\pi}\; \displaystyle\frac{1}{4!}} \cdot (2 1 1 \big| \:-2, 1, -1) =
    \displaystyle\frac{1}{16} \: \sqrt{\displaystyle\frac{5}{2\,\pi}} \cdot (2 1 1 \big| \:-2, 1, -1).
\end{array}
\label{eq.app.4.8}
\end{equation}
Using the found values in (\ref{eq.app.3.3}) for Clebsch-Gordan coefficients:
\[
\begin{array}{ll}
  (2 1 1 \big| \,2,  -1, \,1) = \sqrt{\displaystyle\frac{3}{5}}, &
  (2 1 1 \big| \,0, \,1, \,1) = \sqrt{\displaystyle\frac{1}{10}}, \\
  (2 1 1 \big| \,0,  -1, -1)  = \sqrt{\displaystyle\frac{1}{10}}, &
  (2 1 1 \big| \,-2, \,1, -1) = \sqrt{\displaystyle\frac{3}{5}},
\end{array}
\]
we obtain:
\begin{equation}
\begin{array}{lcccccl}
  C_{112}^{-1 -1} & = &
    \displaystyle\frac{1}{16} \: \sqrt{\displaystyle\frac{5}{2\,\pi}} \cdot (2 1 1 \big| \;2, -1, 1) & = &
    \displaystyle\frac{1}{16} \: \sqrt{\displaystyle\frac{5}{2\,\pi}} \cdot \sqrt{\displaystyle\frac{3}{5}} & = &
    \displaystyle\frac{1}{16} \: \sqrt{\displaystyle\frac{3}{2\,\pi}}, \\

  C_{112}^{-1 1} & = &
    - \displaystyle\frac{1}{8} \sqrt{\displaystyle\frac{15}{\pi}} \cdot (2 1 1 \big| \;0 1 1) & = &
    - \displaystyle\frac{1}{8} \sqrt{\displaystyle\frac{15}{\pi}} \cdot \sqrt{\displaystyle\frac{1}{10}} & = &
    - \displaystyle\frac{1}{8} \sqrt{\displaystyle\frac{3}{2\,\pi}}, \\

  C_{112}^{1 -1} & = &
    - \displaystyle\frac{1}{8} \sqrt{\displaystyle\frac{15}{\pi}} \cdot (2 1 1 \big| \;0, -1, -1) & = &
    - \displaystyle\frac{1}{8} \sqrt{\displaystyle\frac{15}{\pi}} \cdot \sqrt{\displaystyle\frac{1}{10}} & = &
    - \displaystyle\frac{1}{8} \sqrt{\displaystyle\frac{3}{2\,\pi}}, \\

  C_{112}^{11} & = &
    \displaystyle\frac{1}{16} \: \sqrt{\displaystyle\frac{5}{2\,\pi}} \cdot (2 1 1 \big| \:-2, 1, -1) & = &
    \displaystyle\frac{1}{16} \: \sqrt{\displaystyle\frac{5}{2\,\pi}} \cdot \sqrt{\displaystyle\frac{3}{5}} & = &
    \displaystyle\frac{1}{16} \sqrt{\displaystyle\frac{3}{2\,\pi}}.
\end{array}
\label{eq.app.4.9}
\end{equation}

Write the values for the calculated coefficients:
\begin{equation}
\begin{array}{llll}
  \vspace{2mm}
  C_{110}^{-1 -1} = 0, &
  C_{110}^{-1 1} = -\displaystyle\frac{1}{8} \cdot \sqrt{\displaystyle\frac{3}{2\,\pi}}, &
  C_{110}^{1 -1} = -\displaystyle\frac{1}{8} \cdot \sqrt{\displaystyle\frac{3}{2\,\pi}}, &
  C_{110}^{11} = 0; \\

  \vspace{2mm}
  C_{111}^{-1 -1} = 0, &
  C_{111}^{-1 1} = -\displaystyle\frac{3}{8} \cdot \sqrt{\displaystyle\frac{1}{2\,\pi}}, &
  C_{111}^{1 -1} = \displaystyle\frac{3}{8} \cdot \sqrt{\displaystyle\frac{1}{2\,\pi}}, &
  C_{111}^{11} = 0; \\

  \vspace{2mm}
  C_{112}^{-1 -1} = \displaystyle\frac{1}{16} \: \sqrt{\displaystyle\frac{3}{2\,\pi}}, &
  C_{112}^{-1 1} = - \displaystyle\frac{1}{8} \sqrt{\displaystyle\frac{3}{2\,\pi}}, &
  C_{112}^{1 -1} = - \displaystyle\frac{1}{8} \sqrt{\displaystyle\frac{3}{2\,\pi}}, &
  C_{112}^{11} = \displaystyle\frac{1}{16} \sqrt{\displaystyle\frac{3}{2\,\pi}}.
\end{array}
\label{eq.app.4.10}
\end{equation}
%-----------------------------------------------------------------------------------------------------------------------

%-----------------------------------------------------------------------------------------------------------------------
\subsection{Functions $f_{l_{f}n}^{m \mu^{\prime}}(\theta)$
\label{app.5}}

We define function $f_{l_{f}n}^{m \mu^{\prime}}(\theta)$ so:
\begin{equation}
  f_{l_{f} n}^{m \mu^{\prime}}(\theta) =
    P_{l_{f}}^{1} (\cos{\theta}) \;  P_{1}^{1} (\cos{\theta}) \;  P_{n}^{|m+\mu^{\prime}|}(\cos{\theta}).
\label{eq.app.5.1}
\end{equation}
At $l_{f}=1$ and $n=0,1,2$ we obtain:
\begin{equation}
\begin{array}{lcl}
  f_{10}^{m \mu^{\prime}}(\theta) & = &
    P_{1}^{1} (\cos{\theta}) \; P_{1}^{1} (\cos{\theta}) \; P_{0}^{|m+\mu^{\prime}|} (\cos{\theta}), \\
  f_{11}^{m \mu^{\prime}}(\theta) & = &
    P_{1}^{1} (\cos{\theta}) \; P_{1}^{1} (\cos{\theta}) \; P_{1}^{|m+\mu^{\prime}|} (\cos{\theta}), \\
  f_{12}^{m \mu^{\prime}}(\theta) & = &
    P_{1}^{1} (\cos{\theta}) \; P_{1}^{1} (\cos{\theta}) \; P_{2}^{|m+\mu^{\prime}|} (\cos{\theta}).
\end{array}
\label{eq.app.5.2}
\end{equation}
We rewrite at different values of $m = \pm 1$ and $\mu^{\prime} = \pm 1$:
\begin{equation}
\begin{array}{lclcl}
  f_{10}^{-1, -1}(\theta) & = &
    P_{1}^{1} (\cos{\theta}) \;  P_{1}^{1} (\cos{\theta}) \;  P_{0}^{2} (\cos{\theta}) & = & 0, \\
  f_{10}^{-1 1}(\theta) & = &
    P_{1}^{1} (\cos{\theta}) \;  P_{1}^{1} (\cos{\theta}) \;  P_{0}^{0} (\cos{\theta}) & = &
    \sin{\theta} \cdot \sin{\theta} \cdot 1 =
    \sin^{2}{\theta}, \\
  f_{10}^{1 -1}(\theta) & = &
    P_{1}^{1} (\cos{\theta}) \;  P_{1}^{1} (\cos{\theta}) \;  P_{0}^{0} (\cos{\theta}) & = &
    \sin{\theta} \cdot \sin{\theta} \cdot 1 =
    \sin^{2}{\theta}, \\
  \vspace{3mm}
  f_{10}^{11}(\theta) & = &
    P_{1}^{1} (\cos{\theta}) \;  P_{1}^{1} (\cos{\theta}) \;  P_{0}^{2} (\cos{\theta}) & = & 0; \\

  f_{11}^{-1, -1}(\theta) & = &
    P_{1}^{1} (\cos{\theta}) \;  P_{1}^{1} (\cos{\theta}) \;  P_{1}^{2} (\cos{\theta}) & = & 0, \\
  f_{11}^{-1 1}(\theta) & = &
    P_{1}^{1} (\cos{\theta}) \;  P_{1}^{1} (\cos{\theta}) \;  P_{1}^{0} (\cos{\theta}) & = &
    \sin{\theta} \cdot \sin{\theta} \cdot \cos{\theta} =
    \sin^{2}{\theta} \cos{\theta}, \\
  f_{11}^{1 -1}(\theta) & = &
    P_{1}^{1} (\cos{\theta}) \;  P_{1}^{1} (\cos{\theta}) \;  P_{1}^{0} (\cos{\theta}) & = &
    \sin{\theta} \cdot \sin{\theta} \cdot \cos{\theta} =
    \sin^{2}{\theta} \cos{\theta}, \\
  \vspace{3mm}
  f_{11}^{1 1}(\theta) & = &
    P_{1}^{1} (\cos{\theta}) \;  P_{1}^{1} (\cos{\theta}) \;  P_{1}^{2} (\cos{\theta}) & = & 0; \\

  f_{12}^{-1, -1}(\theta) & = &
    P_{1}^{1} (\cos{\theta}) \;  P_{1}^{1} (\cos{\theta}) \;  P_{2}^{2} (\cos{\theta}) & = &
    \sin{\theta} \cdot \sin{\theta} \cdot 3 \sin^{2}{\theta} =
    3 \sin^{4}{\theta}, \\
  f_{12}^{-1 1}(\theta) & = &
    P_{1}^{1} (\cos{\theta}) \;  P_{1}^{1} (\cos{\theta}) \;  P_{2}^{0} (\cos{\theta}) & = &
    \sin{\theta} \cdot \sin{\theta} \cdot \frac{1}{2} (3\cos^{2}{\theta} - 1) =
    \frac{1}{2} \sin^{2}{\theta} \: (3\cos^{2}{\theta} - 1), \\
  f_{12}^{1 -1}(\theta) & = &
    P_{1}^{1} (\cos{\theta}) \;  P_{1}^{1} (\cos{\theta}) \;  P_{2}^{0} (\cos{\theta}) & = &
    \sin{\theta} \cdot \sin{\theta} \cdot \frac{1}{2} (3\cos^{2}{\theta} - 1) =
    \frac{1}{2} \sin^{2}{\theta} \: (3\cos^{2}{\theta} - 1), \\
  f_{12}^{11}(\theta) & = &
    P_{1}^{1} (\cos{\theta}) \;  P_{1}^{1} (\cos{\theta}) \;  P_{2}^{2} (\cos{\theta}) & = &
    \sin{\theta} \cdot \sin{\theta} \cdot 3 \sin^{2}{\theta} =
    3 \sin^{4}{\theta}.
\end{array}
\label{eq.app.5.3}
\end{equation}
\newpage
\bibliography{Bremsstrahlung_full_version_eng}
%-----------------------------------------------------------------------------------------------------------------------

%-----------------------------------------------------------------------------------------------------------------------
\end{document}